\documentclass[12pt]{article}
\usepackage{amssymb}
\usepackage{amsmath}
\usepackage{amsthm}
\usepackage{cancel}
\usepackage{slashed}
\usepackage{fullpage}
\usepackage{color}
\usepackage{braket}
\usepackage{graphicx}
\usepackage{multirow}
\usepackage{verbatim}
\usepackage{cite}
\usepackage{bigints}
\usepackage{simplewick}
\usepackage{authblk}

\usepackage{caption}
\usepackage{subcaption}

\usepackage{feynmp}

\usepackage{hyperref}
\hypersetup{
    linktocpage,
     colorlinks,
     citecolor=darkgreen,
     linkcolor= darkgreen,
     urlcolor=darkgreen
}

\usepackage{pdfsync}
\usepackage{tikz}
\usetikzlibrary{shapes,backgrounds}
\usepackage{verbatim}
\usetikzlibrary{decorations.pathmorphing, decorations.shapes}
\usepackage{xcolor}
\usetikzlibrary{arrows.meta}

\makeatletter

\pgfdeclaredecoration{gluon}{coil}
{
  \state{coil}[switch if less than=%
    0.5\pgfdecorationsegmentlength+
    \pgfdecorationsegmentaspect\pgfdecorationsegmentamplitude+%
    \pgfdecorationsegmentaspect\pgfdecorationsegmentamplitude to last,
               width=+\pgfdecorationsegmentlength]
  {
    \pgfpathcurveto
    {\pgfpoint@oncoil{0    }{ 0.555}{1}}
    {\pgfpoint@oncoil{0.445}{ 1    }{2}}
    {\pgfpoint@oncoil{1    }{ 1    }{3}}
    \pgfpathcurveto
    {\pgfpoint@oncoil{1.555}{ 1    }{4}}
    {\pgfpoint@oncoil{2    }{ 0.555}{5}}
    {\pgfpoint@oncoil{2    }{ 0    }{6}}
    \pgfpathcurveto
    {\pgfpoint@oncoil{2    }{-0.555}{7}}
    {\pgfpoint@oncoil{1.555}{-1    }{8}}
    {\pgfpoint@oncoil{1    }{-1    }{9}}
    \pgfpathcurveto
    {\pgfpoint@oncoil{0.445}{-1    }{10}}
    {\pgfpoint@oncoil{0    }{-0.555}{11}}
    {\pgfpoint@oncoil{0    }{ 0    }{12}}
  }
  \state{last}[next state=final]
  {
    \pgfpathcurveto
    {\pgfpoint@oncoil{0    }{ 0.555}{1}}
    {\pgfpoint@oncoil{0.445}{ 1    }{2}}
    {\pgfpoint@oncoil{1    }{ 1    }{3}}
    \pgfpathcurveto
    {\pgfpoint@oncoil{1.555}{ 1    }{4}}
    {\pgfpoint@oncoil{2    }{ 0.555}{5}}
    {\pgfpoint@oncoil{2    }{ 0    }{6}}
  }
  \state{final}{}
}

\def\pgfpoint@oncoil#1#2#3{%
  \pgf@x=#1\pgfdecorationsegmentamplitude%
  \pgf@x=\pgfdecorationsegmentaspect\pgf@x%
  \pgf@y=#2\pgfdecorationsegmentamplitude%
  \pgf@xa=0.083333333333\pgfdecorationsegmentlength%
  \advance\pgf@x by#3\pgf@xa%
}

\makeatother

\def\blobcolor{gray}

\usetikzlibrary{decorations.shapes}
\tikzset{myarrow/.style 2 args={
        decoration={markings,
            mark= at position #1 with {\arrow{#2}} ,
        },
        postaction={decorate}
    }
}

\tikzset{decorate glaubr/.style= {
 decorate, decoration={shape backgrounds,shape=circle,shape size=1pt,shape sep=2pt}, line width=0.3pt }}

\DeclareRobustCommand{\Fig}[1]{Fig.~\ref{#1}}

\newcommand{\dbar}{d}

\def\nn{\nonumber}

\def\cB{\mathcal{B}}
\def\cC{\mathcal{C}}

\def\cM{\mathcal{M}}

\def\cO{\mathcal{O}}
\def\cP{\mathcal{P}}

\def\tr{{\rm tr}}

\def\dg{\dagger}

\def\slash{\!\!\!/}

\newcommand{\cn}[1]{\cancel{#1}}

\newcommand{\Li}{\text{Li}}
\def\be{\begin{equation}}
\def\ee{\end{equation}}

\def\l{\langle}
\def\r{\rangle}

\newcommand{\Eq}[1]{Eq.~\eqref{#1}}

\allowdisplaybreaks

\def\LPeq{\cong}
\def\LPFeq{\cong_{\text{\tiny{IR}}}}

\definecolor{darkred}{rgb}{0.8,0.0,0.0}
\definecolor{darkblue}{rgb}{0.0,0.0,0.9}
\definecolor{darkerblue}{rgb}{0.0,0.0,0.5}
\definecolor{darkgreen}{rgb}{0.0,0.5,0.0}
\definecolor{black}{rgb}{0.0,0.0,0.0}
\definecolor{brown}{rgb}{0.6,0.4,0.2}
\newcommand{\red}{\color{darkred}}
\newcommand{\blue}{\color{darkblue}}
\newcommand{\green}{\color{darkgreen}}
\newcommand{\black}{\color{black}}

\newcommand{\ccol}{\color{darkblue}}

\newcommand{\spincol}{\color{brown}}
\newcommand{\softcol}{\color{darkred}}

\newcommand{\scs}{ {{\softcol s}} }

\newcommand{\ccTH}{ {{\ccol 3}} }
\newcommand{\ccOT}{ {({\ccol 12})} }
\newcommand{\ccO}{ {{\ccol 1}} }
\newcommand{\ccT}{ {{\ccol 2}} }
\newcommand{\cci}{ {{\ccol i}} }
\newcommand{\ccj}{ {{\ccol j}} }
\newcommand{\cck}{ {{\ccol k}} }
\newcommand{\rN}{ {{\ccol N}} }

\newcommand{\ccn}{ {{\ccol n}} }
\newcommand{\ccN}{ {{\ccol N}} }
\newcommand{\ccM}{ {{\ccol M}} }

\newcommand{\cdt}{ \hspace{-0.5mm}\cdot\hspace{-0.5mm}}

\makeatletter
\newcommand*{\hyperlinkcite}[1]{\hyper@link{cite}{cite.#1\@extra@b@citeb}}
\makeatother

\newcommand{\Sp}{ \mathbf{Sp} }

\newcommand{\TT}{ \mathbf{T} } 
\newcommand{\TX}{ \mathbf{X} }

\newcommand{\Sij}{ S_{\cci \ccj}}

\newcommand{\bI}{ {\mathbf I}}
\newcommand{\bD}{ {\mathbf \Delta}}
\newcommand{\bIb}{ { \overline\bI}}

\def\zero{0}
\def\one{1}

\def\fme{\varepsilon}
\def\pol{\varepsilon}
\def\ep{\epsilon}

\newcommand{\Spnf}{ \Sp^{\text{non-fact}}}
\newcommand{\Spfa}{ \Sp^{\text{fact}}}

\def\cMb{\overline{\mathcal{M}}}

\title{Collinear factorization violation and effective field theory}

\author{Matthew D. Schwartz${}^1$}
\author{Kai Yan${}^1$}
\author{Hua Xing Zhu${}^2$}

\affil{${}^1${\small \emph{Department of Physics,
Harvard University, Cambridge, MA 02138, USA}}}
\affil{${}^2${\small \emph{Center for Theoretical Physics,
Massachusetts Institute of Technology, Cambridge, MA 02139, USA}}
}

\begin{document} 

\begin{fmffile}{feyngraph}
\unitlength = 1mm

\maketitle

\begin{abstract}
The factorization of amplitudes into hard, soft and collinear parts is known to be violated in situations where incoming particles are collinear to outgoing ones. This result was first derived by studying limits where non-collinear particles become collinear. 
We show that through an effective field theory framework with Glauber operators, these factorization-violating effects can  be reproduced from
an amplitude that is factorized before the splitting occurs. We confirm results at one-loop, through single Glauber exchange, and at two-loops, through double Glauber exchange. To approach the calculation, we begin by reviewing the importance of Glauber scaling for factorization. We show that for any situation where initial state and final state particles are not collinear, the Glauber contribution is entirely contained in the soft contribution. The contributions coming from Glauber operators are necessarily non-analytic functions of external momentum, with the non-analyticity arising from the rapidity regulator. The non-analyticity is critical so that Glauber operators can both preserve factorization when it holds and produce factorization-violating effects when they are present.

\end{abstract}

\newpage
\tableofcontents

\newpage

\section{Introduction}
That scattering cross sections factorize is critical to the predictive power of quantum field theory at particle colliders. At proton colliders, such as the Large Hadron Collider currently running at CERN, factorization allows the computation of a cross section to be written as the product or convolution of separate components: a hard part, sensitive only to the net momenta going into particular directions, a soft part, describing the distribution of radiation of parametrically lower energy than going into the hard collision, a collinear part, describing the branching of partons into sets of partons going in nearly the same direction, parton distribution functions, describing the probability of finding a parton with a given fraction of the proton's energy, and fragmentation functions, describing the probability of a quark or gluon to transition into a given color-neutral baryon or meson. Then each of these components can be computed or measured separately.

While the above qualitative description of factorization certainly holds to some extent, it is sometimes violated. 
The violation might contribute only a phase to the amplitude and therefore cancel at cross-section level, or it might contribute
to the magnitude of the amplitude and have observable consequences. An  example such consequence is the appearance of super-leading logarithms~\cite{Forshaw:2006fk,Forshaw:2008cq,Keates:2009dn}. 
These are contributions to a cross section in a region with no hard particles that scale like $\alpha^n \ln^m x$  with $m>n$ for some observable $x$. The prediction from factorization is that only soft radiation should be relevant in the region (since it has no hard particles) and soft contributions scale at most like $\alpha^n \ln^n x$. 
In this paper, we progress towards understanding when factorization fails 
by connecting different approaches to factorization-violation.

To begin, it is important to be precise about what is meant by factorization. There are certain situations where factorization has been proven
to hold.  For example, amplitudes of quarks and gluons in quantum chromodynamics (QCD) are known to factorize when all the particles are in the final state. For this case, a precise statement of factorization is~\cite{FS1,FS2}
\be
\bra{X} \phi^\star \cdots \phi \ket{0}
 \;\LPeq\; 
\cC(\Sij) \, \frac{\bra{X_\ccO} \phi^\star W_\ccO \ket{0}}{\bra{0} Y_\ccO^\dg W_\ccO \ket{0}} \,\cdots\, \frac{\bra{X_\rN} W_\rN^\dg\phi \ket{0}}{\bra{0} W_\rN^\dg Y_\rN \ket{0}}
\;\bra{X_\scs} Y_\ccO^\dg \cdots Y_\rN \ket{0}
\label{sQEDmain}
\ee
More details of this formula, including spin and color indices, can be found in~\cite{FS2}. Briefly,
 $\bra{X}= \bra{X_\scs}\bra{X_\ccO}\cdots\bra{X_\ccN}$ is a state comprising quarks or gluons all of which are either soft, in $\bra{X_\scs}$, 
or collinear to one of $N$ sectors, in $\bra{X_\ccO}$ through $\bra{X_\ccN}$. The objects $W_\ccj$ and $Y_\ccj$ are collinear and soft Wilson lines (path ordered exponentials of gluon fields). The difference between $W_\ccj$ and $Y_\ccj$ is that the $W_\ccj$ represent the radiation from all the non-collinear sectors and hence point away from the $\ccj$ direction  while the $Y_\ccj$ represent sources for eikonal radiation and  point along the $\ccj$ direction. 
The function $\cC(\Sij)$ is an infrared-finite Wilson coefficient depending only on the net momenta going into each collinear sector. 

The symbol $\LPeq$ means that the
 two sides of Eq.~\eqref{sQEDmain} are equal at leading power. In particular, the infrared divergences (soft and collinear) agree on both sides
 and if one computes some infrared safe observable, such as the sum of the masses $\tau = \sum m_i^2$ of all the jets, then all of the singular parts of the observable distribution will also agree (any term which is singular as $\tau\to 0$). The operator $\cO = \phi^\star \cdots \phi$ is chosen to be a local operator made from scalar fields, for simplicity, but a similar formula holds for gauge-invariant operators made from quarks or gluons, or for scattering amplitudes~\cite{FS2}. Color and spin indices are suppressed. 
 
 In Eq.~\eqref{sQEDmain} all the fields are those of full QCD: ordinary quarks and gluons. This formula is closely related to factorization in Soft-Collinear Effective Theory (SCET)~\cite{Bauer:2000ew,Bauer:2001ct,Bauer:2002nz,Beneke:2002ni,Beneke:2002ph}. In SCET, only leading power interactions among fields are kept in the SCET Lagrangian, making the Feynman rules more complicated than those in QCD. Also in SCET,  the soft-collinear overlap is removed through a diagram-by-diagram zero-bin subtraction procedure~\cite{Manohar:2006nz}, rather than through operator-level subtractions, as in Eq.~\eqref{sQEDmain}. The two formulations are equivalent, and also equivalent to factorization formulas in traditional QCD~\cite{Almeida:2014uva}. 
 
 In the derivation of Eq.~\eqref{sQEDmain} in~\cite{FS2}, which is similar to traditional factorization 
 proofs~\cite{Collins:1981uk,Sen:1981sd,Sen:1982bt,Collins:1984kg}, it is essential that the eikonal approximation can be used. The eikonal approximation amounts to equating the limit where  momenta are soft ($k^\mu \to 0$) with the limit where the energies of all the collinear particles are large ($Q \to \infty $). The $Q\to \infty$ limit allows collinear particles to be replaced by classical sources (Wilson lines) so that the soft radiation is insensitive to the structure of the collinear sector. The subtlety is that there are regions of virtual momenta phase-space in which 
all of the components  are soft, $k^\mu  \ll Q$, but $Q$ is still relevant. For example, $k^\mu \to 0$ holding $Q = {\vec{k}_\perp^2}/{k_0}$ fixed is not possible after the eikonal limit ($Q \to \infty$) has been taken. It is phase space regions associated with limits like this 
that are dangerous for factorization, as we will review.

In Section~\ref{sec:contain} we show that the only time the Glauber scaling subtlety might spoil amplitude-level factorization is when there are initial states collinear to final states. This is of course a well-known result, but it is helpful to revisit the derivation to set the stage for investigations into factorization-violation. 
 If we explicitly avoid such configurations, the factorization formula in Eq.~\eqref{sQEDmain} has a natural generalization. Let $\bra{X}=\bra{X_\ccO,\cdots X_\ccN}$ be the collinear sectors of final state and $\ket{Z} = \ket{Z_{\ccN+1} \cdots Z_{\ccM}}$ be the collinear sectors of the initial state and let $\bra{X_\scs}$ and $\ket{Z_\scs}$ be the soft particles in the initial and final states respectively.\footnote{Soft particles in the initial state are not particularly interesting physically, but since factorization holds if they are included, we allow for $\ket{Z_\scs}$ to be nonzero for  completeness}
 We assume no two sectors are collinear. Note that this is not a strong requirement -- for any set of external momenta, one can always define the threshold $\lambda$ for collinearity to be so small that each momentum is in its own sector (the only time this cannot be done is if two momenta are exactly proportional). Then Eq.~\eqref{sQEDmain} generalizes to\footnote{For physical, positive-energy momenta, incoming Wilson lines
 are denoted with a bar, like $\bar{Y}_\ccj$ or $\bar{W}_\ccj$ (see~\cite{FS1}) and have a different $i\fme$ prescription than outgoing, un-barred, Wilson lines. In this paper, our convention is that all momenta are outgoing and so distinguishing incoming and outgoing Wlison lines is unnecssary.} 
 \be
\bra{X} \phi^\star \cdots \phi \ket{Y}
 \;\LPeq\; 
\cC(\Sij) \, \frac{\bra{X_\ccO} \phi^\star W_\ccO \ket{0}}{\bra{0} Y_\ccO^\dg W_\ccO \ket{0}} \,\cdots\, \frac{\bra{0} W_\rN^\dg\phi \ket{Z_\ccM}}{\bra{0} W_\rN^\dg Y_\rN \ket{0}}
\;\bra{X_\scs} Y_\ccO^\dg \cdots Y_\rN \ket{Z_\scs}
\label{sQEDmainXY}
\ee
This is a non-trivial generalization of Eq.~\eqref{sQEDmain}. It requires the Glauber gluon contributions to be identical on both sides, even when there are initial states involved.

The reason the factorization formula in Eq.~\eqref{sQEDmainXY} holds is because for hard scattering with no initial state collinear to any final state, the Glauber contribution is always contained in the eikonal contribution. 
This containment holds in QCD when contours can be deformed out of the Glauber region into the eikonal region (where the eikonal approximation can be used). That is, it holds when there is no pinch in the Glauber region. We review these observations in Section~\ref{sec:contain}. That the eikonal limit contains the Glauber contribution is often assumed  to study Glauber effects through matrix elements of Wilson lines
(see e.g.~\cite{Laenen:2015jia}), so it is important to understand when it holds.

In the context of Soft-Collinear Effective Theory, being able to deform contours from the Glauber region into the
eikonal region is closely related to the ``soft-Glauber correspondence" or the ``Cheshire Glauber" discussed in~\cite{Rothstein:2016bsq}.
To be clear, the soft-Glauber correspondence is a statement about when Glauber operators in SCET reproduce the Glauber limit of soft
graphs in SCET. One expects that the soft-Glauber correspondence holds if there is no pinch in the Glauber region, however this has not
been shown. Indeed, it is not even a precise statement, since the pinches are properties of graphs in QCD and the soft-Glauber correspondence refers to SCET. One goal of our analysis is to clarify the relationship between Glauber operators in SCET and pinches in QCD.

 A corollary of Eq.~\eqref{sQEDmainXY} is that infrared divergences of purely virtual graphs in full QCD, including all Glauber contributions, are exactly reproduced by the factorized expression. Factorization violation can only show up in graphs with emissions, that is, in relating an amplitude to one with an additional collinear or soft particle.  In particular, when there are collinear emissions, it is known that factorization does not hold. To be precise, amplitude-level splittings are often described through an operator $\Sp$ which acts on an $n$-body matrix element $\ket{\cMb}$ turning it into a matrix element with $n+1$ partons $\ket{\cM}$:
 \be
 \ket{\cM} = \Sp \cdot \ket{\cMb}
 \ee
In a $1 \to 2$ splitting a parton with momentum $P^\mu$ splits into two partons with momenta
  $p_\ccO^\mu$ and $p_\ccT^\mu$ with $P^\mu \LPeq p_\ccO^\mu + p_\ccT^\mu$. 
  Factorization implies that this splitting function $\Sp$ should depend only on the momenta and colors of particles collinear to the $P^\mu$ direction. This requirement, called {\it strict factorization}, was shown by Catani, de Florian and Rodrigo to be violated in certain situations~\cite{Catani:2011st}. 
 In particular, when $p_\ccO^\mu$ is incoming,  $p_\ccT^\mu$ is outgoing and there is another incoming colored particle with some momentum $p_\ccTH^\mu$ not collinear to $P^\mu$ and another outgoing colored particle with momentum $p_\ccj^\mu$ not collinear to $p_\ccT^\mu$, then $\Sp$ can depend on the color of the $p_\ccTH^\mu$ and $p_\ccj^\mu$ partons. 
 
 We review this calculation of strict factorization violation in Section~\ref{sec:catani}. The approach of~\cite{Catani:2011st}, and also~\cite{Forshaw:2012bi}, is to calculate $\Sp$ by taking the limit of a matrix element with $n+1$ directions as it reduces to a matrix element with $n$ directions. Then $\Sp$ can be deduced from known results about IR singularities of $n+1$ parton matrix elements.  In other words, one calculates $\Sp$ by turning $\ket{\cM}$ into $\ket{\cMb}$. 
 
 The necessity of using $n+1$-body matrix elements to calculate $\Sp$  is a little counterintuitive. Since the splitting originates from $\ket{\cMb}$  it seems one should not need information about a general $n+1$ body matrix element $\ket{\cM}$ to deduce it. Indeed, for timelike splittings
 (as in $e^+e^- \to $ hadrons),
 one can start from factorized expression, like in Eq.~\eqref{sQEDmain}, and caculate $\Sp$ from within the collinear sector of $\ket{\cMb}$. This
 calculation is done at leading order explicitly in~\cite{FS1}.
 In this paper we show that one can still calculate $\Sp$ from $\ket{\cMb}$ when strict factorization is violated, through the inclusion of Glauber operators in the effective theory.

It is natural to propose including Glauber modes~\cite{Liu:2008cc,Donoghue:2009cq,Bauer:2010cc,Fleming:2014rea} into Soft-Collinear Effective Theory (SCET).
However, since Glauber gluons have transverse components much larger than their energies, they cannot be represented by on-shell fields in a Lagrangian, like soft and collinear modes are. Recently, a framework to incorporate Glauber contributions into SCET was proposed by Rothstein and Stewart~\cite{Rothstein:2016bsq}.  The Glauber gluons are introduced not through new on-shell fields, but as potential interactions among pre-existing collinear and soft fields. We briefly review this approach in Section~\ref{sec:scetg}. One of the main new results of this paper is the direct calculation of 1-loop and (partial) 2-loop factorization violating contributions to $\Sp$ in Sections~\ref{sec:fvscet} and \ref{sec:scettwo} with the SCET Glauber formalism. The calculations are highly non-trivial, depending critically on the effective field theory interactions and the rapidity regulator. They therefore provide a satisfying cross check on both SCET and the factorization-violating splitting amplitudes in~\cite{Catani:2011st,Forshaw:2012bi}.

Stepping back from the technical calculations, we make some general observations about properties that the Glauber operator contributions in SCET must have. For example, they must not spoil factorization when factorization holds (as for all outgoing particles). This forces the Glauber contributions to be non-analytic functions of external momentum. It is impressive that this required non-analyticity is exactly produced through the Glauber operators with the non-analytic rapidity regulator. We summarize some of the features of the SCET calculations that allow this to work in Section~\ref{sec:nona}.
 
In this paper, we use the convention that all momenta are outgoing, so that incoming momenta $p^\mu$ have negative energy, $p^0 <0$. With this convention $p_\ccO\cdt p_\ccT < 0$ for a spacelike splitting (one incoming and one outgoing) and $p_\ccO \cdt p_\ccT >0$ for a timelike splitting (both incoming or both outgoing). It also means energy fractions $z= \frac{E_\ccT}{E_\ccO + E_\ccT}$ will be negative for spacelike splittings and positive for timelike splittings. 
In Section~\ref{sec:review}, we review the various modes appearing in hard-soft-colliner factorization of scattering amplitudes, and their scaling. In Section~\ref{sec:contain}, we show that for large-angle hard scattering, the effects of Glauber exchange is contained in the eikonal limit.
Therefore amplitude-level factorization formula is not modified. In Section~\ref{sec:isolate}, we discuss several approaches to isolating the Glauber contribution. In Section~\ref{sec:catani}, we summarize known results about factorization-violation for timelike splitting.
We review the connection between imaginary terms in 1-loop graphs and factorization violation, and 
how the factorization-violating splittings amplitude is derived from the infrared structure of  $n+1$-parton amplitudes in QCD. In Section~\ref{sec:fvscet}, we compute the 1-loop Glauber contributions to timelike splitting amplitudes that are not contained in the soft contributions
using SCET. Both the IR-divergent and the IR-finite contributions to the 1-loop factorization-violating effects are reproduced. 
 In Section~\ref{sec:scettwo}, we compute the contributions to  two-loop factorization violation in timelike splittings involving 
 double Glauber exchange. These contributions exactly reproduce the real part of the 2-loop splitting amplitude from~\cite{Catani:2011st}. 
 We summarize some rather remarkable non-analytic properties of the Glauber contributions  in Section~\ref{sec:nona}
 and summarize our results in Section~\ref{sec:conclusion}.


\section{Elements of factorization and Glauber scaling \label{sec:review}}
Glauber gluons play a central role in understanding violations of factorization, so we devote this section to explaining their origin and relevance. 
Our goal here is to clarify the concepts of scaling, the relationship between soft and Glauber regions of momentum space, and the reasons that Glauber gluons can spoil factorization.

Broadly speaking, the goal of factorization is to write some scattering amplitude $\cM$, which is a function of lightlike external momenta $p_\ccO \cdots p_\ccn$ as
\be
\ket{\cM(p_\ccO, \cdots, p_\ccn)} \LPeq \ket{\cM_\text{factorized}(p_\ccO, \cdots, p_\ccn)} \label{MMf}
\ee
where $\ket{\cM_\text{factorized}}$ is simpler in some way. Here the symbol $\LPeq$ implies that the two are not exactly equal, but equal up to parametrically small power corrections in some function of the momenta (e.g. in $\lambda = {p_\ccO \cdt p_\ccT}/{Q^2}$ with $Q$ the center of mass energy).

  A necessary condition for Eq.~\eqref{MMf} to hold is that the infrared divergences agree on both sides. Since factorization  involves writing an amplitude as products of simpler amplitudes each of which contain some subset of the infrared divergences, a first step to understanding factorization is to classify infrared divergences. 

Classifying the infrared divergences amounts the following consideration. Take a given Feynman diagram written as an integral over various loop momenta $k_i^\mu$.  We associate given values  $k_{i,0}^\mu$ of these momenta to an infrared divergence if the
integral is infinite when integrated in an infinitesimal compact volume around $k_{i,0}^\mu$. The singularity requires a pole in the integration
region, so at least one of the propagators must go on-shell within this volume. More precisely, 
 the pole must be on the integration contour at $\fme=0$ ($\fme$ here refers to the $i\fme$ in the Feynman propagator). However, at small finite $\fme$, the poles are necessarily off the contour of integration, so the singularity only occurs if there are two coalescing poles on different sides of the contour, or a pole at the end-point of the contour. 
This condition is often described as saying that the integration contour cannot be deformed away from the singular region, so that the singularity becomes {\bf pinched} on the integration contour as $\fme \to 0$. 
This necessary condition is encoded in the {\bf Landau equations}~\cite{Landau:1959fi}. For a theory with massless particles with external momenta $p_\ccj^\mu$, the Landau equations imply that the possible values for the infrared singular regions of integration are either soft, $k_{i,0}^\mu=0$ exactly, or collinear $k_{i,0}^\mu =z p_\ccj^\mu$ for some $z$ and some $p_\ccj^\mu$. Thus, we say that the {\bf pinch surface} for a theory with massless particles comprises the soft region ($k_i=0$) and regions collinear to the direction of each external momentum. 

Landau equations are derived using only the denominators in a Feynman diagram. It can certainly happen that the numerator structure makes the diagram infrared-finite even though the contour is pinched according to the denominators. In addition, whether a diagram is divergent  or not can depend on gauge. So the Landau conditions give a necessary but not sufficient condition for a singularity.

Although knowing the pinch surface, that is, the {\it exactly singular} region of loop momenta is a good start, this
surface does not immediately tell us anything about factorization. It does however refine the problem: to match the infrared divergences of a given amplitude, it is enough to match the integral in all the regions around the pinch surface. Thus it inspires us to look for a factorized expression by Taylor expanding the integrand around the pinch surface.

To expand around the $k^\mu_i=0$ part of the pinch surface, we can equivalently expand around the limit where all of the external momenta become infinitely energetic ($|p_\ccj^0| \to \infty$). 
Generically, this lets us replace  propagator involving a loop momentum $k^\mu$ and an external lightlike momentum $p^\mu$ as
\be
\frac{1}{(p+k)^2 + i\fme} \to \frac{1}{2 p\cdot k+ i\fme}  \label{softlimit}
\ee
This replacement is known as taking the  {\bf eikonal limit}. Treating the momentum $p^\mu$ as infinite allows us to ignore the recoil of $p^\mu$ when $k^\mu$ is emitted, so that the $p^\mu$ is essentially a classical background source. This approximation is at the heart of all factorization proofs. 

The subtlety where Glauber scaling comes in is that in taking the eikonal limit, in deriving Eq.~\eqref{softlimit}, it is not enough to have $k^\mu \ll Q$  for all components of $k^\mu$, where $Q=p^0$ is the external particle's energy. Rather, we must also have $k^2 \ll  p\cdt k$. To appreciate the difference between these two limits, rather than taking the limit, as in Eq. \eqref{softlimit}, let us write the propagator as its eikonal version plus a remainder
\be
\frac{1}{(p+k)^2 + i\fme } = { \frac{1}{2p\cdot k + i\fme}}  - { \frac{k^2}{((p+k)^2 +i\fme)(2 p\cdot k+ i\fme)} }\label{softsum}
\ee
This exact relation lets us write a diagram in the full theory as the sum of diagrams, each of which represents an explicit integral with the original propagator replaced by one of these two terms. The first term generates the soft part 
$\;\bra{X_\scs} Y_\ccO^\dg \cdots Y_\rN \ket{0}$
of factorized expressions like Eq.~\eqref{sQEDmain}, and the second term generates collinear parts. Factorization holds only if the collinear parts do not have infrared divergences associated with soft singularities. 

Scaling arguments are powerful tools for determining whether soft singularities are present. 
The replacement in Eq.~\eqref{softlimit} amounts to taking ${k^2}/{k\cdt p} \to 0$. One can apply this limit by rescaling all the loop momenta as $k_i^\mu \to \kappa^2 k^\mu_i$ and keeping the leading terms as $\kappa \to 0$. 
This guarantees that the remaining integral, that is the difference between the original integral and the one with this replacement, must scale like $\kappa^n$ with $n>0$.

Let us call this type of scaling, where all the components of all the loop momenta scale the same way, {\bf eikonal scaling}. We could take $k^\mu \to \kappa^2 k^\mu$, which we call {\bf ultrasoft scaling} or $k^\mu \to \kappa k^\mu$, which we call {\bf soft scaling}.
Soft and ultrasoft scaling are both examples of eikonal scaling and equivalent for 
  determining whether an integral is superficially divergent.  
To see this, consider
  changing variables from $k^\mu$ to $\{\kappa,\Omega\}$, with $\kappa$ the radial variable and  $\Omega$ representing generically some angular variables. Then a diagram which scales like $\kappa^n$ transforms to a $\int d\Omega \int \kappa^{n-1} d \kappa$ integral which is divergent near $\kappa=0$ if and only if $n \le 0$. Using ultrasoft rather than soft scaling would have the diagram scale like $\kappa^{2n}$ which is still divergent under the same conditions.

Next, we must ask whether a diagram could still be divergent when integrated around $k^\mu=0$ pinch surface even if it is power-counting
finite under eikonal scaling. After transforming $k^\mu$ to  $\{\kappa,\Omega\}$, such a  divergence could come from the angular integrals over the $\Omega$'s. A sufficient condition to guarantee infrared finiteness is if the integral scales like $\kappa^n$ with $n>0$ under any possible scalings of the different components of $k^\mu$~\cite{FS2}.

To examine this possibility, let us go to lightcone coordinates. We can decompose an arbitrary momentum $k^\mu$ with respect to two given lightlike momenta $p^\mu$ and $q^\mu$ as 
\be
k^\mu =\frac{1}{Q} k^- p^\mu + \frac{1}{Q} k^+ q^\mu + k_\perp^\mu
\ee
where
\be
 k^- =\frac{2}{Q} k \cdot q, \quad k^+ = \frac{2}{Q} k \cdot p
\ee
with $Q^2 =2 p\cdot q$. 
We also often use the 2-vector perpendicular component $\vec{k}_\perp$ with  $\vec k_\perp^2 = - (k^\mu_\perp)^2$.
Then Eq.~\eqref{softsum} becomes (ignoring the $i\fme$ prescription temporarily)
\be
\frac{1}{(p+k)^2 } = { \frac{1}{Q k^+}}  \left[ 1
- { \frac{ k^+ k^- - \vec k_\perp^2}{Q k^+ + k^+ k^- - \vec k_\perp^2 } } \right]\label{softsumlc}
\ee
Under ultrasoft (eikonal) scaling, 
\be
(k^+,k^-,\vec k_\perp) \to (\kappa^2\, k^+,\kappa^2\, k^-, \kappa^2 \,\vec k_\perp) 
\ee
and  the second term on the right in Eq.~\eqref{softsumlc} is suppressed by a factor of $\kappa^2$ with respect to the first term.
Since diagrams are at most logarithmically divergent, the strongest a divergence can be is $\kappa^0$, and therefore
integrals involving the second term are finite under eikonal scaling~\cite{FS2}.
 
  What other scalings can we consider? We need to send $k^\mu \to 0$, so let us normalize $k^+ \to \kappa^2 k^+$. Then we can generally write $k^- \to \kappa^{2a} k^-$ and $\vec k_\perp\to \kappa^{b} \vec k_\perp$ with $a >  0$ and $b > 0$.  The second term then scales like
\be
{ \frac{ k^+ k^- - \vec k_\perp^2}{Q k^+ + k^+ k^- - \vec k_\perp^2 } } 
 \to
{ \frac{ \kappa^{2+2a}k^+ k^- -\kappa^{2b} \vec k_\perp^2}{\kappa^{2} Q k^+ +\kappa^{2+2a} k^+ k^- -\kappa^{2b}\vec  k_\perp^2 } } 
\label{kappascale}
\ee
Now, if $b>1$, then the $\vec k_\perp^2$ terms in the denominator can be neglected and the integral scales like $\kappa^{\min(2a,2b-2)}$. However, since $a>0$ (so that we approach the soft pinch), this term is power-counting finite.  Thus we must have $b \le 1$. 
For $b \le 1$, the $\vec k_\perp^2$ term dominates both numerator and denominator and this term scales like  
$\kappa^0$. The scaling is independent of $a$ and $b$ (for $b\le 1)$ so we can take $a=2$ and $b=1$ to represent this case. Thus the only possible approach to $k^\mu=0$ which is not automatically infrared finite is
\be
(k^+,k^-,\vec k_\perp) \to (\kappa^2\, k^+,\kappa^2\, k^-, \kappa \,\vec k_\perp) 
\ee
This is known as {\bf Glauber scaling}. It is the only possible scaling towards the $k^\mu=0$ pinch under which the substitution in
Eq.~\eqref{softsum} might not leave an infrared-finite remainder.
Gluons with momenta that have Glauber scaling are often called {\bf Glauber gluons}. These gluons are spacelike and purely virtual:
as $\kappa \to 0$ they approach the soft singularity from a direction in which $k^2  <0$, in contrast with soft or collinear gluons, which can 
have $k^2=0$ for finite $\kappa$.

As an aside,  note that we are taking $a>0$ and $b>0$ so that we zoom in on the soft region of the pinch surface. If we take $a=0$, then the numerator and denominator of Eq.~\eqref{kappascale} both scale the same way and there is no additional suppression. However, if $a=0$ then $k^\mu$ remains finite as $\kappa \to 0$. In fact, it approaches the direction $p^\mu$ of the line that we are expanding around (as in Eq.~\eqref{softsumlc}). Thus this is collinear scaling. We can represent this scaling with $b=1$ so that under {\bf collinear scaling}
\be
(k^+,k^-,\vec k_\perp) \to (\kappa^2\, k^+,  k^-, \kappa \,\vec k_\perp) 
\ee
That the expansion in Eq.~\eqref{softsumlc} does not improve the convergence under collinear scaling is neither surprising nor a problem. That collinear singularities are completely reproduced in the factorized expression was shown in~\cite{FS2}. 


\section{Glauber containment in hard scattering \label{sec:contain}}
In this section, we will show that, for scattering amplitudes, singularities associated with Glauber scaling are already contained in the soft factor when no incoming momentum is collinear to an outgoing momentum.
Because of the simplicity of the pinch surface with massless external particles, all of the relevant issues already appear at 1-loop in a vertex correction graph. Thus we begin with the 1-loop example, then work out the general argument.

\subsection{1-loop example}

To see whether and when Glauber gluon invalidate Eikonal approximation, let us turn now to an explicit example: the Sudakov form factor in scalar QED. The diagram is
\be
I_{\text{full}} = 
\begin{gathered}
\resizebox{20mm}{!}{
     \fmfframe(0,0)(0,0){
\begin{fmfgraph*}(20,35)
\fmfstraight
	\fmfleft{L1}
	\fmfright{R1,R2}
	\fmf{fermion}{L1,v1}
	\fmf{fermion,label=$p_\ccT \searrow$,l.s=right,l.d=0.5mm}{v1,R1}
	\fmf{fermion,label=$p_\ccO \nearrow$,l.s=right,l.d=0.5mm}{R2,v2}
	\fmf{fermion}{v2,L1}
	\fmf{photon,label=$\uparrow k$,tension=0}{v1,v2}
        \fmfv{d.sh=circle, d.f=30,d.si=0.2w}{L1}
\end{fmfgraph*}
}
}
\end{gathered}
=  i g_s^2 \int \frac{d^4k}{(2\pi)^4} 
\frac{ (2p_\ccO - k)^\mu }{ [ (p_\ccO - k)^2 + i\fme]}
\frac{\Pi_{\mu\nu}(k)}{k^2+i\fme}
\frac{(2p_\ccT + k)^\nu}{[(p_\ccT+ k)^2 + i\fme]}
\label{eq:vfull}
\ee
Here, $i\Pi_{\mu\nu}$ is the numerator of the photon propagator. In Feynman gauge, $i\Pi_{\mu\nu} = - i g_{\mu\nu}$. While Feynman gauge is great for actually computing graphs, for studying factorization physical gauges such as lightcone gauge, in which $\Pi_{\mu\nu}$ represents a sum over physical polarizations, can be simpler. In lightcone gauge
\be
\Pi^{\mu\nu}(k) = -g^{\mu\nu} + \frac{r^\mu k^\nu + r^\nu k^\mu}{r\cdot k}
\label{Pimndef}
\ee
with $r^\mu$ a reference lightlike 4-vector. Note that $r_\mu \Pi^{\mu\nu}=0$ and $k_\mu \Pi^{\mu\nu}=\frac{k^2}{r\cdot k}r^\nu$. 

According to the Landau equations, which ignore the numerator structure, a necessary condition for a singularity is that all of the propagators go on-shell at once. This can happen when $k^\mu = z p_\ccO^\mu$ for some $z$, $k^\mu = z p_\ccT^\mu$ for some $z$ or when $k^\mu =0$. 
 Under ultrasoft scaling, the integration measure scales like $\kappa^8$ and the denominator factors scale like $\kappa^2, \kappa^4$ and $\kappa^2$ respectively. The numerator scales like $\kappa^0$ in either Feynman gauge or lightcone gauge, thus this integral is soft-sensitive. Under collinear scaling, the measure scales like $\kappa^4$ and the denominator factors scale like $\kappa^2, \kappa^2$ and $\kappa^0$ respectively. Thus, in Feynman gauge where $\Pi^{\mu\nu} \sim \kappa^0$, the graph is collinear-sensitive. However in lightcone gauge, since
$k_\mu \Pi^{\mu\nu} \sim \frac{k^2}{r\cdot k}r^\nu \sim \kappa^1$ and  $k^\mu  \propto p_\ccO^\mu$ on the collinear pinch surface, there is extra suppression. Thus this graph is actually collinear-finite in physical gauges. (In general, graphs which involve lines connecting different collinear sectors are collinear finite in physical gauges according to Lemma 3 of~\cite{FS2}.) Thus for this graph, integrations around the collinear regions of the pinch surface, $k^\mu = z p_\ccO^\mu$ and $k^\mu = z p_\ccT^\mu$ for $z \ne 0$ are finite. 

\begin{figure}[t]
	\centering
	\includegraphics[width=.4\textwidth]{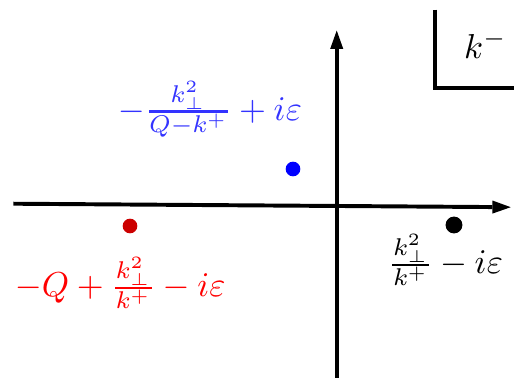}
	\hspace{5mm}
  	\includegraphics[width=.4\textwidth]{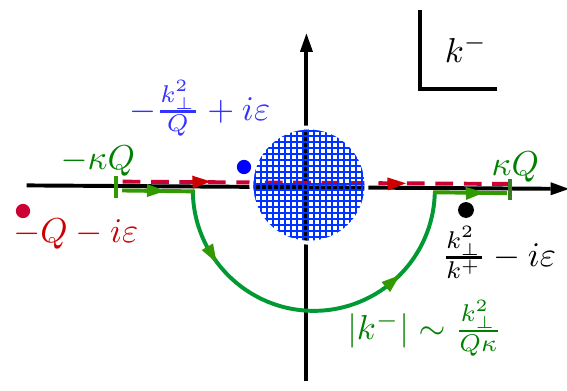}
	\hspace{5mm}
	\caption{
	Left: poles in the complex $k^-$ plane for $I_\text{full}$. Right: the complex $k^-$ plane of $I_\text{full}$ in the region around $k^\mu=0$.
	In this region, all components of $k^\mu$ are less than $\kappa Q$.  The integration contour, $-\kappa Q \le k^- \le \kappa Q$,
	(the dashed red line on the real axis) can be deformed (to the solid green contour) to avoid the Glauber region (hashed blue) for any values
	of $|k^+| < \kappa Q$ and $|\vec{k}_\perp| < \kappa Q$.
	}
\label{fig:sudakovpoles}
\end{figure}

To study the singularity structure of this diagram, it is useful to go to lightcone coordinates with respect to $p_\ccO^\mu$ and $p_\ccT^\mu$, with $k^+ = 2 k \cdot p_\ccT/Q$, $k^- = 2 k \cdot p_\ccO/Q$, and $Q^2 = 2 p_\ccO \cdot p_\ccT$:
\be
I_\text{full}=  i g_s^2 \int \frac{d k^+ d k^- d^2 k_\perp}{2 (2\pi)^4} 
\frac{ (2p_\ccO - k)^\mu }{ [ { -Q k^- + k^+ k^- - \vec k_\perp^2 + i\fme} ]}
\frac{\Pi_{\mu\nu}(k)}{[k^+ k^- - \vec k_\perp^2+i\fme]}
\frac{(2p_\ccT + k)^\nu}{[ { Q k^+ + k^+ k^- - \vec k_\perp^2 + i\fme}]}
\label{Ifull}
\ee
The denominator has zeros in the complex plane at
\be
k^- = {\blue  -\frac{\vec k_\perp^2}{Q-k^+} + i\fme \frac{1}{Q-k^+}}, \qquad
k^- = \frac{\vec k_\perp^2}{k^+} - i\fme \frac{1}{k^+},\qquad
k^- = {\red \frac{\vec k_\perp^2 - Q k^+}{k^+} - i\fme \frac{1}{k^+} }
\label{poles}
\ee
These are on the same side of the real axis unless $0 <  k^+ < Q$. 
Thus the integral vanishes outside of this range and we can restrict $0 \le k^+ \le Q$.  
This configuration is shown on the left side of Fig.~\ref{fig:sudakovpoles}.
Looking at the poles in the $k^+$ plane shows that we must also have $-Q \le k^- \le 0$.

Since we are only interested here in the soft pinch surface, we can restrict the integration region so that all components of $k^\mu$ have magnitude less than $Q$. We also take $\Pi^{\mu\nu} = - g^{\mu\nu}$, since the other terms in lightcone gauge do not affect the power counting around the soft pinch surface. Then,
\be
I_\text{full} \LPeq - i 2 Q^2 g_s^2 \int_{-\kappa Q}^{\kappa Q} \frac{d k^+ d k^- d^2 k_\perp}{2 (2\pi)^4} 
\frac{ 1 }{ [ { -Q k^-  - \vec k_\perp^2 + i\fme} ]}
\frac{1}{[k^+ k^- - \vec k_\perp^2+i\fme]}
\frac{1}{[ { Q k^+ - \vec k_\perp^2 + i\fme}]}
\label{fullhs}
\ee
Here, $\LPeq$ means we are restricting the integral to the soft pinch surface, with $\kappa \ll 1$. 
This has poles in the complex $k^-$ plane at
\be
k^- = {\blue -\frac{\vec k_\perp^2}{Q} + i\fme}, \qquad
k^- = \frac{\vec k_\perp^2}{k^+} - i\fme
\ee
The third pole from the original integral in Eq.~\eqref{poles} has moved off to $k^- = -\infty - i\fme$.  

What we we would like is to drop the $\vec k_\perp^2$ terms compared to $Q k^+$ and $Q k^-$. This can only be justified if it is parametrically true everywhere in the integration region. It cannot be justified in  the Glauber region, that is, in regions of $k^\mu$ where
 $k^- \lesssim {\vec k_\perp^2}/{Q}$. But let us look at the complex $k^-$ plane in more detail at fixed $k_\perp $ and $k^+$, both of which are soft ($ < \kappa Q$). The dangerous Glauber region is shown as the hatched area that nearly touches the pole at $k^- = - \frac{\vec k_\perp^2}{Q} + i\fme$. The integration contour (the real $k^-$ line) goes right through this region. To avoid this region, we note that since $k^+ < \kappa Q$, the pole at $k^- = {\vec k_\perp^2}/{k^+} - i\fme$ is parametrically far away from the Glauber region. Thus we can deform the contour downward into the complex plane avoiding the Glauber region explicity. For example, we could take move onto the arc
  with $|k^-| = {\vec k_\perp^2}/{Q \kappa}$. This arc avoids the Glauber region without crossing the non-Glauber pole. Once the contour is 
  out of the Glauber region, we can use eikonal scaling $\vec k_\perp^2 \ll k^- Q$ to simplify the integrand. Then we can deform the contour back. Note that we can do this deformation for any $k^+$ and $\vec k_\perp$. We can then do the same manipulation for the $k^+$ integral to justify dropping $\vec k_\perp^2 \ll k^+ Q$. The result is
\be
I_\text{soft} = - 2 i Q^2 g_s^2 \int_{-\kappa Q}^{\kappa Q} \frac{d k^- d k^+ d^2 k_\perp}{2 (2\pi)^4} 
\frac{ 1 }{ [ { -Q k^- + i\fme} ]}
\frac{1}{[k^- k^+ - \vec k_\perp^2+i\fme]}
\frac{1}{[ { Q k^++ i\fme}]}
\label{Isoft}
\ee
which is the same integral we would get from taking the eikonal limit of $I_\text{full}$.  Thus all of the soft singularities of $I_\text{full}$, including those from the Glauber region~(the hatched region), are contained in $I_\text{soft}$. 
It is not hard to confirm directly that the Glauber region is contained in the soft integral through direct calculation (see Section~\ref{sec:isolate} below).

An implication of the contour deformation argument is that all the integrals coming from the use of the second (not eikonal) term in the replacement in Eq.~\eqref{softsum} (diagrams with all-blue lines, in the language of~\cite{FS2}) are
 completely IR finite. 
 For example, performing this replacement on the $p_\ccT$ line results in
\be
I_\text{remain}=  2 i Q^2 g_s^2 \int_{-\kappa Q}^{\kappa Q} \frac{d^4k }{(2\pi)^4} 
\frac{ 1 }{ [ { -Q k^- - \vec k_\perp^2 + i\fme} ]
[k^- k^+ - \vec k_\perp^2+i\fme]
[ { Q k^+ + i\fme}]} \frac{k^- k^+ - \vec k_\perp^2}{Q k^+ - \vec k_\perp^2 + i\fme}
\ee
That this integral is IR finite is easy to see -- the $k^2$ pole is canceled by the expansion and the remaining
poles in the $k^+$ plane (or $k^-$ plane) are on the same side of the real axis and so the integral vanishes. A more general argument
is that once the contour is deformed out of the Glauber region, the new term suppresses the IR divergent part of the amplitude in the entire
integration region. Thus the amplitude is power-counting finite (scaling like $\kappa^n$ with $n>0$), as it would be under eikonal
scaling without the contour deformation.

\subsection{Spacelike example \label{sec:TL}}
Next, let us look at a diagram with a particle in the initial state and one in the final state.
\be
I_\text{full}^\text{SL}
= 
\begin{gathered}
\resizebox{40mm}{!}{
     \fmfframe(0,00)(0,0){
     \begin{fmfgraph*}(60,28)
	\fmftop{L1}
	\fmfbottom{R1,R2}
	\fmf{fermion,label=$p_\ccO - k \swarrow$,l.d=0.3,l.s=right}{L1,v1}
	\fmf{fermion,label=$p_\ccO \swarrow$,l.d=0.3,l.s=right}{v1,R1}
	\fmf{fermion,label=$\searrow p_\ccT  $,l.s=right,l.d=0.3}{R2,v2}
	\fmf{fermion,label=$\searrow p_\ccT + k $,l.s=right,l.d=0.3}{v2,L1}
	\fmf{photon,label=$\leftarrow k$,tension=0}{v1,v2}
        \fmfv{d.sh=circle, d.f=30,d.si=0.1w}{L1}
\end{fmfgraph*}
}}
\end{gathered}
 = i g_s^2 \int \frac{d^4k}{(2\pi)^4} 
\frac{ (2p_\ccO - k)^\mu }{ [ (p_\ccO - k)^2 + i\fme]}
\frac{\Pi_{\mu\nu}(k)}{k^2+i\fme}
\frac{(2p_\ccT + k)^\nu}{[(p_\ccT+ k)^2 + i\fme]}
\label{ITL}
\ee
Recall our convention that the incoming momentum, $p_\ccO^\mu$ is treated as outgoing with negative energy. 
Let us assume that $p_\ccO^\mu$ and $p_\ccT^\mu$ are not proportional to each other. Then we define $Q^2 =- 2 p_\ccO \cdt p_\ccT>0$. As $Q$ is the only scale in the problem, we zoom in on the soft singularity again by applying $k^\mu \ll Q$. We can go to lightcone coordinates 
in the $p_\ccO^\mu, p_\ccT^\mu$ frame, as before. The integral then becomes
\be
I_\text{full}^\text{SL} \LPeq  2 i Q^2 g_s^2 \int_{-\kappa Q}^{\kappa Q} \frac{d k^- d k^+ d^2 k_\perp}{2 (2\pi)^4} 
\frac{ 1 }{ [ { Q k^-  - \vec k_\perp^2 + i\fme} ]}
\frac{1}{[k^- k^+ - \vec k_\perp^2+i\fme]}
\frac{1}{[ { Q k^+ - \vec k_\perp^2 + i\fme}]}
\label{fullTL}
\ee
Now the poles are at
\be
k^- = { \frac{\vec k_\perp^2}{Q} - i\fme}, \qquad
k^- = \frac{\vec k_\perp^2}{k^+} -\text{sign}(k^+) i\fme
\ee
As in the spacelike case, one pole is in the Glauber region, but the other is not.  Thus we can deform the contour away from the Glauber 
region and justify the eikonal expansion here as well. As it happens (see Section~\ref{sec:isolate} below), the Glauber contribution to this integral exactly vanishes.  
But the point is that the Glauber contribution is contained in the eikonal limit for either timelike or spacelike kinematics, whether or not it happens to vanish. 

\subsection{General argument \label{sec:general}}
Now let us generalize the argument from the previous section to the $n$-loop case with arbitrary final and initial states. All we will assume is that none of the final states are collinear to any of the initial states. The proof of factorization in~\cite{FS2} did not use any information about initial or final states or about integration contours. What was shown is that all the terms which are power-counting divergent under eikonal scaling and collinear scaling in any of the directions in the full theory are reproduced in the factorized expression. 

What we need to show is that integrals inloving the second term when the substitution in Eq.~\eqref{softsum} is used, do not have any soft singularities despite their being power-counting divergent under Glauber scaling. Since the factorized expression in Eq.~\eqref{sQEDmain}
reproduces the self-energy graphs in the full theory exactly,  the only graphs we need to be concerned about are the ones which would ordinarily be part of the hard vertex. That is, those connecting to more than one collinear direction. We need to show that the
eikonal expansion can be justified for such diagrams.

In order for there to be a singularity in a diagram, according to the Landau criteria, all the propagators have to be either exactly zero or proportional to an external momentum. So let us put all the loop momenta except for one exactly on the singular surface. This leaves a single loop integration variable we call $k^\mu$. We can trace this momentum along the diagram. For example, 
\be
\begin{gathered}
\resizebox{35mm}{!}{
     \fmfframe(0,5)(0,0){
\begin{fmfgraph*}(50,50)	
\fmfstraight
\fmfleft{L1,L2,L3}
	\fmfright{R1,R2,R3}
	\fmf{plain}{L1,v1,C}
    \fmf{plain}{L2,v2,C}
    \fmf{plain,tension=1.6}{L3,v3,C}
    \fmf{plain,tension=1.6}{C,w1,y1,R1}
    \fmf{plain}{C,w2,R3}
    \fmffreeze
    \fmf{plain,fore=red}{C,w1,y1}
    \fmf{plain,fore=red}{C,w2}
    \fmf{photon,tension=0.1,fore=black}{v3,x1}
    \fmf{photon,tension=0.1,fore=red,label={\rotatebox{30}{$\downarrow$}$k$},l.s=left,l.d=0.1mm}{w2,x1}
    \fmf{photon,tension=0.1,fore=red,label=$\downarrow k$,l.s=left}{x1,y1}
    \fmf{photon,tension=0,fore=black}{v1,w1}
    \fmf{phantom,tension=0.5}{R2,x1}
    \fmfv{d.sh=circle, d.f=30,d.si=0.1w}{C}
    \fmf{phantom,tension=0,fore=red,label={$k$\rotatebox{0}{$\nwarrow$}},l.s=right,l.d=0.01w}{C,w1,y1}
    \fmf{phantom,tension=0,fore=red,label=$k$ ,l.s=left,l.d=0.07w}{C,w2}
    \fmf{phantom,tension=0,label={\rotatebox{-10}{$\nearrow$}},l.s=left,l.d=0.01w}{C,w2}
    \fmf{phantom,tension=0,label=$p_\ccO$}{R3,w2}
    \fmf{phantom,tension=0,label=$p_\ccT$}{R1,y1}
\end{fmfgraph*}
}
}
\end{gathered}
= \begin{gathered}
\int\frac{d^4 k}{(2\pi)^4}
 \frac{N}{k^2 + i\fme}
 \frac{1}{k^2 + i\fme}
 \frac{1}{(k-p_\ccT)^2 + i\fme}
 \frac{1}{(k-p_\ccT)^2 + i\fme}
 \frac{1}{(k+p_\ccO)^2 + i\fme}
 \end{gathered}
 \label{Gpinch}
\ee
where $N$ is some numerator structure.
This is almost identical to the Sudakov form factor diagram we studied in the previous subsection. The only difference is that now there are multiple poles at each singular point. Although the other loop integrations that we are neglecting make the actual diagram much less singular than this (no diagram can scale to a negative power of $\kappa$ under any soft scaling), we do not need to make use of the extra cancellations. The only thing to observe is that the poles are all in the same parts of the $k^\mu$ phase space as in the simple vertex correction. Thus the contour deformation works in exactly the same way and the eikonal expansion can be justified. 

What happens if the loop momentum connects to an incoming and an outgoing direction? As long as $p_\ccO^\mu$ and $p_\ccT^\mu$ are not proportional to each other, one can still justify the eikonal expansion, as in Section~\ref{sec:TL}. 

The only time a problem can arise is if an incoming momentum and an outgoing momentum are collinear. For example, consider a configuration like this
\be
\begin{gathered}
\begin{tikzpicture}
 \node at (0,0) {
\parbox{60mm}
{
\resizebox{45mm}{!}{
     \fmfframe(0,5)(0,0){
\begin{fmfgraph*}(80,50)	
	\fmfleft{L1,L2}
	\fmfright{R1,R2,R3}
    \fmf{plain,label={$p_\ccO$\rotatebox{20}{$\nwarrow$}},tension=2}{L2,v2}
    \fmf{plain,label={$p_\ccO+p_\ccT+k$\rotatebox{-15}{$\nwarrow$}},l.s=right,l.d=5}{v2,v}
    \fmf{phantom,tension=2}{L1,v1}
    \fmf{plain,l.s=left,l.d=5}{v1,v}
    \fmf{phantom,l.s=left,l.d=5,tension=0.5}{v,R2}
    \fmf{gluon,label=$p_\ccT + k$,l.s=left}{v2,v3}
    \fmf{gluon,label=$p_\ccT \rightarrow$,l.s=left}{v3,R3}
    \fmf{phantom,l.s=right}{v1,v4}
    \fmf{phantom}{v4,R1}
    \fmf{gluon,tension=0,label=$ \downarrow k$,l.s=left,l.d=10}{v3,v4}
    \fmf{phantom,tension=0.01}{v,x1,x2,x4,R1}
    \fmf{plain,tension=0}{v,x1}
    \fmf{phantom,tension=0.01}{v,y1,y2,y4,R2}
    \fmf{plain,tension=0}{v,y1}
    \fmf{phantom,tension=0.01}{v,z1,z2,z4,R3}
    \fmf{plain,tension=0}{v,z1}
    \fmfv{d.sh=circle, d.f=30,d.si=0.1w}{v}
\end{fmfgraph*}
}
}
}
};
\draw[black,fill=gray!20] (-1,-1.2) ellipse (1.6 and 0.5);
\end{tikzpicture}
\end{gathered}
\hspace{-25mm}
\begin{gathered}
\sim \int_{-\kappa Q}^{\kappa Q} \frac{d^4 k }{(2\pi)^4} 
\frac{1}{[(p_\ccT+k)^2 + i\fme]}
\frac{1}{[(p_\ccO+p_\ccT+k)^2 + i\fme]}
\frac{1}{[k^2 + i\fme]}
\cdots
\end{gathered}
\ee
If $p_\ccO^\mu$ and $p_\ccT^\mu$ are collinear then $2 p_\ccO \cdt k =- Q_\ccO k^-$ and $2 p_\ccT \cdt k = Q_\ccT k^-$ for two energies $Q_\ccO$ and $Q_\ccT$ and the same lightcone component $k^-$ is in both products. Moreover $Q_\ccO \ge Q_\ccT$ by momentum conservation. 
Then
\be
I \LPeq 
\int_{-\kappa Q}^{\kappa Q} \frac{d^4 k }{(2\pi)^4}
\frac{1}{[{ Q_\ccT k^- - \vec k_\perp^2 + i\fme}]}
\frac{1}{[{- (Q_\ccO-Q_\ccT) k^- - \vec k_\perp^2 + i\fme}]}
\frac{1}{[ k^-k^+ - \vec k_\perp^2 + i\fme]}
\cdots
\ee
This type of diagram has poles at
\be
{ k^- = \frac{\vec k_\perp^2}{Q_\ccT} - i\fme},\qquad
{ k^- = -\frac{\vec k_\perp^2}{Q_\ccO - Q_\ccT} + i\fme},\qquad
k^- = \frac{\vec k_\perp^2}{k^+} - i\fme,\qquad
\ee
The first two poles, coming from the two collinear propagators are on opposite sides of the real axis and not parametrically separated so one cannot deform the contour out of the Glauber region. Therefore one cannot justify dropping $\vec k_\perp^2 \ll Q_\cci k^-$ in the integrand  since such a modification may miss infrared divergences. 

From this example, we see that for the Glauber region to deserve special concern, we need to have the momentum $k^\mu$ flowing in opposite directions through two lines that are collinear to each other. Thus a graph like Eq.~\eqref{ITL} is not problematic, even if $p_\ccO^\mu$
 and $p_\ccT^\mu$ are proportional.  Graphs in which the two collinear lines that the loop momenta runs through are  in the same final state sector are also not problematic  -- the collinear sector in the factorized expression is  the same as in full QCD so all of the singularities are necessarily reproduced in this region.

So far, we have considered only virtual graphs. Equivalently, we assumed that each sector has one particle, with no two momenta proportional to each other. We can weaken this requirement and allow for multiple particles in each sector. For particles with momenta $p^\mu$ and
$q^\mu$ to be in the same collinear sector, we require $p \cdt q < \lambda^2 Q^2$ with $Q$ a hard scale (e.g. the center of mass energy).  The parameter $\lambda$ is presumed to be small and the factorization formula is supposed to hold to leading power in $\lambda$. 
Soft external particles can have energies up to  $\lambda^2 Q$. 
The effect of having multiple collinear particles in a sector or soft particles is to make some lines in diagrams like  
Eq.~\eqref{Gpinch}
off-shell, so that $p^2 = \lambda^2 Q^2$. For $\lambda=0$, these lines would be massless and the graph would be infrared divergent, however,
at finite $\lambda$, the IR divergence is regulated. 
That is, a graph which would be logarithmically divergent may now be finite proportional to $\ln \lambda$. 
In this way, the factorization formula which reproduces the infrared singularities at $\lambda =0$ also reproduces 
the amplitude to leading power in $\lambda$ when there are soft particles or multiple collinear particles.

 The key point is that adding soft or collinear particles does not invalidate the argument that justifies the eikonal approximation. On the pinch surface, which characterizes both the IR divegences and the leading power behavior at small $\lambda$, the contours can always be deformed out of the Glauber region. When a graph only involves particles in a single collinear sector with only outgoing (or only incoming) particles, there can be Glauber singularities, however these graphs are identical in the factorized and full theory amplitudes, and so factorization still holds.  
  Thus factorization holds for arbitrary soft and collinear sectors, as long as no final state particles are collinear to any initial state particles.

\subsection{Summary}
In summary, we have analyzed in detail the when the eikonal approximation can be used to prove factorization, as it was for
Eq.~\eqref{sQEDmain}. We find that this equation can be generalized. Take any initial state $\ket{Z}=\ket{Z_\ccO}\cdots\ket{Z_\ccM}\ket{Z_\scs}$, with the momenta in $\ket{Z_\ccj}$ all collinear to a direction $n_\ccj$ and all the momenta in $\ket{Z_\scs}$ soft, and any final state $\bra{X}=\bra{X_\ccO} \cdots \bra{X_\ccN}\bra{X_\scs}$, with similar definitions, and assuming no initial state direction is collinear to any final state direction, then to leading power in $\lambda$
\be
\bra{X} \phi^\star \cdots \phi \ket{Z}
 \;\LPFeq\; 
\cC(\Sij) \, \frac{\bra{X_\ccO} \phi^\star W_\ccO \ket{0}}{\bra{0} Y_\ccO^\dg W_\ccO \ket{0}} \,\cdots\, \frac{\bra{0} W_\rN^\dg\phi \ket{Z_\ccN}}{\bra{0} W_\rN^\dg Y_\rN \ket{0}}
\;\bra{X_\scs} Y_\ccO^\dg \cdots Y_\rN \ket{Z_\scs}
\label{sQEDmainXY2}
\ee
Since $\lambda$ is arbitrary, a corrollary is that for all virtual diagrams for hard scattering in which no final state and initial state momenta are proportional, the complete IR divergences of the full graphs in QCD, including any coming from the Glauber region, are exactly reproduced by the factorized expressions. Another corrollary is that all possible violations of factorization are associated with situations where initial states and final state momenta are collinear. 

\section{Isolating the Glauber contribution \label{sec:isolate}}
We have seen that most of the time, singularities associated with Glauber scaling are automatically contained in the expansion around zero momentum, using homeogenous, eikonal scaling.
Factorization violation is associated with situations where this containment fails, so that the eikonal limit does not reproduce all of the soft singularities. To clarify the role that Glauber gluons play in amplitude-level factorization, it may be helpful to isolate their contributions. In this section, we explore some approaches to identifying the Glauber contribution and we explore the connection between the Glauber limit and the imaginary part of amplitudes. 

In this section we will mostly be concerned, once again, with the 1-loop vertex correction diagram in scalar QED,
Eq.~\eqref{eq:vfull}:
\be
I_\text{full}
=
\begin{gathered}
\resizebox{20mm}{!}{
     \fmfframe(0,0)(0,0){
\begin{fmfgraph*}(20,35)
\fmfstraight
	\fmfleft{L1}
	\fmfright{R1,R2}
	\fmf{fermion}{L1,v1}
	\fmf{fermion,label=$p_\ccT \searrow$,l.s=right,l.d=0.5mm}{v1,R1}
	\fmf{fermion,label=$p_\ccO \nearrow$,l.s=right,l.d=0.5mm}{R2,v2}
	\fmf{fermion}{v2,L1}
	\fmf{photon,label=$\uparrow k$,tension=0}{v1,v2}
        \fmfv{d.sh=circle, d.f=30,d.si=0.2w}{L1}
\end{fmfgraph*}
}
}
\end{gathered}
 \LPeq - i 2 Q^2 g_s^2 \int \frac{d^4k}{(2\pi)^4} 
\frac{ 1 }{ [ { -Q k^-  - \vec k_\perp^2 + i\fme} ]}
\frac{1}{[k^- k^+ - \vec k_\perp^2+i\fme]}
\frac{1}{[ { Q k^+ - \vec k_\perp^2 + i\fme}]}
\label{IfulltoG}
\ee
We have ignore the numerator structure because it's irrelevant to our discussion. The following analysis is very similar for QCD.

 In QED, the amplitude in dimensional regularization  in $4-2\ep$ dimensions and $\overline{\text{MS}}$ scheme is
\be
I_\text{full} = \frac{ \alpha_s }{2 \pi} \left( \frac{\mu^2}{-2 p_\ccO \cdt p_\ccT - i\fme} \right)^\ep \left(  \frac{1}{\ep^2} +
 \text{finite}
  \right) 
\ee
This is an analytic function of $p_\ccO \cdt p_\ccT$ with a branch cut when $p_\ccO \cdt p_\ccT>0$. Expanding around $\ep=0$ gives
\be
I_\text{full} = \frac{ \alpha_s }{2 \pi} \left(  \frac{1}{\ep^2}  - \frac{1}{\ep}  \ln \frac{-2 p_\ccO \cdt p_\ccT - i\fme}{\mu^2}
+\text{finite}
  \right) 
  \label{QEDv}
\ee
This has an imaginary part if and only if $p_\ccO \cdt p_\ccT >0$. 
The Glauber contribution produces only the imaginary part of this result, as we will now see in a number of different ways.

\subsection{Method of regions}

According to the method of regions, we can isolate the Glauber contribution by assuming $k^\pm \sim \kappa^2$, $\vec k_\perp \sim \kappa$ and expanding the integrand to leading order in $\kappa$. This gives
\be
I_\text{Glauber}
= - i 2 Q^2 g_s^2 \int \frac{d^4k}{(2\pi)^4} 
\frac{ 1 }{ [ { -Q k^-  - \vec k_\perp^2 + i\fme} ]}
\frac{1}{[ - \vec k_\perp^2+i\fme]}
\frac{1}{[ { Q k^+ - \vec k_\perp^2 + i\fme}]}
\label{IGregions}
\ee
In this form, one cannot integrate over $k^-$ and $k^+$ using Cauchy's theorem since the integrand does not die off fast enough as $k^\pm \to \infty$. Changing to $k^- = k_0 - k_z$ and $k^+ = k_0 + k_z$ gives
\be
I_\text{Glauber}
=  - i 2 Q^2 g_s^2 \int \frac{d^4k}{(2\pi)^4} 
\frac{ 1 }{ [ { -Q k_0 + Q k_z  - \vec k_\perp^2 + i\fme} ]}
\frac{1}{[ - \vec k_\perp^2+i\fme]}
\frac{1}{[ { Q k_0 + Q k_z- \vec k_\perp^2 + i\fme}]}
\label{Iggz0}
\ee
Ordinarily, the $k^2$ term is quadratic in $k_0$ and $k_z$ so using $k^\pm$ is simpler, but in the Glauber limit the transverse components dominate so there is no advantage.

For timelike seperation (both outgoing) as written, the poles in the $k_0$ plane are at $k_0 = \mp {\vec k_\perp^2}/{Q} \pm k_z \pm i\fme$.  Closing the $k_0$ contour downwards to pick up the 
$k_0 = {\vec k_\perp^2}/{Q} - k_z - i\fme$ pole results in an integral which is divergent at large $|k_z|$. This divergence can be regulated by a rapidity regulator. Following Ref.~\cite{Rothstein:2016bsq}, we use the $\eta$ regulator of Ref.~\cite{Chiu:2012ir}. Adding also a small mass $m$ to regulate the infrared singularity and working in $d=4-2\ep$ dimensions to regulate the UV divergence 
gives
\begin{align}
I_\text{Glauber}
 &= -  2 Q g_s^2 \mu^{4-d}  \int \frac{d^{d-2}k_\perp d k_z}{(2\pi)^{d-1}} 
 \frac{\nu^{2\eta}}{|2k_z|^{2 \eta}}
\frac{ 1 }{ [ { 2 Q k_z - 2\vec k_\perp^2  + i\fme} ]}
\frac{1}{[ - \vec k_\perp^2 - m^2+  i\fme]}
\\
&=  - \frac{\alpha_s}{2 \pi} (i\pi)\left( \frac{1}{\ep_{\text{UV}}} + \ln\frac{\mu^2}{m^2} \right)
\label{Gcont}
\end{align}
Note that the result is independent of $\eta$ (see also Appendix B.2 of~\cite{Rothstein:2016bsq}). Essentially, adding the $|2 k_z|^{2 \eta}$ term allows us to do the $k_z$ integral. The imaginary part of the full soft graph  is identical~\cite{Rothstein:2016bsq} confirming that the Glauber is contained in the soft. 

If we also use dimensional regularization to regulate IR, the integral would be scaleless and vanish. We then deduce that in pure dimensional regularization 
\be
I_\text{Glauber} =  \frac{\alpha_s}{2 \pi }(- i \pi) \left(\frac{1}{\ep_{\text{UV}}} - \frac{1}{\ep_\text{IR}}\right)
\label{IGdr}
\ee
Thus the Glauber contribution is IR divergent and purely imaginary.

Note that if $p_\ccT^\mu$ were incoming, then $Q k^+$ term would fips sign and the poles in the $k_0$ plane would be at $k_0 = -{\vec k_\perp^2}/{Q}  \pm k_z + i\fme$. Since both poles are above the integration contour, we can close the contour downwards
giving zero. In other words, the Glauber limit with spacelike separation gives zero for this diagram.

Thus we find that the Glauber limit gives zero in the spacelike case ($p_\ccO \cdt p_\ccT < 0$) and a purely imaginary number in the timelike case ($p_\ccO \cdt p_\ccT > 0$). Since the amplitude is zero in a compact region of $p_\ccO \cdt p_\ccT$, it cannot be an analytic function of $p_\ccO\cdt p_\ccT$ without vanishing completely. The non-analyticity comes from the rapidity regulator, since $|2 k_z|^{2 \eta}$ is a non-analytic function. This non-analyticity is therefore unavoidable if the Glauber contribution is to give {\it only} the imaginary part of the amplitude.

\subsection{Cut-based approach}

Yet another way to study Glauber gluon is through discontinuity of scattering amplitudes. This is natural from the viewpoint that Glauber gluon is associated with the imaginary part of scattering amplitudes at lowest order. 

Using Cutkosky's cutting rule, the s-channel discontinuity of Eq.~\eqref{IfulltoG} can be written as
\begin{multline}
  \label{eq:1}
\text{Disc}_s I_\text{full} =   i (-2\pi i)^2 g_s^2 \int \frac{d k^- d k^+ d^2 k_\perp}{2 (2\pi)^4} \delta( (p_\ccO - k)^2)\theta((p_\ccO - k)^0) \delta((p_\ccT+k)^2)  (\theta(p_\ccT+k)^0)
\\ 
\times \frac{ (2p_\ccO - k)^\mu \Pi_{\mu\nu}(k) (2p_\ccT + k)^\nu}{[k^- k^+ - \vec k_\perp^2+i\fme]} 
\end{multline}
where the only possible cut is through the top and bottom line in the vertex diagram. 
We shall see that at this order, the Glauber contribution is given by the forward limit of the discontinuity, $\vec k_\perp \ll Q$. The on-shell conditions of $p_\ccO-k$ and $p_\ccT+k$ automatically enforce the Glauber scaling, $k^-\sim k^+ \sim k^2_\perp/Q$. In the forward limit, the two-body phase space measure becomes 
\begin{align}
  \label{eq:2}
   & i (-2\pi i)^2 g_s^2 \int \frac{d k^- d k^+ d^2 k_\perp}{2 (2\pi)^4} \delta( (p_\ccO - k)^2)\theta((p_\ccO - k)^0) \delta((p_\ccT+k)^2)  (\theta(p_\ccT+k)^0)
\nn \\
&\hspace{2cm}
\longrightarrow - \frac{i g_s^2}{2 Q^2}\int \frac{d^2 k_\perp}{(2 \pi)^2}
\end{align}
and the matrix element to the right of the cut becomes forward scattering amplitude
\begin{align}
  \label{eq:3}
  \frac{ (2p_\ccO - k)^\mu \Pi_{\mu\nu}(k) (2p_\ccT + k)^\nu}{[k^- k^+ - \vec k_\perp^2+i\fme]}  
\longrightarrow \frac{2 Q^2}{ \vec k_\perp^2 - i\fme}
\end{align}
The one-loop Glauber contribution is thus given by
\begin{align}
  I_\text{Glauber} \equiv \frac{1}{2} \lim_{k_\perp/Q \to 0} \text{Disc}_s I_\text{full} = - \frac{i \alpha_s}{2 \pi} \int \frac{d^2 k_\perp}{\vec k_\perp^2}   \label{IGlaubercut}
\end{align} 
in agreement with other definition of Glauber contribution. It's interesting to note that defining the Glauber contribution in this way avoid the use of rapidity regulator in intermediate step of the calculation. This is comforting, as the result should be regulator independent. 
 
Isolating the Glauber contribution in momentum space at one-loop level with this approach was introduced in~\cite{Korchemsky:1987wg}, and similar cutting prescription in position space explored more in~\cite{Laenen:2015jia}. This was also nothing but the s-channel unitarity approach for extracting the Reggie trajectory of scattering amplitudes~\cite{Fadin:1995km}. However, it should be noted that only at one-loop level, one can have a clear identification of the leading term in the $\ep$ expansion of  Glauber contribution and discontinuity. At higher loop, there is no direct relation anymore. For example, at two loops, double Galuber-exchanged contribution can contribution $(i\pi)^2 = - \pi^2$, which is real and has no discontinuity. 

\subsection{Poisition space}
One can interpret Eq.~\eqref{IGlaubercut} as describing a potential $\widetilde{V}(k)\sim\frac{g_s^2}{\vec k_\perp^2}$ between the two outgoing particles. Fourier transforming, the potential $V(x) \sim g_s^2 \ln |x_\perp|$ depends only on the transverse separation $x_\perp$ between the  particles. Like the Coulomb potential, $V(r)\sim\frac{g_s^2}{r}$, the Glauber potential is time-independent, but unlike the Coulomb potential, the Glauber potential additionally does not depend on the longitudinal separation $x_L$~\cite{Rothstein:2016bsq}. 

Since the Glauber contribution is contained in the soft contribution for this graph, 
we can simplify the calculation from one in full QCD by taking the energies of $p_\ccO^\mu$ and $p_\ccT^\mu$ to infinity. Following~\cite{Laenen:2015jia}, to regulate some of the divergence, we also take 
$p_\ccO^\mu$ and $p_\ccT^\mu$ timelike. Thus we write $p_\ccO^\mu = Q_\ccO n_\ccO^\mu$ and $p_\ccT^\mu = Q_\ccT n_\ccT^\mu$ with $n_\ccO^2 = n_\ccT^2=1$.  The integral then becomes
\begin{align}
I_\text{soft} &=    g_s^2 \int \frac{d^4k}{(2\pi)^4} \frac{n_{\ccO \ccT}}{(-n_\ccO \cdot k + i\fme)( n_\ccT \cdot k + i\fme)(k^2 + i\fme)}\\
&=    g_s^2 \int \frac{d^4k}{(2\pi)^4}  \int_0^\infty d s_\ccO \int_0^\infty d s_\ccT
\frac{n_{\ccO \ccT}}{k^2 + i\fme} e^{-i s_\ccO (n_\ccO \cdot k) + i s_\ccT (n_\ccT \cdot k)}
\\
&=    g_s^2  \int_0^\infty d s_\ccO \int_0^\infty d s_\ccT \frac{n_{\ccO \ccT}}{(s_\ccO n_q^\mu - s_\ccT n_2^\mu)^2 - i\fme}
\label{Ifullpos}
\end{align}
where $n_{\ccO\ccT} = n_\ccO \cdot n_\ccT$. 
Schwinger parameters $s_\ccO$ and $s_\ccT$ have been introduced on the second line. They represent the proper time that the particles have travelled from the hard vertex. The result of course matches the matrix element of timelike Wilson lines, which we could have written down directly.

A nice observation from~\cite{Laenen:2015jia} is that, in this form, we can see that an imaginary part can only come from  times $s_\ccO$ and $s_\ccT$ for which particles along the $n_\ccO^\mu$ and $n_\ccT^\mu$ direction are lightlike separated.
That is, the Glauber contribution is associated with spacetime points under which particles moving in the two directions can causally influence each other. As an analogy, think of passengers on trains going in two different directions shining a light at each other. The light from one train can be seen on the other train only at the appropriate spacetime point. In contrast, if one particle is incoming and the other outgoing, as in Eq.~\eqref{ITL}, it is impossible for the two to be lightlike separated -- one cannot see light from a train which arrived in the station before your train left (except at the origin of time). Thus, there is no imaginary part in that situation, and no Glauber contribution. Taking the limit where the trajectories become lightlike, from the timelike separation (both outgoing) case, the support of the imaginary part lie on the lightcone, $s_\ccO=0$ or $s_\ccT=0$. In the case of spacelike separation, the lightlike limit is still zero.

The amplitude in Eq.~\eqref{Ifullpos} is the soft amplitude, from reducing the full diagram in QCD in the eikonal limit. As shown in the previous section, it contains the complete Glauber contribution, which is now identified as the imaginary part of the diagram. Defining the cusp angle $\gamma$ through $\cosh \gamma = - n_{\ccO \ccT} $ and changing variables $s_\ccO = s e^\tau$ and $s_\ccT=s$, the integral becomes~\cite{Chien:2011wz}.
\begin{align}
I_\text{soft} &= \frac{\alpha_s}{4\pi}  \int_0^\infty \frac{ds}{s} \int_{-\infty}^\infty d \tau \frac{\cosh \gamma}{\cosh \tau + \cosh \gamma}\\
& =\frac{\alpha_s}{8 \pi} \int_0^\infty \frac{ds}{s} \gamma \cosh \gamma
\end{align}
This is again UV and IR divegent. In the spacelike speparation case (one incoming and one outgoing) $n_{\ccO \ccT} <0$ and $\gamma>0$ is real. Then this integral is real. The timelike case (both outgoing) corresponds to $n_{\ccO \ccT} >0$ whereby $\cosh \gamma <0$ and $\gamma$ is complex. Thus only in the case of timelike sepration does the amplitude have an imaginary part. 

The fact that the Glauber contribution is purely imaginary at 1-loop implies that it will necessarily cancel in cross section calculations at next-to-leading order. More generally, the 1-loop Glauber contribution exponentiates into a phase which cancels in cross sections to all orders. This exponentiation comes about in the same way that the exponentiation of the Coulomb phase comes about. One way to see it is by computing the energy of a moving charge in the potential of another moving charge, either classically or in quantum mechanics~\cite{Jackiw:1991ck} or by mapping to AdS~\cite{Chien:2011wz}. These considerations also lead to a nice shockwave picture of Glauber exchange in fowrard scattering~\cite{Rothstein:2016bsq}. As exponentiation of the Glauber phase corresponds very closely to Abelian exponentiation, it naturally is also limited in the non-Abelian theory. For example, irreducible 2-loop contributions or beyond with both Glauber and soft/collinear loops present may not exponentiate.

\subsection{Effective field theory Glauber operator \label{sec:scetg}}
We have seen that taking the Glauber limit of the integrand, according to the method of regions, gives a result which is purely imaginary and non-vanishing only in the timelike case. The effective field theory (EFT) approach tries to write down a Lagrangian whose Feynman rules generate the integral so that one no longer has to take limits of integrands. This Lagrangian can be derived by matching (writing down all possible operators consistent with symmetries and working out their coefficients to agree with QCD) or by performing a multipole
expansion in the classical theory, keeping the leading interactions according to some specified scalings.
In general, there is not a 1-to-1 correspondence between diagrams in the effective theory and diagrams in QCD, even after those diagrams are expanded according to some scaling. For example, EFT operators often included Wilson lines for gauge invariance. These Wilson lines represent the leading power contribution of many diagrams in QCD. 

The effective field theory that isolates the infrared singular regions of QCD is called Soft-Collinear Effective Theory~\cite{Bauer:2000ew,Bauer:2001ct,Bauer:2002nz,Beneke:2002ni,Beneke:2002ph} (see~\cite{Becher:2014oda} for a review). If a process involves hard directions are $p_\ccO^\mu$ and $p_\ccT^\mu$, 
then in SCET collinear fields denoted by $\xi_\ccj$ are introduced in the $p_\ccO^\mu$ and $p_\ccT^\mu$ direction. These fields have labels fixing the large and perpendicular components of their momenta, with the dynamics determined by fluctuations around these parametrically large components.
The extension of SCET to include Glauber contributions was recently achieved in~\cite{Rothstein:2016bsq}. 
The prescription is to add Glauber operators to the Lagrangian for each pair of directions in the theory. For example, in QED we would add
\be
\cO_{G \ccO \ccT}^{\text{QED}} 
=
\left[ \bar{\xi}_\ccO\frac{ \slashed{n}_\ccT }{2}  \xi_\ccO \right]
 \frac{1}{\cP_\perp^2} 
 \left[  \bar{\xi}_\ccT\frac{ \slashed{n}_\ccO }{2}  \xi_\ccT   \right]
 \label{OpQED}
\ee
 Here ${\cP_\perp}$ is an operator which picks out the $\perp$ component of the collinear fields it acts. 
 In QCD, the operator is sigifigantly more complicated~\cite{Rothstein:2016bsq}
 \begin{multline}
\cO_{G \ccO \ccT}^{\text{QCD}} 
=
\left[ \bar{\xi}_\ccO W_\ccO \frac{ \slashed{n}_\ccT }{2} T^A W_\ccO^\dagger \xi_\ccO \right]
 \frac{1}{\cP_\perp^2}  \\
 \times
\left[
\cP_\perp^\mu Y_\ccO^\dagger Y_\ccT \cP_{\perp \mu}
-g \cP_\mu^{\perp} \cB_{\ccO S\perp}^\mu Y_\ccO^\dagger Y_\ccT
- gY_\ccO^\dagger Y_\ccT  \cB_{\ccT S\perp}^\mu\cP_\mu^{\perp}
-g  \cB_{\ccO S\perp}^\mu Y_\ccO^\dagger Y_\ccT \cB_{\ccT S\perp}^\mu
-\frac{ig }{2} n_\ccO^\mu n_\ccT^\nu Y_\ccO \widetilde{G}_{\mu\nu}^S Y_\ccT
\right]^{AB} \\
\times 
 \frac{1}{\cP_\perp^2} 
\left[ \bar{\xi}_\ccT W_\ccT \frac{ \slashed{n}_\ccO }{2} T^B W_\ccT^\dagger \xi_\ccT \right]
 \label{OpQCD}
\end{multline}
The additional terms involve collinear Wilson lines $W_\ccj$, soft Wilson lines $Y_\ccj$, soft gluon fields $\cB_{\perp \mu}$,
and soft gluon field strengths $\widetilde{G}_{\mu\nu}$. For definitions of all of these objects, see~\cite{Rothstein:2016bsq}.
The collinear Wilson lines are necessary to ensure collinear gauge-invariance. The terms on the second line all involve soft fields and are separately soft-gauge-invariant. The variety of terms in this expression is determined by matching amplitudes in SCET and in full QCD.
Although one might imagine their coefficients get corrected order by order in perturbation theory through this matching, it seems that in fact they do not. Remarkably this operator appears to not receive any corrections and is not renormalized.

 At leading order in $g_s$, the QCD operator reduces to the QED one up to the group theory factors. The leading order interaction of this operator is a 4-point interaction connecting lines in the $n_\ccO^\mu$ direction with lines in the $n_\ccT^\mu$ directions. 
 The Feynman rule produces a factor of $\frac{i g_s^2 }{\vec k_\perp^2}$ wIth $k^\mu$ the momentum transferred between the two lines. 
 Operator insertions are drawn either with a red oval, indicated its pointlike nature, or with a red dotted line, indicating its origin in gluon exchange. For example,
\be
I_{\cO_{G12}}
=
\begin{gathered}
\resizebox{20mm}{!}{
     \fmfframe(0,0)(0,0){
\begin{fmfgraph*}(20,35)
\fmfstraight
	\fmfleft{L1}
	\fmfright{R1,R2}
	\fmf{fermion}{L1,v1}
	\fmf{fermion,label=$p_\ccT \searrow$,l.s=right,l.d=0.5mm}{v1,R1}
	\fmf{fermion,label=$p_\ccO \nearrow$,l.s=right,l.d=0.5mm}{R2,v2}
	\fmf{fermion}{v2,L1}
	\fmf{dbl_dots,fore=red,label=$\uparrow k$,tension=0}{v1,v2}
        \fmfv{d.sh=circle, d.f=30,d.si=0.2w}{L1}
\end{fmfgraph*}
}
}
\end{gathered}
 \LPeq - i 2 Q^2 g_s^2 \int \frac{d^4k}{(2\pi)^4} 
\frac{ 1 }{ [ { -Q k^-  - \vec k_\perp^2 + i\fme} ]}
\frac{1}{[ - \vec k_\perp^2+i\fme]}
\frac{1}{[ { Q k^+ - \vec k_\perp^2 + i\fme}]}
\ee
 This is identical to the Glauber limit of the original diagram, in Eq.~\eqref{IfulltoG}. 

Just adding this operator to the SCET Lagrangian with no other modification will lead to overcounting. As we already observed, the Glauber contribution is contained in the soft diagrams as well. The resolution proposed in~\cite{Rothstein:2016bsq} is to subtract off the overlap, diagram by diagram. The viewpoint of~\cite{Rothstein:2016bsq} is that the Glauber is separate mode from the soft. Thus the true soft contribution, $S$, meaning soft without Glauber, should be defined as the naive soft contribution $\widetilde{S}$ (including Glauber) with its Glauber limit $S^{(G)}$  subtracted: 
$S = \tilde{S} - S^{(G)}$. This subtraction is done using the zero-bin subtraction method~\cite{Manohar:2006nz} -- subtract from the soft diagram its limit where only the terms to leading power in Glauber scaling are kept. 
For those cases where the soft-Glauber contribution is identical to the Glauber contribution, which includes active-active and active-spectator interactions~\cite{Rothstein:2016bsq}, the result is the same as using the naive soft graph only.


\section{Factorization-violation in collinear splittings \label{sec:catani}}
As observed in Section~\ref{sec:general}, the only time hard-soft-collinear factorization can break down is in situations where an incoming particle is collinear to an outgoing one, the space-like collinear limit. In fact, it is known that in the space-like limit, amplitude-level factorization {\it is} violated~\cite{Catani:2011st}.
We now proceed to reproduce and discuss this important result. 

Collinear factorization implies that the matrix element $\ket{\cM}$ for an amplitude with $m$ particles collinear to a particular direction is related to the amplitude $\ket{\cMb}$ with only one particle collinear to that direction.
For simplicity, we consider here only the case with $m=1$, describing $1\to 2$ collinear splittings.
Let us say there are $n$ total particles in $\ket{\cM}$, of which only $p_\ccT^\mu$ is collinear to $p_\ccO^\mu$. In this case, the splitting function $\Sp$ is defined as
\be
\ket{\cM(p_\ccO,\cdots, p_\ccn,)} \LPeq \Sp(p_\ccO,p_\ccT ; p_\ccTH,\cdots, p_\ccn) \cdot \ket{\cMb ( P,p_\ccTH, \cdots p_\ccn) }
\label{gensplit}
\ee
Here, $P^\mu \LPeq p_\ccO^\mu + p_\ccT^\mu$, meaning $P^\mu$ is an onshell momentum ($P^2=0$) which is equal to the sum of the two momenta that split, up to power corrections in $\lambda^2 =2  {p_\ccO\cdt p_\ccT}/{Q^2}$, with $Q$ the center-of-mass energy or some other
hard scale.  The object $\Sp$ is an (amplitude-level) splitting function, or {\it splitting amplitude}. The matrix elements should be thought of as vectors in color space and $\Sp$ as an operator acting on these vectors.

Eq.\eqref{gensplit}, which has the splitting function depending on all the momenta, is called {\bf generalized collinear factorization}~\cite{Catani:2011st}. Even generalized factorization is non-trivial. The non-trivial part is that the splitting function is universal, independent of the short-distance physics encoded in the matrix element $\ket{\cMb}$. 

Generalized factorization is not terribly useful for computing cross sections. For example, since generalized splitting amplitudes depend on all the hard directions and colors in the processes, they do not allow us to use the semi-classical parton-shower simulation method to generate jets substructure to all orders. The parton shower can be justified when the $\Sp$ depends only on the momenta $p_\ccO^\mu$ and $p_\ccT^\mu$ in the relevant collinear sector, $\Sp= \Sp(p_\ccO,p_\ccT)$, and only on the color $\bf{T}_{\ccol 1}$ in that sector. 
When $\Sp$ has this special form, we say, following~\cite{Catani:2011st} that {\bf strict collinear factorization} holds.

Using results from the previous sections, we first confirm that strict factorization holds when there are zero or one colored particles in the initial state. We then discuss the case with two or more initial state particles where strict factorization may fail. We review the calculation of the 1-loop factorization-violating effect in the IR-divergent part of $\Sp$ from~\cite{Catani:2011st} and summarize other results from QCD. In the next sections we will reproduce these results from SCET with Glauber operators.

\subsection{Strict factorization}
Let us start with the situation where $p_\ccO^\mu$ is the momentum of an outgoing quark, $p_\ccT^\mu$ is the momentum of an outgoing gluon 
and none of the other $n-2$ momenta are collinear to $p_\ccO^\mu$ and $p_\ccT^\mu$.
We also take our matrix elements to be of operators with $n-1$ fields, e.g. $\cO = \bar\psi_1 \cdots \psi_{n-1}$.  
 In this situation, the hard-soft-collinear factorization formula in Eq.~\eqref{sQEDmainXY} holds for 
$\ket{\cMb}$ and for $\ket{\cM}$. We can write
\be
\ket{ \cMb} \LPeq
\dfrac
{\bra{P} \bar\psi \, W_\ccO \ket{0}
}
{ \tr\bra{0} Y^\dg_\ccO W_\ccO \ket{0} } \cdot 
\ket{   \cM_{{\black {\text{rest}}}}}
\;,
\qquad
\ket{ \cM }\LPeq
\dfrac
{\bra{p_\ccO,p_\ccT} \bar\psi \, W_\ccO \ket{0}
}
{ \tr\bra{0} Y^\dg_\ccO W_\ccO \ket{0} }
\cdot 
\ket{   \cM_{{\black {\text{rest}}}}}
\;
\ee
Here, spin and color indices are suppressed (see Section 12 of~\cite{FS2} for more details) and $\tr$ indicates a color trace. 
$\ket{\cM_\text{rest}}$ represents the product of the Wilson coefficient with the matrix element of soft Wilson lines and the collinear matrix elements involving momenta in other directions. Critically, the form of $\ket{\cM_\text{rest}}$ is identical for both factorization formulas. Thus the splitting function is the ratio of the two:
\be
\Sp = \frac{
\bra{p_\ccO,p_\ccT} \bar\psi \, W_\ccO \ket{0}
}
{
\bra{P} \bar\psi \, W_\ccO \ket{0}
}
\label{Spscet}
\ee
To be explicit, at tree-level, the splitting function for a right-handed quark and a negative  helicity gluon can be derived in this way~\cite{FS2}
\be
\Sp_{{\spincol R-} }^\zero
(p_\ccO,p_\ccT) =  g_s\, \frac{ \sqrt{2}}{[p_\ccT p_\ccO]} \frac{z}{\sqrt{1-z}}  
\TT_\ccO
\ee
Which agrees with the well-known tree-level QCD splitting amplitudes~\cite{Mangano:1990by}.
Derivations for other amplitude-level splitting functions in this way can be found in~\cite{FS2}. 

Note that in this case (both $p_\ccO^\mu$ and $p_\ccT^\mu$ outgoing), the splitting function depends only on these two momenta and also only on the net color in the $\ccO$ direction. This is what is referred to as strict factorization.

\subsection{Strict factorization violation from $i\pi/\ep$ terms}
Next, let us consider a general process where  $p_\ccO^\mu$ and $p_\ccT^\mu$ are to become collinear and the other momenta point in generic directions. We do not yet specify which particles are incoming or outgoing and want to see what sufficient conditions are for strict factorization to be violated. 

Since $p_\ccO^\mu$ or $p_\ccT^\mu$ may be incoming, we cannot assume the hard-soft-collinear factorization formula in Eq.~\eqref{sQEDmainXY} is correct at leading power in $\lambda^2 = p_\ccO\cdot p_\ccT/Q^2$. However, as long as $p_\ccO^\mu$ and $p_\ccT^\mu$ are not pointing in exactly the same direction (so $\lambda \ne 0$), then the factorization formula guarantees that all of the infrared divergences of the full theory are reproduced in the factorized expression. That is, the Glauber contribution is contained in the soft except at the exceptional point in phase space where $p_\ccO^\mu \propto p_\ccT^\mu$. This is very powerful, as we can then determine the IR divergences of $\ket{\cM}$ and $\ket{\cMb}$ separately and then explore the limit where $p_\ccO^\mu$ becomes collinear to $p_\ccT^\mu$~\cite{Catani:2011st}.

Consider the matrix element $\ket{\cMb}$ of a hard operator with $n-1$ fields. $\ket{\cMb}$ is the matrix element before the emission, where the parton in the $\ccO$ direction has momentum $P^\mu \LPeq p_\ccO^\mu + p_\ccT^\mu$.
\be
\ket{\cMb} = 
\begin{gathered}
\resizebox{50mm}{!}{{
\fmfframe(0,-15)(0,5){
\begin{fmfgraph*}(60,40)
	\fmfleft{L1,L2,L3}
	\fmfright{R1,R2,R3,R4,R5}
    \fmf{fermion,label=$\ccOT$,tension=2,l.s=left}{L2,v}
  	\fmf{gluon,tension=0}{L2,v}
	\fmf{fermion,label=$\cdots~~~n-1$,l.s=right}{v,R1}
	\fmf{phantom,l.s=left,l.d=3}{v,R2}
    \fmf{fermion,label=$n$,l.s=left}{v,R3}
	\fmf{fermion,label=$3$,l.s=left,tension=2}{L1,v}
    \fmf{phantom,l.s=left}{L3,v}
    \fmf{phantom,l.s=right}{v,R5}
    \fmfv{d.sh=circle, d.f=30,d.si=0.1w}{v}
\end{fmfgraph*}
}
}
}
\end{gathered}
\ee
The color associated with $P^\mu$ we write as $\ccOT$. This means that the color operator acts as the sum of the color operators for $p_\ccO^\mu$ and $p_\ccT^\mu$:  $\TT^{\ccOT} \cdot \TX = (\TT^1 + \TT^2)\cdot \TX$.

Now, as long as $p_\ccO^\mu$ and $p_\ccT^\mu$ are not proportional, nothing stops us from treating them as separate hard directions. Then the $\ket{\cM}$ is also given by a the matrix element of a local operator whose infrared divergences are reproduced by the factorized expression
\be
\ket{\cM} = 
\begin{gathered}
\resizebox{50mm}{!}{
\fmfframe(0,-15)(0,5){
     \begin{fmfgraph*}(60,40)
	\fmfleft{L1,L2,L3}
	\fmfright{R1,R2,R3,R4,R5}
	\fmf{fermion,label=$1$,tension=2,l.s=left}{L2,v}
	\fmf{fermion,label=$\cdots~~~n-1$,l.s=right}{v,R1}
	\fmf{gluon,label=$~~~~~2$,l.s=left,l.d=5}{v,R2}
        \fmf{fermion,label=$n$,l.s=left}{v,R3}
	\fmf{fermion,label=$3$,l.s=left,tension=2}{L1,v}
        \fmf{phantom,l.s=left}{L3,v}
        \fmf{phantom,l.s=right}{v,R5}
        \fmfv{d.sh=circle, d.f=30,d.si=0.1w}{v}
    \end{fmfgraph*}
}}
\end{gathered}
\ee
For strict factorization to hold, these two expressions should be related by a splitting function $\Sp$ that depends only on the $\ccOT$ system, independent of the rest of the processes.

Let us now look at the 1-loop corrections to $\ket{\cMb}$ and $\ket{\cM}$. In particular, we are interested in the Glauber contribution since that is where factorization violation might come from. The Glauber contribution is contained in the soft contribution and at 1-loop is purely imaginary, as shown in Section~\ref{sec:isolate}. Indeed, at 1-loop the Glauber contribution is particular simple: it gives a factor of
$\frac{\alpha_s}{2\pi} (\frac{i \pi}{\ep_\text{IR}} - \frac{i \pi}{\ep_\text{UV}})$, as in Eq.~\eqref{IGdr} if the two lines the loop connects are both outgoing or both incoming. Here we keeps the IR poles only as the UV poles will be renormalized away.  In QCD, one gets this 1-loop contribution multiplied by a group theory factor of $\TT_\cci \cdot \TT_\ccj$. For example, a gluon connecting lines $\ccO$ and $\ccTH$ gives
\be
\ket{\cM^{G1} }= 
\begin{gathered}
\resizebox{41mm}{!}{
\fmfframe(0,-15)(0,5){
     \begin{fmfgraph*}(50,40)
	\fmfleft{L1,L2,L3}
	\fmfright{R1,R2,R3,R4,R5}
	\fmf{plain,tension=5,l.s=left}{L2,vg1,v}
	\fmf{fermion,l.s=right}{v,R1}
	\fmf{gluon,l.s=left,l.d=5}{v,R2}
        \fmf{fermion,l.s=left}{v,R3}
	\fmf{plain,l.s=left,tension=5}{L1,vg2,v}
        \fmf{phantom,l.s=left}{L3,v}
        \fmf{phantom,l.s=right}{v,R5}
	\fmf{dbl_dots, fore=red,tension=0}{vg1,vg2}
        \fmfv{d.sh=circle, d.f=30,d.si=0.1w}{v}
	\fmfv{l=$\cci$}{L2}
	\fmfv{l=$\ccj$}{L1}
    \end{fmfgraph*}
}}
\end{gathered}
=
\sum_{\substack{ \cci < \ccj \\
\text{both in or both out}}}
 \TT_\cci \cdot \TT_\ccj\, \frac{\alpha_s}{2 \pi}\frac{i\pi}{\ep} 
\cdot \ket{\cM^0}
\label{eq:oneglauber}
\ee
where $G1$ stands for the 1-loop Glauber contribution to $\ket{\cM}$ and $\ket{\cM^0}$ is the $\ket{\cM}$ to lowest
order in $g_s$. We have neglected the UV pole in Eq.~\eqref{eq:oneglauber}. Although we draw the contribution as a dotted red line, as in the SCET approach, all we are
using here is that the Glauber contribution is identified with the imaginary part at 1-loop which follows from any method of computation. 
The 1-loop Glauber corrections to $\ket{\cMb}$ is given by the same formula, but summed over the $n-1$ partons. 

Now let us first consider the case where all the particles are outgoing or all incoming. We can use color conservation to simplify the expression,
$\sum_\ccj \TT_j = 0$ and $\TT_i \cdot \TT_i = C_i$ where $C_i$ is the group casimir (a number not an operator). This gives
\be
\ket{\cM^{G1}} =   
 \frac{\alpha_s}{2 \pi}\frac{i\pi}{\ep}
\frac{1}{2} \sum_{\cci = 1}^n  \TT_\cci \cdot(-\TT_\cci)  \cdot \ket{\cM^0}= 
- \frac{\alpha_s}{4 \pi}\frac{i\pi}{\ep}
\sum_{\cci=1}^n C_\cci 
 \cdot \ket{\cM^0}
 \hspace{10mm} \text{(all outgoing)}
\ee
For $\ket{\cMb}$ we have
\be
\ket{  {\cMb}^{G1}} =  
 \frac{\alpha_s}{4 \pi}\frac{i\pi}{\ep}
\left[ -\TT_{\ccOT}\cdot \TT_{\ccOT} -  \sum_{\cci = 3}^n  \TT_\cci \cdot \TT_\cci \right]
\cdot \ket{\cMb^0} = -
 \frac{\alpha_s}{4 \pi}\frac{i\pi}{\ep}
  \left[ C_{\ccOT} + \sum_{\cci=3}^n C_\cci \right] \cdot \ket{\cMb^0}
\ee
The splitting function has to reproduce the ratio of these two. Thus
\be
i\text{Im}\,\Sp^\one
= 
\frac{\alpha_s}{4 \pi}\frac{i\pi}{\ep}
 \left[ C_{\ccOT} - C_\ccO -  C_\ccT \right]\cdot \Sp^\zero  + \text{finite terms} \hspace{10mm} \text{(all outgoing)} 
\label{Spout}
\ee
This color factor only depends on the colors of the particles in the $\ccO\ccT$ sector, independent of the rest of the event, consistent with strict collinear factorization.

Next, suppose that $p_\ccO^\mu$ is in the initial state but all other particles are in the final state. Then there is no Glauber contribution from anything connecting to $p_\ccO^\mu$ nor from anything connecting to $P^\mu = p_{\ccOT}^\mu$ in the case of $\ket{\cMb}$. With this arrangement,
\be
\ket{\cM^{G1}} =    \frac{\alpha_s}{4 \pi}\frac{i\pi}{\ep}
\sum_{\cci = 2}^n  \TT_\cci \cdot(-\TT_\cci - \TT_\ccO ) 
\cdot \ket{\cM^0}
=
 - \frac{\alpha_s}{4 \pi}\frac{i\pi}{\ep}
\left[- C_\ccO + \sum_{\cci=2}^n C_\cci   \right]
\cdot \ket{\cM^0}
\hspace{10mm} \text{(one incoming)}
\ee
where $\sum_{\cci=2}^n \TT_i = - \TT_1$ has been used twice. Similarly,
\be
\ket{ \cMb^{G1}} =    \frac{\alpha_s}{4 \pi}\frac{i\pi}{\ep}
\sum_{\cci = 3}^n  \TT_\cci \cdot(-\TT_\cci - \TT_{\ccOT})
\cdot \ket{\cMb^0}
= - \frac{\alpha_s}{4  \pi } \frac{i \pi}{\ep} \left[ - C_\ccOT + \sum_{\cci=3}^n C_\cci \right]
\cdot \ket{\cMb^0}
\hspace{3mm}(\ccO~\text{incoming)}
\ee
Thus, with one incoming colored particle
\be
i \text{Im}\,\Sp^\one
=  \frac{\alpha_s}{4 \pi}\frac{i\pi}{\ep}
 \left[ - C_{\ccOT} + C_\ccO  - C_\ccT \right]  \cdot \Sp^\zero  + \text{finite terms}  \hspace{10mm} \text{(one incoming)}
 \ee
Although this splitting function is different from the all outgoing case, Eq.~\eqref{Spout}, both depend only on the colors of the particles in the $\ccO \ccT$ sector. Thus with one incoming particle, we cannot conclude that strict collinear factorization is violated.

Now suppose $p_\ccO^\mu$ and another particle, $p_\ccTH^\mu$ are both in the initial state, with $p_\ccTH^\mu$ still generic (not collinear to any other direction). For $\cM$, which has $p_\ccT^\mu$ outgoing, we get
\begin{align}
\ket{\cM^{G1}} &=   
 \frac{\alpha_s}{2 \pi}\frac{i\pi}{\ep}
\left[ \TT_\ccO \cdot \TT_\ccTH +\frac{1}{2} \sum_{\cci = 2,4\cdots n}  \TT_\cci \cdot(-\TT_\cci - \TT_\ccO - \TT_\ccTH ) \right] 
\cdot \ket{\cM^0}\\
&= 
 \frac{\alpha_s}{2 \pi}\frac{i\pi}{\ep}
  \left[2 \TT_\ccO \cdot \TT_\ccTH+ \frac{1}{2} C_\ccO 
 + \frac{1}{2} C_\ccTH   
-\frac{1}{2} C_\ccT -\frac{1}{2}\sum_{\cci = 4}^n C_\cci  
 \right]
\cdot \ket{\cM^0}
\hspace{2mm} (\ccO~\text{and}~\ccTH~\text{incoming})
\label{twoin}
\end{align}
The matrix element $\ket{\cMb}$, correspondingly has $\ccOT$ and $\ccTH$ incoming. Its 1-loop Glauber contribution is
\be
\ket{\cMb^{G1}} = 
 \frac{\alpha_s}{2 \pi}\frac{i\pi}{\ep}
  \left[2 \TT_\ccOT \cdot \TT_\ccTH+ \frac{1}{2} C_\ccOT 
 + \frac{1}{2} C_\ccTH   
 -\frac{1}{2}\sum_{\cci = 4}^n C_\cci  
 \right]
 \cdot \ket{\cMb^0}
\ee
So we find
\be
\Sp^{1,\text{non-fact}}=
i \text{Im}\,\Sp^\one
= 
\frac{\alpha_s}{4 \pi}\frac{i\pi}{\ep}
\left[ - C_{\ccOT} + C_\ccO  - C_\ccT - 4\TT_\ccT \cdot \TT_\ccTH \right] \cdot  \Sp^\zero
 + \text{finite terms} 
\hspace{3mm} (\ccO~\text{and}~\ccTH~\text{incoming})
\label{SPG31}
 \ee
 where $\Spnf$ denotes strict-factorization-violating contributions.
  This form indicates a violation of strict collinear factorization: the splitting function depends on the color of particles in the matrix element other than those involved in the splitting ($\TT_\ccTH$ in this case). 
The result in Eq.~\eqref{SPG31} was first derived in Ref.~\cite{Catani:2011st} by examining IR singularities of full QCD amplitudes~\cite{Catani:1998bh,Becher:2009qa,Gardi:2009qi} in different kinematical regions. 
authors of Ref.~\cite{Catani:2011st} also derived the 1-loop finite part and the 2-loop IR singular part of factorization violating splitting amplitudes.  

\subsection{Strict-factorization violation from full QCD}
Next, we summarize some known additional results about factorization-violation from full QCD, including the 1-loop finite parts and the 2-loop divergent parts of $\Spnf$.

The IR divergent part of a 1-loop amplitude is defined relative to the tree-level amplitude as
  \be
 \ket{\cM^1} = \bI^1(\ep) \ket{\cM^0} + \ket{\cM^{1,\text{fin.}}} \label{Mrat}
 \ee
 where $ \ket{\cM^{1,\text{fin.}}}$ is a finite, analytic function of the momenta. 
 The general expression for  $\bI^1(\ep)$ follows from Eq.~\eqref{QEDv} with the appropriate sum over pairs of external legs to which the virtual gluon can attach and appropriate color factors. 
 The ${1}/{\ep^2}$ poles are color diagonal.
 Using color conservation to simplify the result, an amplitude in QCD with $n$ external partons with colors $\TT_i$ has IR divergences
 given by~\cite{Catani:1998bh}
 \be
 \bI^1(\ep)
   = \frac{\alpha_s}{2 \pi} \frac{1}{2} \left[
   -\sum_{i=1}^n \left(\frac{C_i}{\ep^2} + \frac{\gamma_i}{\ep}\right) -
   \frac{1}{\ep}\sum_{i\ne j }^n
     \TT_i \cdot \TT_j \ln  \frac{- s_{\cci\ccj}-i\fme  }{\mu^2}
   \right]
   \label{div1loop}
 \ee
 Here $\gamma_i$ is the regular (non-cusp) anomalous dimension: $\gamma_q =3 C_F/2$ for quarks and $\gamma_g = \beta_0 = \frac{11}{6} C_A - \frac{2}{3} T_F n_f$ for gluons.  Although anomalous dimensions are usually associated with UV divergences, they appear in this expression because they can be extracted using properties of scaleless integrals in dimensional regularization, in which the UV and IR divergences cancel~\cite{Becher:2009qa}.

We are interested here in the order-by-order expansion of the splitting amplitudes. We write
 \be
 \ket{\cM^0 + \cM^1 + \cdots} \LPeq (\Sp^0 + \Sp^1 + \cdots)\ket{\cMb^0 +\cMb^1 + \cdots}
 \ee
 So that $\ket{\cM^0} = \Sp^0 \ket{\cMb^0}$ at tree-level, $\ket{\cM^1} = \Sp^0 \ket{\cMb^1} + \Sp^1 \ket{\cMb^0}$ at 1-loop, 
 and so on. It can be helpful to separate out the divergent parts of the splitting function too. We define,
 \be
 \Sp^1 = \bI_C^1 \cdot \Sp^0 + \Sp^{1,\text{fin.}}
 \label{Sp1}
 \ee
 where $\Sp^{1,\text{fin.}}$ is an IR-finite analytic function of momenta. All the IR divergences are absorbed in $\bI_C^1$.  It is not hard to show that~\cite{Catani:2011st,Forshaw:2012bi}
\be
 \bI_C^1 = \bI^1 - \bIb^1
 \label{IC1}
 \ee
 where $\bIb^1$ is the divergent part of $\ket{\cMb^1}$, as in Eq.~\eqref{Mrat}. 

Plugging in Eq.~\eqref{div1loop}  gives
\begin{multline}
\bI_C^1 = 
\frac{\alpha_s}{2 \pi}
 \frac{1}{2} \left[
\frac{C_\ccOT - C_\ccO - C_\ccT}{\ep^2} 
-\frac{\gamma_\ccOT - \gamma_\ccO - \gamma_\ccT}{\ep}
-\frac{2}{\ep} 
 \TT_\ccO \cdot \TT_\ccT \ln \frac{-s_{\ccO \ccT} -i\fme}{\mu^2} 
\right.\\
\left.
- \frac{2}{\ep}\sum_{j=3}^n \left(
  \TT_\ccO \cdot \TT_\ccj \ln \frac{-s_{\ccO \ccj} -i\fme}{\mu^2}
 +  \TT_\ccT \cdot \TT_\ccj \ln \frac{- s_{\ccT \ccj} - i\fme }{\mu^2} 
 - \TT_\ccOT \cdot \TT_\ccj \ln \frac{- s_{\ccOT \ccj} - i\fme }{\mu^2} 
 \right)
\right]
\label{IC1gen}
\end{multline}
The factorization violation in $\Sp^1$ is contained in the imaginary part of $\bI_C^1$. Explicity,
when there are two incoming momenta,
  \be
i \text{Im}\,\bI_C^1
= 
\frac{\alpha_s}{4 \pi}\frac{i\pi}{\ep}
\left[ - C_{\ccOT} + C_\ccO  - C_\ccT - 4\TT_\ccT \cdot \TT_\ccTH \right]   
\hspace{10mm} (\ccO~\text{and}~\ccTH~\text{incoming})
\label{SPG31b}
 \ee
 in agreement with Eq.~\eqref{SPG31}. 

The same approach can be used to extract the finite parts of the 1-loop splitting functions. 
The expression for $\bI_C^1$ from~\cite{Catani:2011st}, including terms that are IR-finite is
 \begin{multline}
 \bI_C^1 =  c_\Gamma \left(\frac{-s_{\ccO\ccT} - i\fme}{\mu^2} \right)^{-\ep} \frac{\alpha_s}{2 \pi} \frac{1}{2} \left\{ \frac{C_\ccOT - C_\ccO - C_\ccT}{\ep^2} + \frac{\gamma_\ccOT - \gamma_1 -\gamma_2 + \beta_0}{\ep}
 \right.\\
 \left.
 +\frac{2}{\ep} \left[
  \sum_{j=3}^n \TT_\ccO \cdot \TT_\ccj  f(\ep,1-z)
 + \sum_{j=3}^n \TT_\ccT \cdot \TT_\ccj  f(\ep,z - i\fme s_{\ccj \ccT})
 \right]
 \right\}
 \label{ICfine}
 \end{multline}
 where $ c_\Gamma \equiv    \frac{\Gamma(1+\ep)\Gamma^2(1-\ep)}{(4\pi)^{2-\ep}\Gamma(1-2\ep)} $.
 The function $f(\ep,z)$ is a hypergeometric function, defined by
 \be
 f(\ep,1/x) \equiv \frac{1}{\ep} \Big[ {}_2F_1(1,-\ep; 1-\ep; 1-x) -1\Big] = \ln x - \ep \Li_2(1-x) + \cO(\ep^2)
 \ee
 The convention taken is that $1-z>0$. This can be assumed for timelike or spacelike splittings  without loss of generality since $z+(1-z) = 1$ (i.e. $z$ can still be positive or negative).  Thus the $f(\ep,1-z)$ factor in Eq.~\eqref{ICfine} is real and unambiguous.
  The function $f(\ep,z)$ has a cut for negative real $z$, i.e. for spacelike splittings. 
 Writing $f(\ep,z - i\fme p_\ccj \cdot p_\ccT)$ in Eq.~\eqref{ICfine} makes the result well-defined. In particular, the sign of the imaginary part of the function is determined by the sign of $p_\ccj \cdot p_\ccT$. This dependence of the analytic continuition on the momentum $p_\ccj$ obstructs the reduction  of Eq.~\eqref{ICfine} to a form that obeys strict factorization.

To see the factorization violating part of the $\ep^0$ term, we can use
 \be
 \text{Im} \, \Li_2 (1-\frac{1}{z \pm i\fme  } ) =  \pm \pi \ln(1-\frac{1}{z})
 \ee
 Then the factorization violating contribution is seen to be
   \begin{multline}
\Sp^{1,\text{non-fact}} = i \text{Im}\,\Sp^1  \\
 = 
c_\Gamma \,\frac{\alpha_s}{4 \pi} (i \pi) \left( \frac{1}{\ep} + \ln\frac{z-1}{z} + \ln \frac{ \mu^2 }{ -   s_{\ccO  \ccT }  } \right)  
\left[  - 4\TT_\ccT \cdot \TT_\ccTH \right] \,\Sp^0 + \cdots 
\hspace{5mm} (\ccO~\text{and}~\ccTH~\text{incoming})
\label{SPG31f}
 \end{multline}
where the $\cdots$ include terms that do not violate strict factorization, such as the $C_\ccO$ and $C_\ccT$ terms in Eq.~\eqref{SPG31b}.

The factorization-violating IR-divergent part of the 2-loop splitting function is also presented in~\cite{Catani:2011st}. 
The 2-loop splitting amplitude can be written as
\be
\Sp^{2} = \left[ \bD_C^2(\ep)  + 
\frac{1}{2}  (\bI_C^1)^2 \right] \cdot \Sp^0 + \frac{1}{\ep}~\text{terms} + \text{finite}
\ee
where
 \begin{multline}
 {\bD}^{2}_C(\ep) 
  =  \left(  \frac{\alpha_s}{2 \pi} \right)^2 \left( \frac{- s_{\ccO \ccT} - i\fme}{\mu^2} \right)^{-2 \ep} \pi f_{abc} \sum_{\cci=1,2}  \sum_{\ccj,\cck =3}^{n}
 \TT^a_{\cci} \TT^b_\ccj \TT^c_\cck \Theta(-z_i) \text{sign}(s_{\cci \ccj}) \Theta(-s_{\ccj \cck})  \\
   \times \ln \left(  - \frac{s_{\ccj P} s_{\cck P} z_1 z_2}{ s_{\ccj \cck} s_{\ccO \ccT}}  - i\fme \right) \left[ - \frac{1}{2 \ep^2} + \frac{1}{\ep} \ln\left(  \frac{-z_i}{ 1- z_i} \right) \right]  
  \end{multline}     
 This contribution is non-vanishing only if the amplitude involves  both incoming and outing colored partons.  
 The most important part of this 2-loop splitting function is the real component, since it can contribute to the cross section. 
 The real part is contained in the anti-Hermitian combination of $\bD^{2}_C (\ep)$  is
 \begin{multline}  \label{delta2real}
 \frac12 \left( {\bD}^{2}_C(\ep) - \widetilde{\bD}^{2, \dg}_C(\ep) \right)   = 
   -i  \frac{\alpha_s^2}{4}  \left( \frac{- s_{\ccO \ccT} - i\fme}{\mu^2} \right)^{-2 \ep}  f_{abc} \sum_{i=1,2}  \sum_{j,k =3}^{n}
 \TT^a_\cci \TT^b_\ccj \TT^c_\cck \Theta(-z_i) \text{sign}(s_{ij}) \Theta(-s_{jk})  \\
   \times  \left[ - \frac{1}{2 \ep^2} + \frac{1}{\ep} \ln\left(  \frac{-z_i}{ 1- z_i} \right) \right]  
 \end{multline}  
 In a representation where the $\TT_\ccj^a$ are purely imaginary, this contribution is purely real.

\section{Factorization violation from SCET \label{sec:fvscet}}
We have seen that splitting functions violate strict factorization starting at 1-loop. The condition for strict-factorization violation is that there be more than one colored particle in both the initial and final state. In such a situation the amplitude for producing a final-state particle collinear to one of of the initial-state particles (a spacelike splitting) depends on the colors and momenta of particles not collinear to it. The factorization-violating contribution was derived in~\cite{Catani:2011st,Forshaw:2012bi} by taking limits of the full $n+1$-body matrix elements in QCD. We reviewed the procedure for the 1-loop IR divergent part and discussed the extension to also include the finite part and to 2-loops.  

In SCET the IR divergences of both $\ket{\cM}$ and $\ket{\cMb}$ agree with full QCD, so the derivation of Eq.~\eqref{IC1gen}, which encodes 1-loop factorization violation, could be done in SCET by taking limits of $n+1$ amplitudes, as in QCD. Note that {\it all} the IR divergences of QCD are reproduced in SCET without any special consideration of the Glauber contribution (i.e. because Glauber modes are contained in soft modes for hard scattering, as discussed in Section~\ref{sec:contain}). However, SCET should also be able to produce the splitting functions using an effective field theory constructed using only the $n$ collinear directions of $\ket{\cM}$, i.e. not by taking the limit of an effective field theory with $n+1$ collinear directions. In that case, we {\it do} need the Glauber operators.
We we would like to know  is whether the generalized splitting function can be derived without knowing features of the full $n+1$ body amplitudes by the addition of Glauber operators to SCET.

\subsection{Tree-level splitting amplitudes in SCET}
To begin, let's discuss how a splitting amplitude would be calculated  in SCET with soft and collinear modes, but no Glauber operators.
SCET without Glaubers can produce the strict-factorization-preserving splitting amplitude $\Spfa$, but not the factorization-violating parts in 
$\Spnf$. 
Recall our notation that $P^\mu\LPeq p_\ccO^\mu + p_\ccT^\mu$ is mother parton momentum. We take $P^\mu$ and the  daughter momentum $p_\ccO^\mu$ to be incoming and $p_\ccT^\mu$ to be outgoing.  The strictly-factorizing splitting amplitude is then as in Eq.~\eqref{Spscet}:
\be
\Spfa = \frac{\bra{p_\ccT} W^\dg_\ccO \,  \psi \ket{p_\ccO}}
{\bra{0}W^\dg_\ccO \,  \psi  \ket{P}}
\label{Spscet2}
\ee
Here $\bar{\psi}$ is an ordinary Dirac fermion and $W_\ccO$ is a collinear Wilson line pointing in a direction $t^\mu$ not collinear to $P^\mu$. In the traditional formulation of SCET $\bar{\psi}$ carries a label specifying the large and perpendicular components of its momentum, and the interactions of $\psi$ with collinear gluons are power expanded. However it is simpler to use the full QCD Feynman rules as we do here.

The tree-level splitting amplitudes are easily computed from Eq.~\eqref{Spscet2} (see~\cite{FS1}):
\be
\Sp^{0} = \frac{\bra{p_\ccT}W^\dg_\ccO \,  \psi \ket{p_\ccO}^\text{tree}}
{\bra{0} W^\dg_\ccO \,  \psi  \ket{P}^\text{tree}}
=~~~
\begin{gathered}
\resizebox{25mm}{!}{
     \fmfframe(0,0)(0,0){
\begin{fmfgraph*}(40,15)
   \fmfstraight
   \fmfleft{L1,L2}
   \fmfright{R1,R2}
   \fmf{fermion}{L1,v,R1}
   \fmf{gluon,tension=0,label=$p_\ccT\nearrow$,l.s=left}{v,R2}
   \fmf{phantom,tension=0,label=$p_\ccO \leftarrow$,l.s=left}{L1,v}
   \fmfv{d.sh=circle, d.f=30,d.si=0.2w}{R1}
\end{fmfgraph*}
}
}
\end{gathered}
~~~
+
~~~
\begin{gathered}
\resizebox{25mm}{!}{
     \fmfframe(0,0)(0,0){
\begin{fmfgraph*}(40,15)
   \fmfstraight
   \fmfleft{L1,L2}
   \fmfright{R1,R2}
   \fmf{phantom}{L1,v,R1}
   \fmf{gluon,tension=0}{v,R2}
   \fmf{fermion,tension=0}{v,R1}
   \fmfv{d.sh=circle, d.f=30,d.si=0.2w}{v}
\end{fmfgraph*}
}
}
\end{gathered}
-
~~~
\begin{gathered}
\resizebox{25mm}{!}{
     \fmfframe(0,10)(0,0){
\begin{fmfgraph*}(40,15)
   \fmfstraight
   \fmfleft{L1}
   \fmfright{R1}
   \fmf{fermion}{L1,R1}
   \fmfv{d.sh=circle, d.f=30,d.si=0.2w}{R1}
\end{fmfgraph*}}}
\end{gathered}
\ee

The first diagram has the gluon coming off the fermion from a Lagrangian interaction, the second diagram has the gluon coming out of the Wilson line. These graphs evaluate to
\be
 \bra{p_\ccT} W^\dg_\ccO \,  \psi  \ket{p_\ccO}^{\text{tree}}
=
g_s\,\TT_\ccO\ \bigg[
\frac{t \cdot\pol}{t \cdot p_\ccT}
-
\frac{\cn{\pol}(p\slash_\ccO+  p\slash_\ccT)}
{(p_\ccO+p_\ccT)^2}
\bigg] v(p_\ccO) 
\label{Mte}
\ee
The dependence on the Wilson line direction $t^\mu$ drops out for physical polarizations.
The result is~\cite{FS1}
\be
\Sp_{{\spincol R-} }^0 =
  g_s\, \frac{ \sqrt{2}}{[p_\ccT p_\ccO]} \frac{z}{\sqrt{1-z}}  
\TT_\ccO,
\qquad
\Sp_{{\spincol R+} }^0 =
  g_s\, \frac{ \sqrt{2}}{\l p_\ccO p_\ccT \r } \frac{1}{\sqrt{1-z}}  
\TT_\ccO,
\label{SpRL}
\ee
$\Sp_{{\spincol L\pm} }^0 $ are related to $\Sp_{{\spincol R\mp} }^0$ by parity conjugation. 
Here ${\spincol R}$ and  ${\spincol L}$ refer to the spin of the fermion (right or left) and ${\spincol \pm}$ refer to the helicities of the emitted gluon. These tree-level splitting functions hold for any kinematical configuration, as strict-factorization holds at tree level.

In Eq.\eqref{Mte}, we used QCD Feynman rules, as is appropriate for evaluating the splitting function defined in Eq.~\eqref{Spscet2}. This is based on the reformulation of SCET in~\cite{FS1} and~\cite{FS2}. In traditional SCET, the splitting function is computed from 
$\bra{p_\ccT} W_{\ccol \bar{n}}^\dg \, \xi_n  \ket{p_\ccO}$ with $\xi_n$ a collinear quark, i.e. one whose interactions are truncated to leading power. The denominator in Eq.~\eqref{Spscet2} is replaced by the diagram-level zero-bin subtraction procedure outlined in~\cite{Manohar:2006nz}. 
The SCET Feynman rules~\cite{Bauer:2000yr}  then give
\be
\bra{p_\ccT}   W_{\ccol \bar{n}}^\dg \, \xi_n  \ket{p_\ccO}^\text{tree}   
= 
- g_s \TT_\ccO \left[ \frac{n\slash}{ 2}  \frac{ n \cdot (p_\ccO + p_\ccT) }{ (p_\ccO + p_\ccT)^2}  \left(  n \cdot \pol   + \frac{ p\slash_{\ccT,\perp} \pol\slash_{\perp} }{ \bar{n} \cdot ( p_\ccO + p_\ccT )} \right)\frac{\bar{n}\slash }{2}+ \frac{\bar n \cdot \pol }{ \bar n \cdot p_\ccT }  \right] v_n (p_\ccO)
  \label{SCETsplit}
\ee
where $p_\ccO^\mu \propto n^\mu$ has been used and the Wilson line direction $t^\mu$ is set to $\bar{n}^\mu$. Simplifying this expression for the various helicity/spin combinations gives the same splitting functions as in Eq.~\eqref{SpRL}.

To compute the 1-loop corrections to $\Sp^{\text{fact}}$ we need to evaluate Eq.~\eqref{Spscet2} to next order. 
non-collinear sectors encapsulated by the Wilson lines. 
Evaluating the relevant graphs should produce a result equivalent to known results about 1-loop splitting amplitudes from full QCD. This calculation has not been done in SCET, to our knowledge, and would certainly be a interesting check on the formalism that we leave to future work.


\subsection{Factorization violating contributions \label{sec1loop}}
To compute the contributions to the generlized splitting function that violate strict-factorization, we obviously cannot start from the factorized expression Eq.~\eqref{sQEDmainXY} which leads to Eq.~\eqref{Spscet2}. The advantages of writing a factorized expression as in Eq.~\eqref{sQEDmainXY} include first, that it involves only QCD fields and the familiar QCD Feynman rules, and second, that the soft-collinear overlap is removed through an operator matrix element. The inclusion of Glauber effects has so far only been formulated in the traditional presentation of SCET~\cite{Rothstein:2016bsq}, with collinear and soft fields and their associated SCET Feynman rules and with the overlap removed by a diagram-by-diagram zero-bin subtraction procedure. In this approach, one writes matrix elements in the effective theory as one big operator product
\be
\ket{\cM} \LPeq
\cC(\Sij) \, 
\bra{p_\ccT; \cdots X_\ccN; X_S} \phi^\star W_\ccO Y_\ccO^\dagger \cdots W_\ccN^\dagger Y_\ccN \phi \ket{p_\ccO; X_\ccTH}
\label{Mscet}
\ee
Similarly, 
\be
\ket{\cMb} \LPeq
\cC(\Sij) \, 
\bra{ \cdots X_\ccN; X_S} \phi^\star W_\ccO Y_\ccO^\dagger \cdots W_\ccN^\dagger Y_\ccN \phi \ket{P; X_\ccTH}
\label{Mbscet}
\ee
The generalized splitting function can be computed as the ratio of these matrix elements using SCET Feynman rules and appropriate zero-bin subtractions. 

As indicated in Section~\ref{sec:scetg}, Glauber effects are included in SCET through the addition of potential operators that couple pairwise all possible fields in all possible collinear sectors. These operators are schematically of the form $\cO_{G} \sim \bar{\psi}_\cci\psi_\cci \frac{1}{\cP_\perp^2}\bar{\psi}_\ccj \psi_\ccj$, as in 
Eq.~\eqref{OpQED}, with a plethora of terms in QCD, as in Eq.~\eqref{OpQCD}. There are different operators coupling quarks to quarks, gluons to gluons and quarks to gluons. See~\cite{Rothstein:2016bsq} for all the details.

Most of the time, the effects of these Glauber operators are identical to the effects of the Glauber limit of the soft diagrams (connecting the soft Wilson lines $Y_\ccj$ in Eq.~\eqref{Mscet}). Thus one must either subtract the overlap, as is done in~\cite{Rothstein:2016bsq}, or more simply compute the full soft graphs without the soft-Glauber subtraction, and not bother including the Glauber operator contribution when it is not needed. Thus for example, the following graphs contribute to $\ket{\cM}$:
\be
I_\text{Sa} =
\begin{gathered}
\resizebox{30mm}{!}{
     \fmfframe(0,10)(0,10){
\begin{fmfgraph*}(50,40)
    \fmfstraight
    \fmfleft{L1,La,Lb,Lc,L2}
    \fmfright{R1,Ra,Rb,Rc,Rd,R2}
    \fmf{plain}{L1,v1,v11,C}
    \fmf{plain}{L2,v2,v22,C}
    \fmf{plain}{C,w1,w11,R1}
    \fmf{plain}{C,w2,w22,R2}
    \fmf{gluon,tension=0}{v2,Rd}
    \fmf{plain,tension=0}{v2,Rd}
    \fmf{gluon,tension=0,fore=(0,,0.5,,0),label=$~~~\text{{\green soft}}$}{w1,w2}
    \fmfv{d.sh=circle, d.f=30,d.si=0.1w}{C}
    \fmfv{l=$\ccO$}{L2}
    \fmfv{l=$\ccTH$}{L1}
    \fmfv{l=$p_\ccT$}{Rd}
\end{fmfgraph*}
}
}
\end{gathered}
,
\hspace{2cm}
I_\text{Ga} =
\begin{gathered}
\resizebox{30mm}{!}{
     \fmfframe(0,10)(0,10){
\begin{fmfgraph*}(50,40)
    \fmfstraight
    \fmfleft{L1,La,Lb,Lc,L2}
    \fmfright{R1,Ra,Rb,Rc,Rd,R2}
    \fmf{plain}{L1,v1,v11,C}
    \fmf{plain}{L2,v2,v22,C}
    \fmf{plain}{C,w1,w11,R1}
    \fmf{plain}{C,w2,w22,R2}
    \fmf{gluon,tension=0}{v2,Rd}
    \fmf{plain,tension=0}{v2,Rd}
    \fmf{dbl_dots,fore=red,tension=0,label=$~~~\text{{\red Glauber}}$}{w1,w2}
    \fmfv{d.sh=circle, d.f=30,d.si=0.1w}{C}
    \fmfv{l=$\ccO$}{L2}
    \fmfv{l=$\ccTH$}{L1}
    \fmfv{l=$p_\ccT$}{Rd}
\end{fmfgraph*}
}
}
\end{gathered} 
\label{ISG}
\ee
Here the horizontal gluon with a line through it is a collinear emission, the vertical gluon is soft, and the dots are Glauber.  
Since the Glauber limit of $I_\text{Sa}$ gives exactly $I_\text{Ga}$, we can simply compute $I_\text{Sa}$ without zero-bin subtracting and not include $I_\text{Ga}$. Moreover, $I_\text{Sa}$ gives the same result as the analogous contribution to $\ket{\cMb}$ (the graph is the same without the emitted gluon), thus we can ignore both of these graphs when computing the 1-loop splitting amplitude. 

The Glauber graphs which cannot be ignored are those for which there is not a corresponding soft graph. In SCET, the interactions of soft gluons with collinear fields are completely removed from the Lagrangian; they only come from the Wilson lines in Eqs.~\eqref{Mscet} and~\eqref{Mbscet}. These can be drawn coming out of the blob, since that represents the operator containing the Wilson lines, or slightly shifted away from the blob, as in the diagrams in Eq.~\eqref{ISG}.  Thus thus graphs like
\be
I_\text{Sb} =
\begin{gathered}
\resizebox{30mm}{!}{
     \fmfframe(0,10)(0,10){
\begin{fmfgraph*}(50,40)
    \fmfstraight
    \fmfleft{L1,La,Lb,Lc,L2}
    \fmfright{R1,Ra,Rb,Rc,Rd,R2}
    \fmf{plain}{L1,v1,v11,C}
    \fmf{plain}{L2,v2,v22,C}
    \fmf{plain}{C,w1,w11,R1}
    \fmf{plain}{C,w2,w22,R2}
    \fmf{gluon,tension=0}{v2,Rd}
    \fmf{plain,tension=0}{v2,Rd}
    \fmf{gluon,tension=0,fore=(0,,0.5,,0)}{v11,v22}
    \fmfv{d.sh=circle, d.f=30,d.si=0.1w}{C}
    \fmfv{l=$\ccO$}{L2}
    \fmfv{l=$\ccj$}{L1}
    \fmfv{l=$p_\ccT$}{Rd}
\end{fmfgraph*}
}
}
\end{gathered}
,
\hspace{2cm}
I_\text{Gb} =
\begin{gathered}
\resizebox{30mm}{!}{
     \fmfframe(0,10)(0,10){
\begin{fmfgraph*}(50,40)
    \fmfstraight
    \fmfleft{L1,La,Lb,Lc,L2}
    \fmfright{R1,Ra,Rb,Rc,Rd,R2}
    \fmf{plain}{L1,v1,v11,C}
    \fmf{plain}{L2,v2,v22,C}
    \fmf{plain}{C,w1,w11,R1}
    \fmf{plain}{C,w2,w22,R2}
    \fmf{gluon,tension=0}{v2,Rd}
    \fmf{plain,tension=0}{v2,Rd}
    \fmf{dbl_dots,fore=red,tension=0}{v11,v22}
    \fmfv{d.sh=circle, d.f=30,d.si=0.1w}{C}
    \fmfv{l=$\ccO$}{L2}
    \fmfv{l=$\ccj$}{L1}
    \fmfv{l=$p_\ccT$}{Rd}
\end{fmfgraph*}
}
}
\end{gathered} 
\label{ISG2}
\ee
give identical Glauber contributions to $\ket{\cM}$. We can therefore ignore both due to the soft-Glauber zero-bin subtractions. 
Note that while the soft graph factorizes into the product of a soft matrix element and a collinear emission, the Glauber graph does
not factorize. Thus we cannot claim that $I_\text{Gb}$ is identical to the contribution from the analagous graph contributing to $\ket{\cMb}$. 
Instead, we need to look at the graphs contributing to $\ket{\cMb}$:
\be
\bar{I}_\text{Sb} =
\begin{gathered}
\resizebox{30mm}{!}{
     \fmfframe(0,10)(0,10){
\begin{fmfgraph*}(50,40)
    \fmfstraight
    \fmfleft{L1,La,Lb,Lc,L2}
    \fmfright{R1,Ra,Rb,Rc,Rd,R2}
    \fmf{plain}{L1,v1,C}
    \fmf{plain}{L2,v2,C}
    \fmf{plain}{C,w1,R1}
    \fmf{plain}{C,w2,R2}
    \fmf{gluon,tension=0,fore=(0,,0.5,,0)}{v1,v2}
    \fmfv{d.sh=circle, d.f=30,d.si=0.1w}{C}
    \fmfv{l=$\ccO$}{L2}
    \fmfv{l=$\ccj$}{L1}
\end{fmfgraph*}
}
}
\end{gathered}
,
\hspace{2cm}
\bar{I}_\text{Gb} =
\begin{gathered}
\resizebox{30mm}{!}{
     \fmfframe(0,10)(0,10){
\begin{fmfgraph*}(50,40)
    \fmfstraight
    \fmfleft{L1,La,Lb,Lc,L2}
    \fmfright{R1,Ra,Rb,Rc,Rd,R2}
    \fmf{plain}{L1,v1,v11,C}
    \fmf{plain}{L2,v2,v22,C}
    \fmf{plain}{C,w1,w11,R1}
    \fmf{plain}{C,w2,w22,R2}
    \fmf{dbl_dots,fore=red,tension=0}{v11,v22}
    \fmfv{d.sh=circle, d.f=30,d.si=0.1w}{C}
    \fmfv{l=$\ccO$}{L2}
    \fmfv{l=$\ccj$}{L1}
\end{fmfgraph*}
}
}
\end{gathered} 
\ee
These have identical Glauber contributions and can be dropped by the zero-bin subtraction. 

The Glauber graphs that have no corresponding soft graph are:
\be
I_\text{Gc}^{\ccj} =
\begin{gathered}
\resizebox{30mm}{!}{
     \fmfframe(0,10)(0,10){
\begin{fmfgraph*}(50,40)
    \fmfstraight
    \fmfleft{L1,La,Lb,Lc,L2}
    \fmfright{R1,Ra,Rb,Rc,Rd,R2}
    \fmf{plain}{L1,v1,v11,C}
    \fmf{plain}{L2,v2,v22,C}
    \fmf{plain}{C,w1,w11,R1}
    \fmf{plain}{C,w2,w22,R2}
    \fmf{gluon,tension=0}{v2,Rd}
    \fmf{plain,tension=0}{v2,Rd}
    \fmf{dbl_dots,tension=0,fore=red}{v1,v2}
    \fmfv{d.sh=circle, d.f=30,d.si=0.1w}{C}
    \fmfv{l=$\ccO$}{L2}
    \fmfv{l=$\ccj$}{L1}
    \fmfv{l=$p_\ccT$}{Rd}
\end{fmfgraph*}
}
}
\end{gathered}
,
\hspace{1.5cm}
I_\text{Gd}^{\ccj} =
\begin{gathered}
\resizebox{30mm}{!}{
     \fmfframe(0,10)(0,10){
\begin{fmfgraph*}(50,40)
    \fmfstraight
    \fmfleft{L1,La,Lb,Lc,L2}
    \fmfright{R1,Ra,Rb,Rc,Rd,R2}
    \fmf{plain}{L1,v1,v11,C}
    \fmf{plain}{L2,v2,v22,C}
    \fmf{plain}{C,w1,w11,R1}
    \fmf{plain}{C,w2,w22,R2}
    \fmf{gluon,tension=0}{v22,Rd}
    \fmf{plain,tension=0}{v22,Rd}
    \fmf{dbl_dots,fore=red,tension=0}{v1,v2}
    \fmfv{d.sh=circle, d.f=30,d.si=0.1w}{C}
    \fmfv{l=$\ccO$}{L2}
    \fmfv{l=$\ccj$}{L1}
    \fmfv{l=$p_\ccT$}{Rd}
\end{fmfgraph*}
}
}
\end{gathered} 
,
\hspace{1.5cm}
I_\text{Ge}^{\ccj} =
\begin{gathered}
\resizebox{30mm}{!}{
     \fmfframe(0,10)(0,10){
\begin{fmfgraph*}(50,40)
    \fmfstraight
    \fmfleft{L1,La,Lb,Lc,L2}
    \fmfright{R1,Ra,Rb,Rc,Rd,R2}
    \fmf{plain}{L1,v1,v11,C}
    \fmf{plain}{L2,v2,v22,C}
    \fmf{plain}{C,w1,w11,R1}
    \fmf{plain}{C,w2,w22,R2}
    \fmf{gluon,tension=0}{v2,Rd}
    \fmf{plain,tension=0}{v2,Rd}
    \fmf{phantom,tension=0.1}{v2,x1,x2,Rd}
    \fmf{dbl_dots,tension=0,fore=red}{v1,x1}
        \fmfv{d.sh=circle, d.f=30,d.si=0.1w}{C}
    \fmfv{l=$\ccO$}{L2}
    \fmfv{l=$\ccj$}{L1}
    \fmfv{l=$p_\ccT$}{Rd}
\end{fmfgraph*}
}
}
\end{gathered} 
\ee
The upper Glauber vertex in $I_\text{Gc}^{\ccj}$ comes from the expansion of the collinear Wilson line in the quark-quark Glauber operator connecting the $\ccO$ and $\ccj$ directions. One must consider these three graphs for any direction $\ccj=\ccTH\cdots \ccN$. 

To evaluate $I_\text{Gc}^{\ccj}$, we first note that the collinear Wilson line direction $t_\ccO^\mu$ can be anything not collinear to the $p_\ccO^\mu$ direction. We can therefore choose a basis of polarization vectors $\pol_\pm(p_\ccT)$ for the outgoing collinear gluon with momentum $p_\ccT^\mu$ so that 
$t_\ccO \cdot\pol_\pm=0$. Doing so makes graph $I_\text{Gc}^{\ccj}=0$ for any $\ccj$.

Next, we turn to $I_\text{Gd}^{\ccj}$. In position space, this graph describes a Glauber exchange that takes place earlier in time than the collinear emission. That is, the emission interrupts the Glauber loop. In such a situation, a general argument as given in~\cite{Rothstein:2016bsq} that the graph must vanish. We can also see it directly from the integral itself. Working in lightcone coordinates 
so that $p_\ccO^\mu = - \frac12 Q n^\mu$ and $ p_\ccT^\mu =  \frac12 p_\ccT^{+} n^\mu +   \frac12 p_\ccT^{-} \bar{n}^\mu + p_{\ccT,\perp}^\mu$. For $j \geq 3$,  we have $p_\ccj^\mu =  \frac12 Q_\ccj n_\ccj^\mu$ (for outgoing $p_\ccj^\mu$) or $p_\ccj^\mu = - \frac12 Q_\ccj  n_\ccj^\mu$ (for incoming $p_\ccj^\mu$). And we decompose the Glauber momentum $k^\mu$ with respect to $n^\mu$ and $n_\ccj^\mu$ so that for each diagram $k^- = n \cdot k , \; k^+ = n_\ccj \cdot k $. 
 This gives
\be
I_\text{Gd}^{\ccj} =
\begin{gathered}
\resizebox{35mm}{!}{
     \fmfframe(0,10)(0,10){
\begin{fmfgraph*}(50,40)
    \fmfstraight
    \fmfleft{L1,La,Lb,Lc,L2}
    \fmfright{R1,Ra,Rb,Rc,Rd,R2}
    \fmf{plain}{L1,v1,v11,C}
    \fmf{plain}{L2,v2,v22,C}
    \fmf{plain}{C,w1,w11,R1}
    \fmf{plain}{C,w2,w22,R2}
    \fmf{gluon,tension=0,label=$\rightarrow p_\ccT$,l.s=left,l.d=5}{v22,Rd}
    \fmf{plain,tension=0}{v22,Rd}
    \fmf{dbl_dots,fore=red,tension=0,label=$k\downarrow$,l.s=left}{v1,v2}
        \fmfv{d.sh=circle, d.f=30,d.si=0.1w}{C}
    \fmfv{l=$\ccO$}{L2}
    \fmfv{l=$\ccj$}{L1}
    \fmf{phantom,tension=0,label=$p_\ccj \swarrow$,l.s=left,l.d=2}{L1,v1}
    \fmf{phantom,tension=0,label=$ p_\ccO \nwarrow$,l.d=2}{L2,v2}
 \end{fmfgraph*}
}
}
\end{gathered} 
\sim
\begin{gathered}
\hspace{-1cm}g_s^2   \int  \frac{d^4k}{(2\pi)^4}
\;
 \frac{1}{{ \vec k_\perp^2 }}  \;
  \frac{n \cdot n_\ccj}{2} \,
\frac{  \mp 1  }{\mp  Q_\ccj  k^+ - \vec k_\perp^2 + i\fme}
\hspace{3cm} \\
\times \quad 
\frac{1}{Q k^-  - \vec k_\perp^2 +  i\fme} \;
\frac{1}{Q k^-   - (\vec k_\perp+\vec p_{\ccT,\perp})^2 -Q p_\ccT^++  i\fme}
\times \cdots
\end{gathered}
\ee
with the $\cdots$ at the end representing the numerator and spin structure that are irrelevant here. Here the $\mp$ sign denotes either outgoing ($-$) or incoming ($+$) $p_\ccj$.
We note that only two propagators depend on $k^-$, and both of the corresponding poles in the complex $k^-$ plane are below the
real $k^-$ axis.  Thus we can close the $k^-$ contour downward and the integral vanishes.\footnote{This argument only works if the integral
is convergent, which requires the rapidity regulator. See the longer discussion in Section~\ref{sec:nona}.}

 Note that if there were no emission, the graph would reduce to the Glauber vertex correction in Eq.~\eqref{IfulltoG} which does not vanish. The extra emission adds a propagator that causes the integral to be convergent at $|k^-|=\infty$ allowing us to evaluate it using Cauchy's theorem.
Adding more propagators interrupting the Glauber loop will only add more poles on the same side of the real $k^+$ axis. This is the momentum-space version of the argument in~\cite{Rothstein:2016bsq} that Glauber exchanges cannot be interrupted. 

Finally, we turn to graph $I_\text{Ge}^{\ccj}$. This one will not vanish, so we have to work out the full numerator structure. It is
\be
I_\text{Ge}^{\ccj} =
\begin{gathered}
\resizebox{30mm}{!}{
     \fmfframe(0,10)(0,10){
\begin{fmfgraph*}(50,40)
    \fmfstraight
    \fmfleft{L1,La,Lb,Lc,L2}
    \fmfright{R1,Ra,Rb,Rc,Rd,R2}
    \fmf{plain}{L1,v1,v11,C}
    \fmf{plain}{L2,v2,v22,C}
    \fmf{plain}{C,w1,w11,R1}
    \fmf{plain}{C,w2,w22,R2}
    \fmf{gluon,tension=0,label=$\rightarrow p_\ccT$,l.s=left,l.d=5}{v2,Rd}
    \fmf{plain,tension=0}{v2,Rd}
    \fmf{phantom,tension=0.1}{v2,x1,x2,Rd}
    \fmf{dbl_dots,tension=0,fore=red,label={${\red k}$\rotatebox{15}{$\swarrow$}},l.s=left}{v1,x1}
        \fmfv{d.sh=circle, d.f=30,d.si=0.1w}{C}
    \fmfv{l=$\ccO$}{L2}
    \fmfv{l=$\ccj$}{L1}
    \fmfv{l=$\ccT$}{Rd}
    \fmf{phantom,tension=0,label=$p_\ccj \swarrow$,l.s=left,l.d=2}{L1,v1}
    \fmf{phantom,tension=0,label=$ p_\ccO \nwarrow$,l.d=2}{L2,v2}
\end{fmfgraph*}
}
}
\end{gathered} 
= 
\begin{gathered}
 2 i  g_s^3 
 (\TT_\ccT \cdot \TT_\ccj ) \TT_\ccO \overline{\cM}^\zero
 \int \frac{d^d k}{(2\pi)^d}  \frac{1}{|2k^z|^\eta} \frac{n \cdot n_\ccj}{2} \hspace{4cm}
 \\
\times 
\frac{ \mp 1 }{  \mp k^+ -\delta_\ccj + i\fme}  
 \frac{N^\mu(p_\ccO, p_\ccT, k)}{{ \vec k_\perp^2} }\,
  \frac{1}{  k^- - \delta_\ccT +  i\fme} \,
   \,\frac{ 1  }{  - k^-  +  \delta_{\ccO} + i\fme }
    \pol_\mu(p_\ccT) 
   \end{gathered}
\label{IGej}
\ee
with the  upper (lower) signs on the second line corresponding to $p_\ccj^\mu$ outgoing (incoming). 
Here, the $|2 k^z|^{-\eta}$ factor comes from the rapidity regulator. 
The denominator factors are
\be
 \delta_\ccj= \frac{\vec k_{\perp}^2}{Q_\ccj}, \quad
 \delta_{\ccT} =  \frac{  (\vec p_{\ccT,\perp} +\vec k_\perp )^2 }{ p_\ccT^{+}} -p_\ccT^-  , \quad
 \delta_{\ccO} = - \frac{  (\vec p_{\ccT,\perp} + \vec k_\perp )^2 }{ Q- p_\ccT^{+}} -p_\ccT^- 
 .
  \ee
Recall that $p_\ccO = - \frac12 Q n^\mu$, which explains how $\delta_\ccT$ becomes $\delta_\ccO$ under $p_\ccT^\mu \to p_\ccT^\mu + p_\ccO^\mu$. 
 The numerator factor is
\be
N^\mu (p_\ccO, p_\ccT, k) =
   \frac{n\slash}{2}    \Big[  - \frac{2 (p_{\ccT,\perp} + k_\perp)^\mu}{ p_\ccT^+ }  + \frac{( p\slash_{\ccT,\perp} + k \slash_\perp) \gamma_{\perp}^{\mu}}{ -Q + p_\ccT^+ } \Big] \frac { \bar{n}\slash }{ 2 }    
 v (p_\ccO)
\ee
where $\gamma_\perp^\mu$ is the perpendicular components of $\gamma^\mu$, projected out as with a 4-vector: $\gamma_\perp^\mu = \gamma^\mu - \frac{1}{2} \bar{n} \slash n^\mu - \frac{1}{2} n \slash \bar{n}^\mu$. 
In the numerator expression, the $n \slash$ and $\bar{n} \slash$ factors at the beginning and the end project onto the collinear spinors. The part in bracket comes from expanding the QCD vertex at leading power, according to the SCET Feynman rules, using the vertex coming
from the quark-gluon Glauber operator, and simplifying. That this numerator factor depends only on $k_\perp^\mu$, not on $k^+$ or $k^-$ greatly simplifies the calculation. This simplification comes from keeping only the leading-power interactions, as in SCET.

To evaluate this graph we first write $k^\pm = k^0 \pm k^z$ and perform the $k^0$ integration by contours. The poles
at $k^0 = k^z + \delta_\ccT - i\fme$ and $k^0 = k^z + \delta_\ccO + i\fme$ pinch the contour in the Glauber region~\cite{Collins:1988ig}. 
If we take $p_\ccj^\mu$ to be outgoing (upper signs in Eq.~\eqref{IGej}), then the pole from the first propagator is at
 $k^0 = -k^z- \delta_\ccj +i\fme$. 
 We then close the contour downwards setting
$k^0 = k^z + \delta_\ccT$ so that $k^- = \delta_\ccT$ and $k^+ = 2 k^z + \delta_\ccT$. 
 This gives
 \begin{multline}
I_\text{Ge}^{\ccj} =
 - 2 g_s^3 
 (\TT_\ccT \cdot \TT_\ccj ) \TT_\ccO  \overline{\cM}^\zero
 \int \frac{d^{d-2} k_\perp}{(2\pi)^{d-1}} 
 \frac{N^\mu}{{ \vec k_\perp^2} }\,
   \,\frac{ 1  }{   \delta_\ccO - \delta_\ccT  }
    \pol_\mu(p_\ccT) 
\int d k^z \frac{1}{|2k^z|^\eta}
 \frac{ 1 }{ -2 k^z - \delta_\ccT -\delta_\ccj + i\fme}  
 \label{IOgeccj}
\end{multline}
The $k^z$ integral is straightforward to evaluate (see. Eq. (B.4) of~\cite{Rothstein:2016bsq}):
\begin{align}
\int_{-\infty}^{\infty} \frac{d k^z}{2 \pi} \frac{1}{|2 k^z|^{\eta}}
 \frac{1}{2k^z + 2\Delta + i \fme} &= \frac{1}{4\pi}\left[(-2\pi i) \csc(2\pi \eta) \sin (\pi \eta) (-i \Delta)^{-2\eta}\right]\\
 &=\frac{1}{4\pi}(-i \pi) + \cO(\eta)
\end{align}
Note that the $k^z$ integral cares only about the discontinuity of $[-2k^z- \delta_\ccT -\delta_\ccj + i\fme]^{-1}$, 
which is independent of the value of $\delta_\ccT$ and $\delta_\ccj$. 
Even though the result is independent of $\eta$ as $\eta\to 0$, one still needs the rapidity regulator to make the integral well-defined.
Indeed, the rapidity regulator imparts critical non-analyticity allowing the graph  to vanish for $p_\ccT^\mu$ incoming but not
for $p_\ccT^\mu$ outgoing.  

If the non-collinear leg $p_\ccj^\mu$ that the Glauber gluon connects to is incoming  ($+$ sign in Eq.\eqref{IGej}) ,
 then the pole for from the first propagator is at
$k^0 = -k^z + \delta_\ccj - i \fme$.
 We then close the contour upwards setting
 $k^0 = k^z + \delta_\ccO$ so that $k^- = \delta_\ccO$ and $k^+ = 2 k^z + \delta_\ccO$.
This gives
\begin{multline}
I_\text{Ge}^{\ccTH} =
 2 g_s^3
(\TT_\ccT \cdot \TT_\ccj ) \TT_\ccO \overline{\cM}^\zero
 \int \frac{d^{d-2} k_\perp}{(2\pi)^{d-2}} 
 \frac{N^\mu}{{ \vec k_\perp^2 } }\,
   \,\frac{ 1  }{   \delta_\ccO - \delta_\ccT }
    \pol_\mu(p_\ccT)
\int \frac{d k^z}{2 \pi} \frac{1}{|2k^z|^\eta}
 \frac{ 1 }{  2 k^z + \delta_\ccO -\delta_\ccj + i\fme}
  \label{IGe3}
\end{multline}
Compared to Eq.~\eqref{IOgeccj}, we have a relative minus sign from the  integrand. The result is the same as Eq.~\eqref{IOgeccj} with a $-$ out front. That is.
\begin{multline}
I_\text{Ge}^{\ccj} = \mp
 2 g_s^3 
 (\TT_\ccT\cdot  \TT_\ccj ) \TT_\ccO \cdot
\cMb^\zero \frac {n \slash }{ 2 } 
 \left[  - \frac{2 \pol_{\perp,\mu}}{  p_\ccT^+  } + \frac{ \gamma_{\perp,\mu}  \slashed{\pol}_\perp}{Q - p_\ccT^+ }  \right]  \frac { \bar{n}\slash }{ 2 }  u (p_\ccO)  \\
\times \frac{1}{4\pi}(i \pi) \frac{p_\ccT^+(Q - p_\ccT^+)}{Q} 
\int \frac{d^{d-2} k_\perp}{(2\pi)^{d-2}} 
 \frac{   p_{\ccT,\perp}^\mu + k_\perp^\mu  }{  (\vec p_{\ccT,\perp} + \vec k_\perp )^2  } \frac{1}{\vec k_\perp^2} 
\end{multline}
Here the $\mp$ sign denotes either outgoing ($-$) or incoming ($+$) $p_\ccj$.
The $k_\perp$ integral is regulated in $d-2 = 2-2\ep$ dimensions:
 \begin{align}
 \mu^{4-d} \int   \frac{ d^{d-2} k_\perp}{(2 \pi)^{d-2}} \frac{ p_{\ccT,\perp}^\mu + k_\perp^\mu }{ \vec k_{\perp}^2\,  (\vec p_{\ccT,\perp} +\vec k_\perp )^2 }  
 & =  \frac{1}{4 \pi}  \frac{  p_{\ccT,\perp}^\mu  }{\vec p_{\ccT,\perp}^2} \left(\frac{  4 \pi \mu^2 }{ \vec p_{\ccT,\perp}^2} \right)^{\ep}   \,   \frac{ \Gamma(- \ep) \Gamma(1 + \ep)}{ \Gamma(1- 2\ep)} 
  \nn \\
 & =  \frac{1}{4 \pi}    \frac{  p_{\ccT,\perp}^\mu  }{ \vec p_{\ccT,\perp}^2}  \, \left(\frac{  4 \pi e^{-\gamma_{E}} \mu^2 }{ \vec p_{\ccT,\perp}^2} \right)^{\ep}   \left(-  \frac{1}{\ep} + \cO(\ep)  \right)
 \end{align}
Putting things together,  this diagram is 
\begin{multline}
I_\text{Ge}^{\ccj} =\pm
\frac{\alpha_s}{2 \pi}   \, \left(\frac{  4 \pi e^{-\gamma_{E}} \mu^2 }{ p_{\ccT,\perp}^2} \right)^{\ep}   \left( \frac{ i \pi}{\ep} + \cO(\ep)  \right)
( \TT_\ccT \cdot \TT_\ccj ) \TT_\ccO \cdot \cMb^\zero \\
\times g_s \frac{p_\ccT^+(Q - p_\ccT^+)}{Q   \vec p_{\ccT,\perp}^2}   \frac {n \slash }{ 2 }  \left[  - \frac{2 p_{\ccT,\perp} \cdt \pol_\perp }{  p_\ccT^+  } + \frac{ p\slash_{\ccT,\perp}  \slashed{\pol}_\perp}{Q + p_\ccT^+ }  \right] \frac { \bar{n}\slash }{ 2 }   u (p_\ccO)
\label{OffshellSpin}
\end{multline}
where the $+(-)$ sign corresponds to $p_\ccj$ outgoing (incoming).
Using the on-shell condition: $  p_{\ccT,\perp} \cdt \pol_\perp = - \frac{1}{2} p_\ccT^{+} n \cdt \pol -   \frac{1}{2} p_\ccT^{-} \bar{n} \cdt \pol  \overset{ \tiny r = \bar n }{=} - \frac{1}{2} p_\ccT^{+} n \cdt \pol  $, we recognize the spin structure in \Eq{OffshellSpin} to be the same as the tree-level splitting amplitude, as in Eq.~\eqref{SCETsplit}. Thus,
 \begin{align}
 I_\text{Ge}^{\ccj} &=\pm
\frac{\alpha_s}{2 \pi}  \, \left(\frac{  4 \pi e^{-\gamma_{E}} \mu^2 }{\vec p_{\ccT,\perp}^2} \right)^{\ep}   \left(  \frac{i \pi}{\ep} + \cO(\ep)  \right)
( \TT_\ccT \cdot  \TT_\ccj )\,  \Sp^\zero\cdot \cMb^\zero 
\\
 & = \pm \frac{\alpha_s}{2 \pi} ( 4 \pi e^{-\gamma_{E}})^\ep ( i \pi)    \left(   \frac{1}{\ep} + \ln \frac{ \mu^2 }{ \vec p_{\ccT,\perp}^2}  + \cO(\ep) \right)
( \TT_\ccT \cdot  \TT_\ccj )\,  \Sp^\zero\cdot \cMb^\zero \label{GOnej}
\end{align}
Recall that
\be
\frac{z}{1 - z} = \frac{p_\ccT^+}{p_\ccO^+} 
\ee
we have $\vec p_{\ccT,\perp}^2 = (- 2 p_\ccO \cdt p_\ccT ) \frac{ z}{(z-1)}$, and
\be
 I_\text{Ge}^{\ccj} 
 = \pm\frac{\alpha_s}{2 \pi} (4\pi e^{-\gamma_E})^\ep (i \pi)
 \left(   \frac{1}{\ep} + \ln \frac{ \mu^2 }{ -2  p_\ccO \cdot
     p_\ccT } + \ln{\frac{z - 1 }{ z } }  + \cO(\ep) \right)
( \TT_\ccT \cdot  \TT_\ccj )\,  \Sp^\zero \cdot \cMb^\zero
\ee
Summing over all $\ccj$, and let $p_\ccTH$ be an incoming parton, we then get
\begin{align}
\Sp^{1,\text{non-fact}} &=\frac{\alpha_s}{2 \pi} (4\pi e^{-\gamma_E})^\ep ( i \pi )   \left( \frac{1}{\ep} + \ln \frac{ \mu^2 }{ -2  p_\ccO \cdot p_\ccT } + \ln \frac{z-1}{z} \right) 
\left( - \TT_\ccT \cdot \TT_\ccTH +  \sum_{j >3}  \TT_\ccT \cdot \TT_\ccj \right) \Sp^0  \label{GOnePhase}\\
&= i \alpha_s (4\pi e^{-\gamma_E})^\ep \left( \frac{1}{\ep} + \ln \frac{ \mu^2 }{ -2  p_\ccO \cdot p_\ccT  } + \ln \frac{z-1}{z} \right) 
\left( - \TT_\ccT \cdot \TT_\ccTH \right) \Sp^0  + \cdots
\end{align}
where the $\cdots$ respect strict factorization. This reproduces both the singular and the finite parts of the 1-loop factorization-violating splitting function, as in Eq.~\eqref{SPG31f}. We note that with the absent of $\TT_\ccTH$, there would have been no factorization violating contribution at one loop.

\section{Two-loop factorization-violation from SCET \label{sec:scettwo}}
In SCET, the 2-loop splitting function comes from expanding the ratio of $\ket{\cM}$ to $\ket{\cMb}$ as in Eqs.~\eqref{Mscet} and~\eqref{Mbscet} to 2-loop order. 
Physical effects of factorization violation must occur at the cross-section level, thus the most important effect we are looking for is 
a {\it real} contribution to $\Sp^{2,\text{non-fact.}}$. We will therefore focus on isolating this real part, which should match Eq.~\eqref{delta2real}.
Factorization violating effects at 2-loops will necessarily involve insertions of the Glauber operator, and at 2-loops there
can be 1 or 2 exchanged Glauber gluons. 

\begin{figure*}[t!]
\centering 
  \begin{subfigure}[b]{0.24\textwidth}
        \centering
        \includegraphics[height=1.2in]{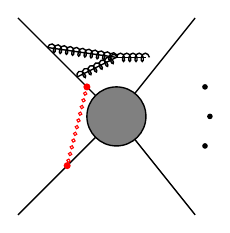}
       \caption{} 
    \end{subfigure}
    \begin{subfigure}[b]{0.24\textwidth}
        \centering
        \includegraphics[height=1.2in]{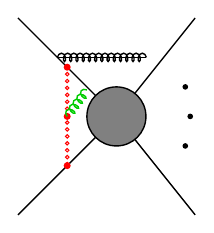}
       \caption{}
    \end{subfigure}  
  \begin{subfigure}[b]{0.24\textwidth}
        \centering
        \includegraphics[height=1.2in]{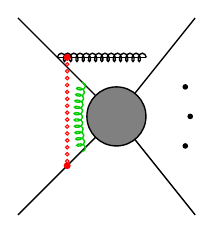}
       \caption{} 
    \end{subfigure}
    \begin{subfigure}[b]{0.24\textwidth}
        \centering
        \includegraphics[height=1.2in]{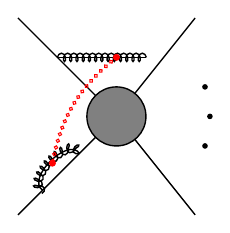}
     \caption{}
    \end{subfigure}
   \caption{ Example Glauber diagrams that can be ignored for the 2-loop splitting amplitude. } \label{GCancel} 
\end{figure*}

Not all 2-loop diagrams involving Glauber gluons can contribute to factorization-violation. For example, none
of the diagrams in  \Fig{GCancel} are relevant. 
\Fig{GCancel}(a) and \Fig{GCancel}(b) describe Glauber exchange right next to the hard interaction.
For these graphs, as at 1-loop, the Glauber is contained in the soft contribution and does not generate factorization violation.
 Diagrams like \Fig{GCancel}(c) with a disconnected soft loop cancel with the product of a 1-loop Glauber exchange and a 1-loop
 soft contribution. 
\Fig{GCancel}(d) is an example of diagram with collinear loops in the $n_\ccj$ sector, with $\ccj \neq 1$.
Since the flow of  Glauber momentum follows the direction of energy flow in the $n_\ccj-$collinear propagators,  
the $n_\ccj-$component of Glauber momentum is not pinched. Thus the Glauber contribution is contained in the soft
contribution, or, more physically, the Glauber contribution acts coherently on the 
on the $n_\ccj-$collinear fields.
The sum of such diagrams will cancel with the product of the 1-loop Glauber contribution and the 1-loop collinear contribution.
It is not hard to see that that loop corrections  due to interactions between fields in the non-collinear sectors 
do not contribute to the factorization-violating part of the splitting amplitude. 

\begin{figure*}[t]
\centering 
  \begin{subfigure}[b]{0.4\textwidth}
        \centering
        \includegraphics[height=1.2in]{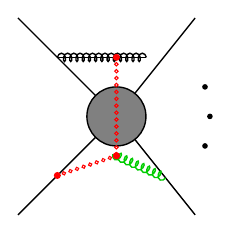}
       \caption{} 
    \end{subfigure}
  \quad 
    \begin{subfigure}[b]{0.4\textwidth}
        \centering
        \includegraphics[height=1.2in]{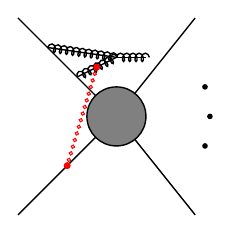}
       \caption{}
    \end{subfigure}  
   \caption{Example Glauber-soft and Glauber-collinear mixing diagrams that may break strict collinear factorization. } \label{GSCol} 
\end{figure*}

Two-loop diagrams  that can contribution to factorization violation must  
involve color exchange between the daughter gluon and hard non-collinear partons. 
They  can be categorized as
double Glauber exchange, Glauber-soft mixing  and Glauber-collinear mixing diagrams. 
The splitting amplitude is thus given by 
\begin{align}
\label{Sp2GlauberDiagrams}
 \Sp^{2, \text{non-fac.}}  \cdot \ket{\cMb^0}
&=  \ket{\cMb^2}_\text{double Glauber}   -  \Sp^{1,\text{non-fac.}} \cdot \ket{\cMb^1}_{\text{Glauber loops}}   \nn \\
 &+\ket{\cMb^2}_\text{Glauber-soft}^\text{non-fac.} + \ket{\cMb^2}_\text{Glauber-coll.}^\text{non-fac.}  -  \Sp^{1,\text{non-fac.}} \cdot  \ket{\cMb^1}_{n_\ccO-\text{coll. loops}}   
\end{align} 
The first term on the right-hand-side of Eq.~\eqref{Sp2GlauberDiagrams} corresponds to double Glauber diagrams, with examples 
 given in \Fig{GBox2} and \Fig{GBox3}. 
 Since each Glauber sub-loop produces a factor of $(i\pi)$, these terms are purely real.
Similar to the 1-loop Glauber diagrams, we expect that double Glauber diagrams to be rapidity-finite and  have no logarithmic dependence
 on the momenta of non-collinear partons. 
The second term in Eq.~\eqref{Sp2GlauberDiagrams}   takes away the 1-loop factorization-breaking effect. 
 The second line of Eq.~\eqref{Sp2GlauberDiagrams}  
comes from Glauber-soft and Glauber-collinear diagrams that violates factorization.  Representive diagrams for each set are shown in \Fig{GSCol}.  
These diagrams have highly non-trivial  kinematic dependence on all external partons and 
involve two-loop multi-leg loop integrals with rapidity divergences. 
A full explicit calculation of these diagram is beyond the scope of our paper.
In the following section we will carry out the calculation of  double Glauber diagrams and show that the first line  of
Eq.~\eqref{Sp2GlauberDiagrams} reproduces the 
leading pole of the real part of $ \Sp^{2, \text{non-fac.}} $. 


\subsection{Double-Glauber diagrams} 

In this section,  we give explicit results for double-Glauber exchange diagrams that can violate strict factorization.
To evaluate the diagrams, we use the rapidity-regularization scheme given in \cite{Rothstein:2016bsq},
which adds a convergence factor of $\frac{1}{|k_z|^\eta}$ to the integrand for both Glauber momenta.
We refer to $p_\ccO$ and $p_\ccT$ as collinear partons and all other partons as non-collinear. 
We discuss and summarize the calculation here, leaving the details of some representative calculations to Appendix~\ref{app}.

\begin{figure*}[t!]
\centering 
  \begin{subfigure}[b]{0.25\textwidth}
        \centering
        \includegraphics[height=1.1in]{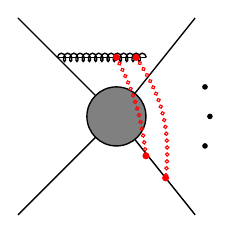}
       \caption{} 
    \end{subfigure}
    \begin{subfigure}[b]{0.25\textwidth}
        \centering
        \includegraphics[height=1.1in]{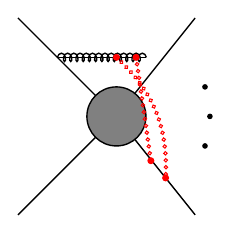}
     \caption{}
    \end{subfigure}
  \begin{subfigure}[b]{0.25\textwidth}
        \centering
        \includegraphics[height=1.1in]{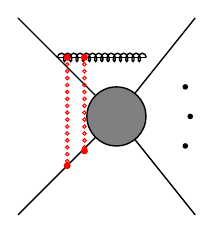}
       \caption{} 
    \end{subfigure}
    \begin{subfigure}[b]{0.25\textwidth}
        \centering
        \includegraphics[height=1.1in]{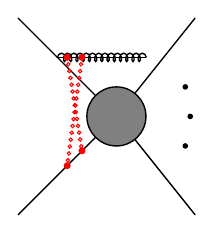}
       \caption{}
    \end{subfigure}
   \begin{subfigure}[b]{0.25\textwidth}
        \centering
        \includegraphics[height=1.1in]{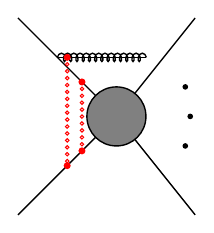}
     \caption{} 
    \end{subfigure}
    \begin{subfigure}[b]{0.25\textwidth}
        \centering
        \includegraphics[height=1.1in]{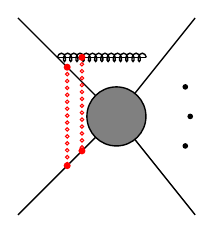}
       \caption{}
    \end{subfigure}  
   \caption{ Double Glauber exchange diagrams that  involve two collinear sectors. }  \label{GBox2}
\end{figure*}

\begin{figure*}[t]
\centering 
  \begin{subfigure}[b]{0.2\textwidth}
        \centering
        \includegraphics[height=1.1in]{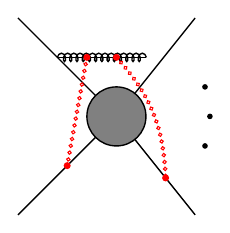}
       \caption{} 
    \end{subfigure}
    \begin{subfigure}[b]{0.2\textwidth}
        \centering
        \includegraphics[height=1.1in]{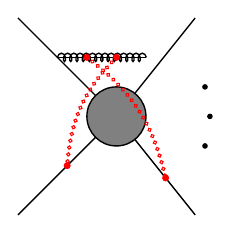}
       \caption{}
    \end{subfigure}  
  \begin{subfigure}[b]{0.2\textwidth}
        \centering
        \includegraphics[height=1.1in]{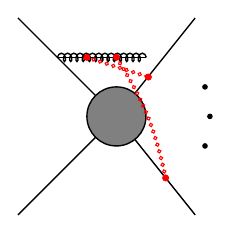}
       \caption{} 
    \end{subfigure}
    \begin{subfigure}[b]{0.2\textwidth}
        \centering
        \includegraphics[height=1.1in]{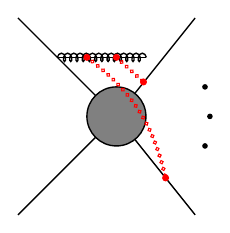}
     \caption{}
    \end{subfigure}
    \begin{subfigure}[b]{0.2\textwidth}
        \centering
        \includegraphics[height=1.1in]{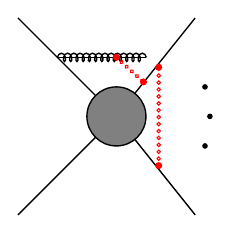}
      \caption{}
    \end{subfigure}  
  \begin{subfigure}[b]{0.2\textwidth}
        \centering
        \includegraphics[height=1.1in]{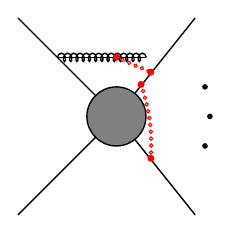}
      \caption{}
    \end{subfigure}
      \begin{subfigure}[b]{0.2\textwidth}
        \centering
        \includegraphics[height=1.1in]{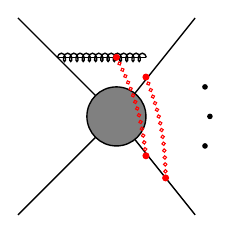}
      \caption{}
    \end{subfigure}  
  \begin{subfigure}[b]{0.2\textwidth}
        \centering
        \includegraphics[height=1.1in]{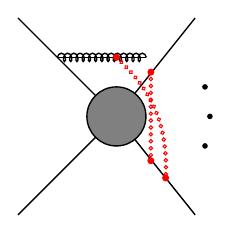}
      \caption{}
    \end{subfigure}       
   \caption{  Double Glauber diagrams  involving three collinear sectors. } \label{GBox3} 
\end{figure*}

The relevant double Glauber diagrams can be divided into two categories, 
those involving two hard-collinear directions (Fig.~\ref{GBox2}),  and those involving three hard-collinear directions (Fig.~\ref{GBox3}). 
 All of these diagrams have at least one Glauber attached to the $p_\ccT$ gluon. Diagrams where neither Glauber attaches to $p_\ccT$ either vanish 
 or contained in  $\ket{\overline{\cM}^1}_\text{Glauber loops}$.   
 None of the relevant diagrams have Glauber gluons attached to $p_\ccO$; when the Glauber gluon attaches to $p_\ccO$, the Glauber loop is interrupted by the real emission and the diagram will vanish (just asr $I_\text{Gd}^\ccj=0$  in Section~\ref{sec1loop}).

We will start with diagrams in \Fig{GBox2} involving two hard-collinear directions.  
We can have two Glauber vertices on $p_\ccT$, the other two on a non-collinear parton $p_\ccj$. 
Or we can have one Glauber vertex on $p_\ccT$, one on the internal collinear line labeled as $p_{(12)}$, and the other two vertices on $p_\ccj$.  

\Fig{GBox2}(a) and \Fig{GBox2}(b) describe Glauber exchange with outgoing non-collinear partons. 
 \Fig{GBox2}(a)  has two parallel Glauber rungs  which  can be ordered in time. 
We find
\be
\begin{tikzpicture}[baseline={([yshift=-.5ex]current bounding box.center)},scale=1.2]
\draw (-1,-1) -- (0,0);
\draw (-1,1) -- (0,0);
\draw (0,0) -- (0.8,1);
\draw (0,0) -- (0.8,-1);
\draw [fill=\blobcolor] (0,0) circle [radius = 0.3];
\draw [fill=black] (0.9,0.3) circle [radius = 0.6pt];
\draw [fill=black] (0.95,0.0) circle [radius = 0.6pt];
\draw [fill=black] (0.9,-0.3) circle [radius = 0.6pt];
\draw[color=black,decorate,decoration={gluon, amplitude=1.2pt,
    segment length=1.8pt, aspect=0.6}] (-0.6,0.6) -- (0.3,0.6);
    \draw (-0.6,0.6) -- (0.3,0.6);
\draw [red,fill=red] (0.0,0.6) circle [radius=0.8pt];
\draw [red,fill=red] (0.2,0.6) circle [radius=0.8pt];
\draw [red,fill=red] (0.3,-0.4) circle [radius=0.8pt];
\draw [red,decorate glaubr] (0.0,0.6) to [bend left=10] (0.3,-0.4);
\draw [red,decorate glaubr] (0.2,0.6) to [bend left=20] (0.5,-0.625);
\draw [red,fill=red] (0.5,-0.625) circle [radius=0.8pt];
\draw [-{Latex[length=3pt]}] (-0.8,1) node[above,  scale=0.5]{$\ccO$}--(-0.6,0.8);
\draw [-{Latex[length=3pt]}] (-0.8,-1) node[below,  scale=0.5]{$\ccTH$}--(-0.6,-0.8);
 \draw [-{Latex[length=3pt]}] (0.8,-0.75)--(1,-1)node[below,  scale=0.5]{$\ccj$};
 \draw (-0.5, 0.4) node[below,  scale=0.5]{$(\ccO \ccT)$};
 \draw [-{Latex[length=3pt]}] (0.1,0.7)--(0.4,0.7) node[above, scale=0.5]{$\ccT$};
\end{tikzpicture}
\hspace{0.5cm}
= 
\begin{gathered}
  (\TT_\ccT^b   \TT_\ccT^c ) ( \TT_\ccj^b \TT_\ccj^c  ) \, 
\Sp^{\zero}\, \overline{ \cM}^{\zero}    
\hspace{6cm}\\
\hspace{2cm}
 \times \frac{1}{2!} \Big( \frac{ \alpha_s}{2 \pi}  \Big)^2 
\left( i \pi \right)^2   \left(\frac{  4 \pi \mu^2 }{ \vec{p}_{\ccT,\perp}^2} \right)^{2 \ep}    
 [\Gamma(- \ep)]^2 \frac{  \Gamma(1-\ep) \Gamma(1 + 2 \ep)}{ \Gamma(1- 3\ep)} 
\end{gathered}
\ee
The $1/2!$ is a symmetry factor coming from time-ordering of the two Glaubers. \Fig{GBox2}(b) has two crossed Glauber rungs such that the ordering of the Glauber vertices 
in light-cone time are the opposite on each line.  
 The integral vanishes since all poles lie on the same side of $k^0-$complex contour:
\be 
\begin{tikzpicture}[baseline={([yshift=-.5ex]current bounding box.center)},scale=1.2]
\draw (-1,-1) -- (0,0);
\draw (-1,1) -- (0,0);
\draw (0,0) -- (0.8,1);
\draw (0,0) -- (0.8,-1);
\draw [fill=\blobcolor] (0,0) circle [radius = 0.3];
\draw [fill=black] (0.9,0.3) circle [radius = 0.6pt];
\draw [fill=black] (0.95,0.0) circle [radius = 0.6pt]; 
\draw [fill=black] (0.9,-0.3) circle [radius = 0.6pt];
\draw[color=black,decorate,decoration={gluon, amplitude=1.2pt,
    segment length=1.8pt, aspect=0.6}] (-0.6,0.6) -- (0.3,0.6);
    \draw (-0.6,0.6) -- (0.3,0.6);
\draw [red,fill=red] (0.0,0.6) circle [radius=0.8pt];
\draw [red,fill=red] (0.2,0.6) circle [radius=0.8pt];
\draw [red,fill=red] (0.35,-0.45) circle [radius=0.8pt];
\draw [red,decorate glaubr] (0.2,0.6) to [bend left=0] (0.35,-0.45);
\draw [red,decorate glaubr] (0.0,0.6) to [bend left=25] (0.5,-0.625);
\draw [red,fill=red] (0.5,-0.625) circle [radius=0.8pt];
\end{tikzpicture}  =0 
\ee
The sum of these two diagrams gives 
\be
\begin{tikzpicture}[baseline={([yshift=-.5ex]current bounding box.center)},scale=1.2]
\draw (-1,-1) -- (0,0);
\draw (-1,1) -- (0,0);
\draw (0,0) -- (0.8,1);
\draw (0,0) -- (0.8,-1);
\draw [fill=\blobcolor] (0,0) circle [radius = 0.3];
\draw [fill=black] (0.9,0.3) circle [radius = 0.6pt];
\draw [fill=black] (0.95,0.0) circle [radius = 0.6pt];
\draw [fill=black] (0.9,-0.3) circle [radius = 0.6pt];
\draw[color=black,decorate,decoration={gluon, amplitude=1.2pt,
    segment length=1.8pt, aspect=0.6}] (-0.6,0.6) -- (0.3,0.6);
    \draw (-0.6,0.6) -- (0.3,0.6);
\draw [red,fill=red] (0.0,0.6) circle [radius=0.8pt];
\draw [red,fill=red] (0.2,0.6) circle [radius=0.8pt];
\draw [red,fill=red] (0.3,-0.4) circle [radius=0.8pt];
\draw [red,decorate glaubr] (0.0,0.6) to [bend left=10] (0.3,-0.4);
\draw [red,decorate glaubr] (0.2,0.6) to [bend left=20] (0.5,-0.625);
\draw [red,fill=red] (0.5,-0.625) circle [radius=0.8pt];
\end{tikzpicture} 
\quad + 
\begin{tikzpicture}[baseline={([yshift=-.5ex]current bounding box.center)},scale=1.2]
\draw (-1,-1) -- (0,0);
\draw (-1,1) -- (0,0);
\draw (0,0) -- (0.8,1);
\draw (0,0) -- (0.8,-1);
\draw [fill=\blobcolor] (0,0) circle [radius = 0.3];
\draw [fill=black] (0.9,0.3) circle [radius = 0.6pt];
\draw [fill=black] (0.95,0.0) circle [radius = 0.6pt]; 
\draw [fill=black] (0.9,-0.3) circle [radius = 0.6pt];
\draw[color=black,decorate,decoration={gluon, amplitude=1.2pt,
    segment length=1.8pt, aspect=0.6}] (-0.6,0.6) -- (0.3,0.6);
    \draw (-0.6,0.6) -- (0.3,0.6);
\draw [red,fill=red] (0.0,0.6) circle [radius=0.8pt];
\draw [red,fill=red] (0.2,0.6) circle [radius=0.8pt];
\draw [red,fill=red] (0.35,-0.45) circle [radius=0.8pt];
\draw [red,decorate glaubr] (0.2,0.6) to [bend left=0] (0.35,-0.45);
\draw [red,decorate glaubr] (0.0,0.6) to [bend left=25] (0.5,-0.625);
\draw [red,fill=red] (0.5,-0.625) circle [radius=0.8pt];
\end{tikzpicture} 
\quad \stackrel{\text{double pole}}{=}  \frac{1}{2!} \, (\TT_\ccT  \cdot  \TT_\ccj )^2  \, 
\Sp^{\zero} \, \overline{ \cM}^{\zero}  
 \, \Big( \frac{ \alpha_s}{2 \pi}  \Big)^2 
\left( i \pi \right)^2  \, \frac{1}{\ep^2}  
\ee
Note that this is exactly half of the 1-loop Glauber exchange diagram squared (see \Eq{GOnej} ). This contribution must be summed over all outgoing legs $\ccj$. 

 \Fig{GBox2}(c) and   \Fig{GBox2}(d) describe Glauber exchanges between $p_\ccT$ and an incoming parton $p_\ccTH$. 
\be
 \begin{tikzpicture}[baseline={([yshift=-.5ex]current bounding box.center)},scale=1.3]
\draw (-1,-1) -- (0,0);
\draw (-1,1) -- (0,0);
\draw (0,0) -- (0.8,1);
\draw (0,0) -- (0.8,-1);
\draw [fill=\blobcolor] (0,0) circle [radius = 0.3];
\draw [fill=black] (0.7,0.3) circle [radius = 0.6pt];
\draw [fill=black] (0.75,0.0) circle [radius = 0.6pt];
\draw [fill=black] (0.7,-0.3) circle [radius = 0.6pt];
\draw[color=black,decorate,decoration={gluon, amplitude=1.2pt,
    segment length=1.8pt, aspect=0.6}] (-0.6,0.6) -- (0.3,0.6);
    \draw (-0.6,0.6) -- (0.3,0.6);
\draw [red,fill=red] (-0.5,0.6) circle [radius=0.8pt];
\draw [red,fill=red] (-0.5,-0.5) circle [radius=0.8pt];
\draw [red,decorate glaubr] (-0.5,0.6) -- (-0.5,-0.5);
\draw [red,fill=red] (-0.325,0.6) circle [radius=0.8pt];
\draw [red,fill=red] (-0.325,-0.35) circle [radius=0.8pt];
\draw [red,decorate glaubr] (-0.325,0.6) -- (-0.325,-0.35);
\end{tikzpicture} 
  =
  \begin{gathered}   (\TT_\ccT^b   \TT_\ccT^c ) ( \TT_\ccTH^c \TT_\ccTH^b  )\,
\Sp^{\zero} \, \overline{ \cM}^{\zero} 
\hspace{6cm}    \\
\hspace{1cm} \times \frac{1}{2!} \Big( \frac{ \alpha_s}{2 \pi}  \Big)^2 
\left( i \pi \right)^2   \left(\frac{  4 \pi \mu^2 }{ \vec{p}_{\ccT,\perp}^2} \right)^{2 \ep}    
 [\Gamma(- \ep)]^2 \frac{  \Gamma(1-\ep) \Gamma(1 + 2 \ep)}{ \Gamma(1- 3\ep)}
 \end{gathered}
\ee
Compared with \Fig{GBox2}(a), the order of color generators switches on the non-collinear leg.
Similar to \Fig{GBox2}(b), the cross Glauber graph vanishes:
\be 
\begin{tikzpicture}[baseline={([yshift=-.5ex]current bounding box.center)},scale=1.2]
\draw (-1,-1) -- (0,0);
\draw (-1,1) -- (0,0);
\draw (0,0) -- (0.8,1);
\draw (0,0) -- (0.8,-1);
\draw [fill=\blobcolor] (0,0) circle [radius = 0.3];
\draw [fill=black] (0.7,0.3) circle [radius = 0.6pt];
\draw [fill=black] (0.75,0.0) circle [radius = 0.6pt];
\draw [fill=black] (0.7,-0.3) circle [radius = 0.6pt];
\draw[color=black,decorate,decoration={gluon, amplitude=1.2pt,
    segment length=1.8pt, aspect=0.6}] (-0.6,0.6) -- (0.3,0.6);
    \draw (-0.6,0.6) -- (0.3,0.6);
\draw [red,fill=red] (-0.5,0.6) circle [radius=0.8pt];
\draw [red,fill=red] (-0.5,-0.5) circle [radius=0.8pt];
\draw [red,decorate glaubr] (-0.5,0.6) -- (-0.35,-0.35);
\draw [red,fill=red] (-0.35,0.6) circle [radius=0.8pt];
\draw [red,fill=red] (-0.35,-0.35) circle [radius=0.8pt];
\draw [red,decorate glaubr] (-0.35,0.6) -- (-0.5,-0.5);
\end{tikzpicture}
\quad =  0
\ee
Now we write  $(\TT_\ccT^b   \TT_\ccT^c ) ( \TT_\ccTH^c \TT_\ccTH^b  )$ as $ (\TT_\ccT  \cdot  \TT_\ccTH )^2 +  C_A  \TT_\ccT  \cdot  \TT_\ccTH $, and
focus on the leading pole. The sum of these two diagrams is 
\be
 \begin{tikzpicture}[baseline={ ([yshift=-.5ex]current bounding box.center) }, scale=1.2]
\draw (-1,-1) -- (0,0);
\draw (-1,1) -- (0,0);
\draw (0,0) -- (0.8,1);
\draw (0,0) -- (0.8,-1);
\draw [fill=\blobcolor] (0,0) circle [radius = 0.3];
\draw [fill=black] (0.7,0.3) circle [radius = 0.6pt];
\draw [fill=black] (0.75,0.0) circle [radius = 0.6pt];
\draw [fill=black] (0.7,-0.3) circle [radius = 0.6pt];
\draw[color=black,decorate,decoration={gluon, amplitude=1.2pt,
    segment length=1.8pt, aspect=0.6}] (-0.6,0.6) -- (0.3,0.6);
    \draw (-0.6,0.6) -- (0.3,0.6);
\draw [red,fill=red] (-0.5,0.6) circle [radius=0.8pt];
\draw [red,fill=red] (-0.5,-0.5) circle [radius=0.8pt];
\draw [red,decorate glaubr] (-0.5,0.6) -- (-0.5,-0.5);
\draw [red,fill=red] (-0.325,0.6) circle [radius=0.8pt];
\draw [red,fill=red] (-0.325,-0.35) circle [radius=0.8pt];
\draw [red,decorate glaubr] (-0.325,0.6) -- (-0.325,-0.35);
\end{tikzpicture} 
 \quad +  
\begin{tikzpicture}[baseline={([yshift=-.5ex]current bounding box.center)},scale=1.2]
\draw (-1,-1) -- (0,0);
\draw (-1,1) -- (0,0);
\draw (0,0) -- (0.8,1);
\draw (0,0) -- (0.8,-1);
\draw [fill=\blobcolor] (0,0) circle [radius = 0.3];
\draw [fill=black] (0.7,0.3) circle [radius = 0.6pt];
\draw [fill=black] (0.75,0.0) circle [radius = 0.6pt];
\draw [fill=black] (0.7,-0.3) circle [radius = 0.6pt];
\draw[color=black,decorate,decoration={gluon, amplitude=1.2pt,
    segment length=1.8pt, aspect=0.6}] (-0.6,0.6) -- (0.3,0.6);
    \draw (-0.6,0.6) -- (0.3,0.6);
\draw [red,fill=red] (-0.5,0.6) circle [radius=0.8pt];
\draw [red,fill=red] (-0.5,-0.5) circle [radius=0.8pt];
\draw [red,decorate glaubr] (-0.5,0.6) -- (-0.35,-0.35);
\draw [red,fill=red] (-0.35,0.6) circle [radius=0.8pt];
\draw [red,fill=red] (-0.35,-0.35) circle [radius=0.8pt];
\draw [red,decorate glaubr] (-0.35,0.6) -- (-0.5,-0.5);
\end{tikzpicture}
\quad  \stackrel{\text{double pole}}{=} 
\frac{1}{2!} \left[ (\TT_\ccT  \cdot  \TT_\ccTH )^2 +  C_A  \TT_\ccT  \cdot  \TT_\ccTH   \right] \, 
\Sp^{\zero} \, \overline{ \cM}^{\zero} 
\, \Big( \frac{ \alpha_s}{2 \pi}  \Big)^2 
\left( i \pi \right)^2   \, \frac{1}{\ep^2}   
\ee
Note that this result is similar to half the square of the leading pole from the corresponding 1-loop diagram;
it contains, however, an additional $C_A$ term that comes from the switching the order of color generators as
compared to \Fig{GBox2}(a).

 So far we considered diagrams with two Glauber vertices on $p_\ccT$.  Now we move on to diagrams  
 with one Glauber vertex on $p_\ccT$ and 
 the other one on the internal collinear line, namely $p_{(\ccO \ccT)}$. 
 The glauber connecting $p_{(\ccO \ccT)}$ with non-collinear parton $p_\ccj$ forms a loop
  that looks like the vertex diagram $I_\text{Glauber}$ discussed in Section~\ref{sec:isolate}.  It vanishes 
if $p_\ccj$ is outgoing such that  $p_{(\ccO\ccT)} \cdot  p_\ccj <0$. 
When $p_\ccj$ is incoming, we can write down 
two such diagrams \Fig{GBox2}(e) and \Fig{GBox2}(f), both are non-vanishing and they only differ in color structure:
\be
\begin{tikzpicture}[baseline={([yshift=-.5ex]current bounding box.center)},scale=1.2]
\draw (-1,-1) -- (0,0);
\draw (-1,1) -- (0,0);
\draw (0,0) -- (0.8,1);
\draw (0,0) -- (0.8,-1);
\draw [fill=\blobcolor] (0,0) circle [radius = 0.3];
\draw [fill=black] (0.7,0.3) circle [radius = 0.6pt];
\draw [fill=black] (0.75,0.0) circle [radius = 0.6pt];
\draw [fill=black] (0.7,-0.3) circle [radius = 0.6pt];
\draw[color=black,decorate,decoration={gluon, amplitude=1.2pt,
    segment length=1.8pt, aspect=0.6}] (-0.6,0.6) -- (0.3,0.6);
    \draw (-0.6,0.6) -- (0.3,0.6);
\draw [red,fill=red] (-0.5,0.6) circle [radius=0.8pt];
\draw [red,fill=red] (-0.5,-0.5) circle [radius=0.8pt];
\draw[red,decorate glaubr] (-0.5,0.6) -- (-0.5,-0.5);
\draw [red,fill=red] (-0.35,0.35) circle [radius=0.8pt];
\draw [red,fill=red] (-0.35,-0.35) circle [radius=0.8pt];
\draw[red,decorate glaubr] (-0.35,0.35) -- (-0.35,-0.35);
\end{tikzpicture}
\hspace{0.5cm}=
\begin{gathered}
 (\TT_\ccT^b  ) ( \TT_\ccTH^b \TT_\ccTH^c  ) \, 
\Sp^{\zero}\, (-\TT_\ccOT^c) \overline{ \cM}^{\zero}  
\hspace{7cm}\\
\hspace{1cm}
  \times \frac{1}{2!} \Big( \frac{ \alpha_s}{2 \pi}  \Big)^2 
\left( i \pi \right)^2   \left(\frac{  4 \pi \mu^2 }{ \vec{p}_{\ccT,\perp}^2} \right)^{2\ep} \, \Gamma(- \ep) \,  
 \frac{ \Gamma(1 - \ep) \Gamma(  1 + \ep)  }{  \Gamma(1- 2\ep) } \, 
 \left(  \frac{1}{\ep_\text{UV}} - \frac{1}{\ep} \right)     
 \end{gathered}
\ee
and
\be
\begin{tikzpicture}[baseline={([yshift=-.5ex]current bounding box.center)},scale=1.2]
\draw (-1,-1) -- (0,0);
\draw (-1,1) -- (0,0);
\draw (0,0) -- (0.8,1);
\draw (0,0) -- (0.8,-1);
\draw [fill=\blobcolor] (0,0) circle [radius = 0.3];
\draw [fill=black] (0.7,0.3) circle [radius = 0.6pt];
\draw [fill=black] (0.75,0.0) circle [radius = 0.6pt];
\draw [fill=black] (0.7,-0.3) circle [radius = 0.6pt];
\draw[color=black,decorate,decoration={gluon, amplitude=1.2pt,
    segment length=1.8pt, aspect=0.6}] (-0.6,0.6) -- (0.3,0.6);
    \draw (-0.6,0.6) -- (0.3,0.6);
\draw [red,fill=red] (-0.5,0.5) circle [radius=0.8pt];
\draw [red,fill=red] (-0.5,-0.5) circle [radius=0.8pt];
\draw [red,decorate glaubr] (-0.5,0.5) -- (-0.5,-0.5);
\draw [red,fill=red] (-0.35,0.6) circle [radius=0.8pt];
\draw [red,fill=red] (-0.35,-0.35) circle [radius=0.8pt];
\draw [red,decorate glaubr] (-0.35,0.6) -- (-0.35,-0.35);
\end{tikzpicture}
\hspace{0.5cm}=
\begin{gathered}
   (\TT_\ccT^b  ) ( \TT_\ccTH^c \TT_\ccTH^b  ) \, 
\Sp^{\zero}\, (-\TT_\ccOT^c) \overline{ \cM}^{\zero} 
\hspace{7cm}\\
\hspace{1cm}
 \times \frac{1}{2!} \Big( \frac{ \alpha_s}{2 \pi}  \Big)^2 
\left( i \pi \right)^2   \left(\frac{  4 \pi \mu^2 }{ \vec{p}_{\ccT,\perp}^2} \right)^{2\ep} \, \Gamma(- \ep) \,  
 \frac{ \Gamma(1 - \ep) \Gamma(  1 + \ep)  }{  \Gamma(1- 2\ep) } \, 
 \left(  \frac{1}{\ep_\text{UV}} - \frac{1}{\ep} \right)  
\end{gathered}
\ee
where $1/\ep$ is associated with the IR divergence.
Using color conservation relation: $ \TT_{(\ccO \ccT)} = - \TT_\ccTH - \cdots - \TT_m$, 
the sum of these two diagrams can be written as 
\begin{multline} \label{TPGlauber}
\begin{tikzpicture}[baseline={([yshift=-.5ex]current bounding box.center)},scale=1.2]
\draw (-1,-1) -- (0,0);
\draw (-1,1) -- (0,0);
\draw (0,0) -- (0.8,1);
\draw (0,0) -- (0.8,-1);
\draw [fill=\blobcolor] (0,0) circle [radius = 0.3];
\draw [fill=black] (0.7,0.3) circle [radius = 0.6pt];
\draw [fill=black] (0.75,0.0) circle [radius = 0.6pt];
\draw [fill=black] (0.7,-0.3) circle [radius = 0.6pt];
\draw[color=black,decorate,decoration={gluon, amplitude=1.2pt,
    segment length=1.8pt, aspect=0.6}] (-0.6,0.6) -- (0.3,0.6);
    \draw (-0.6,0.6) -- (0.3,0.6);
\draw [red,fill=red] (-0.5,0.5) circle [radius=0.8pt];
\draw [red,fill=red] (-0.5,-0.5) circle [radius=0.8pt];
\draw [red,decorate glaubr] (-0.5,0.5) -- (-0.5,-0.5);
\draw [red,fill=red] (-0.35,0.6) circle [radius=0.8pt];
\draw [red,fill=red] (-0.35,-0.35) circle [radius=0.8pt];
\draw [red,decorate glaubr] (-0.35,0.6) -- (-0.35,-0.35);
\end{tikzpicture}
\quad 
+ \quad
\begin{tikzpicture}[baseline={([yshift=-.5ex]current bounding box.center)},scale=1.2]
\draw (-1,-1) -- (0,0);
\draw (-1,1) -- (0,0);
\draw (0,0) -- (0.8,1);
\draw (0,0) -- (0.8,-1);
\draw [fill=\blobcolor] (0,0) circle [radius = 0.3];
\draw [fill=black] (0.7,0.3) circle [radius = 0.6pt];
\draw [fill=black] (0.75,0.0) circle [radius = 0.6pt];
\draw [fill=black] (0.7,-0.3) circle [radius = 0.6pt];
\draw[color=black,decorate,decoration={gluon, amplitude=1.2pt,
    segment length=1.8pt, aspect=0.6}] (-0.6,0.6) -- (0.3,0.6);
    \draw (-0.6,0.6) -- (0.3,0.6);
\draw [red,fill=red] (-0.5,0.5) circle [radius=0.8pt];
\draw [red,fill=red] (-0.5,-0.5) circle [radius=0.8pt];
\draw [red,decorate glaubr] (-0.5,0.5) -- (-0.5,-0.5);
\draw [red,fill=red] (-0.35,0.6) circle [radius=0.8pt];
\draw [red,fill=red] (-0.35,-0.35) circle [radius=0.8pt];
\draw [red,decorate glaubr] (-0.35,0.6) -- (-0.35,-0.35);
\end{tikzpicture}
\quad \stackrel{\text{double pole}}{=}  
 \Big\{  (-\TT_\ccTH \cdot \TT_\ccT)   \, 
\Sp^{\zero}\, (\TT_\ccOT \cdot \TT_\ccTH )  \, \overline{ \cM}^{\zero}
 \\
 \quad + \Big( - \frac12\, C_A\,  \TT_\ccT \cdot \TT_\ccTH       +   \sum_{\ccj =4}^m \frac{i}{2}  f_{bcd} \,  \TT_\ccT^b \,  \TT_\ccTH^c  \,  \TT_\ccj^d \Big)   \Sp^{\zero}   \,  \overline{ \cM}^{\zero}  \Big\} \, 
  \Big( \frac{ \alpha_s}{2 \pi}  \Big)^2 (i \pi)^2 \, \frac{1}{\ep^2}
\end{multline} 
 The first line in \Eq{TPGlauber}  will  be removed by the corresponding  term in 
$ \Sp^{(1)\text{non-fac.}}  \cMb^\one_\text{Glauber loops} $
 in \Eq{Sp2GlauberDiagrams}.

Now we turn to the diagrams with three different collinear
directions, as shown in Fig.~\ref{GBox3}.
 Here we omitted diagrams where four Glauber vertices are on different legs, since those diagrams trivially factorize into the product of two one-loop results,
 which are contained in $\Sp^{1,\text{non.fac}} \ket{\cMb^1}_{\text{Glauber loop}}$. 
 Therefore we focus on diagrams where two Glauber vertices are on the same leg. 
The first line of \Fig{GBox3} are diagrams with two Glauber vertices on $p_\ccT$. 
 Diagrams in the second line have only one Glauber vertex on $p_\ccT$.

First consider diagrams in the first line with two Glaubers exchanged between  $p_\ccT$ and two  non-collinear partons 
 $p_\cci$  and  $p_\ccj$ with $\cci \neq \ccj$ and $( \cci, \ccj  = 3 ,\cdots ,m) $.  
 One immediate concern is that with 3 directions one cannot choose lightcone coordinates aligned with all of them. 
 Fortunately, this is not a problem for Glauber graphs -- at leading power in the Glauber expansion the $\vec k_\perp$ components
 dominate over the projection of $k^\mu$ on any lightcone direction. Thus we can choose one lightcone direction $n_1^\mu$ alligned with the collinear $p_\ccO^\mu$ direction  and the other $n_2^\mu$ algned with any other direction and the calculation is basically the same
 as if there are only two directions involved. More details of the decomposition are given in Appendix~\ref{app}.
The result is:
\be 
\begin{tikzpicture}[baseline={([yshift=-.5ex]current bounding box.center)},scale=1.2]
\draw (-1,-1) -- (0,0);
\draw (-1,1) -- (0,0);
\draw (0,0) -- (0.8,1);
\draw (0,0) -- (0.8,-1);
\draw [fill=\blobcolor] (0,0) circle [radius = 0.3];
\draw [fill=black] (0.9,0.3) circle [radius = 0.6pt];
\draw [fill=black] (0.95,0.0) circle [radius = 0.6pt];
\draw [fill=black] (0.9,-0.3) circle [radius = 0.6pt];
\draw[color=black,decorate,decoration={gluon, amplitude=1.2pt,
    segment length=1.8pt, aspect=0.6}] (-0.6,0.6) -- (0.3,0.6);
    \draw (-0.6,0.6) -- (0.3,0.6);
\draw [red,fill=red] (0.0,0.6) circle [radius=0.8pt];
\draw [red,fill=red] (0.5,-0.625) circle [radius=0.8pt];
\draw [red, decorate glaubr] (0.0,0.6) to [bend left=20] (0.5,-0.625);
\draw [red,fill=red] (-0.3,0.6) circle [radius=0.8pt];
\draw [red,fill=red] (-0.5,-0.5) circle [radius=0.8pt];
\draw [red, decorate glaubr] (-0.3,0.6) to [bend right=0] (-0.5,-0.5);
\end{tikzpicture}
=
\begin{gathered}
  - (\TT_\ccT \cdot \TT_\ccj)  (\TT_\ccT \cdot \TT_\ccTH)  \, \Sp^{\zero} \,  \overline{ \cM}^{\zero} \nn 
\hspace{6cm}\\
\hspace{1cm}
\times \Big( \frac{ \alpha_s}{2 \pi}  \Big)^2 
\left( i \pi \right)^2   \left(\frac{  4 \pi \mu^2 }{ \vec{p}_{\ccT,\perp}^2} \right)^{2 \ep}    
 [\Gamma(- \ep)]^2 \frac{  \Gamma(1-\ep) \Gamma(1 + 2 \ep)}{ \Gamma(1- 3\ep)} 
\end{gathered}
\ee
In \Fig{GBox3}(b)  the Glauber vertices on the outgoing gluon line are switched. The $k^-$ integral then vanishes since all three poles are on the same side of the contour:
\be 
\begin{tikzpicture}[baseline={([yshift=-.5ex]current bounding box.center)},scale=1.2]
\draw (-1,-1) -- (0,0);
\draw (-1,1) -- (0,0);
\draw (0,0) -- (0.8,1);
\draw (0,0) -- (0.8,-1);
\draw [fill=\blobcolor] (0,0) circle [radius = 0.3];
\draw [fill=black] (0.9,0.3) circle [radius = 0.6pt];
\draw [fill=black] (0.95,0.0) circle [radius = 0.6pt];
\draw [fill=black] (0.9,-0.3) circle [radius = 0.6pt];
\draw[color=black,decorate,decoration={gluon, amplitude=1.2pt,
    segment length=1.8pt, aspect=0.6}] (-0.6,0.6) -- (0.3,0.6);
    \draw (-0.6,0.6) -- (0.3,0.6);
\draw [red,fill=red] (0.0,0.6) circle [radius=0.8pt];
\draw [red,fill=red] (0.5,-0.625) circle [radius=0.8pt];
\draw [red,decorate glaubr] (0.0,0.6) to [bend right=20] (-0.5,-0.5);
\draw [red,fill=red] (-0.3,0.6) circle [radius=0.8pt];
\draw [red,fill=red] (-0.5,-0.5) circle [radius=0.8pt];
\draw [red,decorate glaubr] (-0.3,0.6) to [bend left=20] (0.5,-0.625);
\end{tikzpicture}
 \quad =0
\ee
The sum of the two diagram is 
\begin{multline} 
\begin{tikzpicture}[baseline={([yshift=-.5ex]current bounding box.center)},scale=1.2]
\draw (-1,-1) -- (0,0);
\draw (-1,1) -- (0,0);
\draw (0,0) -- (0.8,1);
\draw (0,0) -- (0.8,-1);
\draw [fill=\blobcolor] (0,0) circle [radius = 0.3];
\draw [fill=black] (0.9,0.3) circle [radius = 0.6pt];
\draw [fill=black] (0.95,0.0) circle [radius = 0.6pt];
\draw [fill=black] (0.9,-0.3) circle [radius = 0.6pt];
\draw[color=black,decorate,decoration={gluon, amplitude=1.2pt,
    segment length=1.8pt, aspect=0.6}] (-0.6,0.6) -- (0.3,0.6);
    \draw (-0.6,0.6) -- (0.3,0.6);
\draw [red,fill=red] (0.0,0.6) circle [radius=0.8pt];
\draw [red,fill=red] (0.5,-0.625) circle [radius=0.8pt];
\draw [red,decorate glaubr] (0.0,0.6) to [bend left=20] (0.5,-0.625);
\draw [red,fill=red] (-0.3,0.6) circle [radius=0.8pt];
\draw [red,fill=red] (-0.5,-0.5) circle [radius=0.8pt];
\draw [red,decorate glaubr] (-0.3,0.6) to [bend right=0] (-0.5,-0.5);
\end{tikzpicture}
\quad 
+
\begin{tikzpicture}[baseline={([yshift=-.5ex]current bounding box.center)},scale=1.2]
\draw (-1,-1) -- (0,0);
\draw (-1,1) -- (0,0);
\draw (0,0) -- (0.8,1);
\draw (0,0) -- (0.8,-1);
\draw [fill=\blobcolor] (0,0) circle [radius = 0.3];
\draw [fill=black] (0.9,0.3) circle [radius = 0.6pt];
\draw [fill=black] (0.95,0.0) circle [radius = 0.6pt];
\draw [fill=black] (0.9,-0.3) circle [radius = 0.6pt];
\draw[color=black,decorate,decoration={gluon, amplitude=1.2pt,
    segment length=1.8pt, aspect=0.6}] (-0.6,0.6) -- (0.3,0.6);
    \draw (-0.6,0.6) -- (0.3,0.6);
\draw [red,fill=red] (0.0,0.6) circle [radius=0.8pt];
\draw [red,fill=red] (0.5,-0.625) circle [radius=0.8pt];
\draw [red,decorate glaubr] (0.0,0.6) to [bend right=20] (-0.5,-0.5);
\draw [red,fill=red] (-0.3,0.6) circle [radius=0.8pt];
\draw [red,fill=red] (-0.5,-0.5) circle [radius=0.8pt];
\draw [red,decorate glaubr] (-0.3,0.6) to [bend left=20] (0.5,-0.625);
\end{tikzpicture}
\quad  \stackrel{\text{double pole}}{=}   \\
  \left(  - \frac{1}{2} (\TT_\ccT \cdot \TT_\ccj)  (\TT_\ccT \cdot \TT_\ccTH)  - \frac{i}{2} f_{abc} \TT_\ccT^a \, \TT_\ccTH^b \, \TT_\ccj^c   \right) \Sp^{\zero} \,  \overline{ \cM}^{\zero} \, \Big( \frac{ \alpha_s}{2 \pi}  \Big)^2 
\left( i \pi \right)^2  \frac{1}{\ep^2}
\end{multline} 
We have broken up the color factor   into  terms that are symmetric and antisymmetric under $ \ccTH \leftrightarrow \ccj$ ,  
The first term can be identified with the 
cross terms coming from the exponentiation of  the sum of $I_\text{Ge}^\ccTH$ and $ I_\text{Ge}^\ccj$
computed in Section~\ref{sec1loop}.

Graphs with two outgoing legs as in \Fig{GBox3}(c) are similar with an extra factor of 1/2 from the time-ordering:
\be
\begin{tikzpicture}[baseline={([yshift=-.5ex]current bounding box.center)},scale=1.2]
\draw (-1,-1) -- (0,0);
\draw (-1,1) -- (0,0);
\draw (0,0) -- (0.8,1);
\draw (0,0) -- (0.8,-1);
\draw [fill=\blobcolor] (0,0) circle [radius = 0.3];
\draw [fill=black] (0.9,0.3) circle [radius = 0.6pt];
\draw [fill=black] (0.95,0.0) circle [radius = 0.6pt];
\draw [fill=black] (0.9,-0.3) circle [radius = 0.6pt];
\draw[color=black,decorate,decoration={gluon, amplitude=1.2pt,
    segment length=1.8pt, aspect=0.6}] (-0.6,0.6) -- (0.3,0.6);
    \draw (-0.6,0.6) -- (0.3,0.6);
\draw [red,fill=red] (0.0,0.6) circle [radius=0.8pt];
\draw [red,fill=red] (0.5,-0.625) circle [radius=0.8pt];
\draw [red,decorate glaubr] (0.0,0.6) to [bend left=10] (0.5,-0.625);
\draw [red,fill=red] (-0.3,0.6) circle [radius=0.8pt];
\draw [red,fill=red] (0.325,0.4) circle [radius=0.8pt];
\draw [red,decorate glaubr] (-0.3,0.6) to  (0.325,0.4);
\end{tikzpicture}
=
\begin{gathered}
    \frac{1}{2!} \Big(  (\TT_\ccT \cdot \TT_\ccj)  (\TT_\ccT \cdot \TT_\cck)  \Big) \Sp^{\zero} \,  \overline{ \cM}^{\zero}  
\hspace{6cm}\\
\hspace{1cm}
 \times \Big( \frac{ \alpha_s}{2 \pi}  \Big)^2 
\left( i \pi \right)^2   \left(\frac{  4 \pi \mu^2 }{ \vec{p}_{\ccT,\perp}^2} \right)^{2 \ep}    
 [\Gamma(- \ep)]^2 \frac{  \Gamma(1-\ep) \Gamma(1 + 2 \ep)}{ \Gamma(1- 3\ep)} 
\end{gathered}
\ee
\Fig{GBox3}(d) can be obtained from \Fig{GBox3}(c) by switching $\ccj$ and $\cck$:
\begin{multline} 
\begin{tikzpicture}[baseline={([yshift=-.5ex]current bounding box.center)},scale=1.2]
\draw (-1,-1) -- (0,0);
\draw (-1,1) -- (0,0);
\draw (0,0) -- (0.8,1);
\draw (0,0) -- (0.8,-1);
\draw [fill=\blobcolor] (0,0) circle [radius = 0.3];
\draw [fill=black] (0.9,0.3) circle [radius = 0.6pt];
\draw [fill=black] (0.95,0.0) circle [radius = 0.6pt];
\draw [fill=black] (0.9,-0.3) circle [radius = 0.6pt];
\draw[color=black,decorate,decoration={gluon, amplitude=1.2pt,
    segment length=1.8pt, aspect=0.6}] (-0.6,0.6) -- (0.3,0.6);
    \draw (-0.6,0.6) -- (0.3,0.6);
\draw [red,fill=red] (0.0,0.6) circle [radius=0.8pt];
\draw [red,fill=red] (0.5,-0.625) circle [radius=0.8pt];
\draw [red,decorate glaubr] (0.0,0.6) to [bend left=10] (0.5,-0.625);
\draw [red,fill=red] (-0.3,0.6) circle [radius=0.8pt];
\draw [red,fill=red] (0.325,0.4) circle [radius=0.8pt];
\draw [red,decorate glaubr] (-0.3,0.6) to  (0.325,0.4);
\end{tikzpicture}
\quad 
+
\begin{tikzpicture}[baseline={([yshift=-.5ex]current bounding box.center)},scale=1.2]
\draw (-1,-1) -- (0,0);
\draw (-1,1) -- (0,0);
\draw (0,0) -- (0.8,1);
\draw (0,0) -- (0.8,-1);
\draw [fill=\blobcolor] (0,0) circle [radius = 0.3];
\draw [fill=black] (0.9,0.3) circle [radius = 0.6pt];
\draw [fill=black] (0.95,0.0) circle [radius = 0.6pt];
\draw [fill=black] (0.9,-0.3) circle [radius = 0.6pt];
\draw[color=black,decorate,decoration={gluon, amplitude=1.2pt,
    segment length=1.8pt, aspect=0.6}] (-0.6,0.6) -- (0.3,0.6);
    \draw (-0.6,0.6) -- (0.3,0.6);
\draw [red,fill=red] (0.0,0.6) circle [radius=0.8pt];
\draw [red,fill=red] (0.5,-0.625) circle [radius=0.8pt];
\draw [red,decorate glaubr] (0.0,0.6) to [bend right=0] (0.275,0.35);
\draw [red,fill=red] (-0.3,0.6) circle [radius=0.8pt];
\draw [red,fill=red] (0.275,0.35) circle [radius=0.8pt];
\draw [red,decorate glaubr] (-0.3,0.6) to [bend left=20] (0.5,-0.625);
\end{tikzpicture}  
 \quad 
  \stackrel{\text{double pole}}{=} \\
    \frac{1}{2!} \Big(  (\TT_\ccT \cdot \TT_\ccj)  (\TT_\ccT \cdot \TT_\cck) + (\TT_\ccT \cdot \TT_\cck)  (\TT_\ccT \cdot \TT_\ccj)  \Big) \Sp^{\zero} \,  \overline{ \cM}^{\zero}  
 \times \Big( \frac{ \alpha_s}{2 \pi}  \Big)^2 
\left( i \pi \right)^2 \, \frac{1}{\ep^2}
\end{multline}  
These diagrams produce the cross term from the exponentiation of $I_\text{Ge}^\ccj$ and $I_\text{Ge}^\cck$. 

The remaining diagrams  
on the second line of \Fig{GBox3} can be computed in the same way.  We find for the sum:
\begin{multline}
\begin{tikzpicture}[baseline={([yshift=-.5ex]current bounding box.center)},scale=1.2]
\draw (-1,-1) -- (0,0);
\draw (-1,1) -- (0,0);
\draw (0,0) -- (0.8,1);
\draw (0,0) -- (0.8,-1);
\draw [fill=\blobcolor] (0,0) circle [radius = 0.3];
\draw [fill=black] (0.9,0.3) circle [radius = 0.6pt];
\draw [fill=black] (0.95,0.0) circle [radius = 0.6pt];
\draw [fill=black] (0.9,-0.3) circle [radius = 0.6pt];
\draw[color=black,decorate,decoration={gluon, amplitude=1.2pt,
    segment length=1.8pt, aspect=0.6}] (-0.6,0.6) -- (0.3,0.6);
    \draw (-0.6,0.6) -- (0.3,0.6);
\draw [red,fill=red] (0.0,0.6) circle [radius=0.8pt];
\draw [red,fill=red] (0.3,0.4) circle [radius=0.8pt];
\draw [red,fill=red] (0.3,-0.4) circle [radius=0.8pt];
\draw [red,decorate glaubr] (0.0,0.6) to [bend left=10] (0.3,-0.4);
\draw [red,decorate glaubr] (0.3,0.4) to [bend left=10] (0.5,-0.625);
\draw [red,fill=red] (0.5,-0.625) circle [radius=0.8pt];
\end{tikzpicture} 
+ 
\begin{tikzpicture}[baseline={([yshift=-.5ex]current bounding box.center)},scale=1.2]
\draw (-1,-1) -- (0,0);
\draw (-1,1) -- (0,0);
\draw (0,0) -- (0.8,1);
\draw (0,0) -- (0.8,-1);
\draw [fill=\blobcolor] (0,0) circle [radius = 0.3];
\draw [fill=black] (0.9,0.3) circle [radius = 0.6pt];
\draw [fill=black] (0.95,0.0) circle [radius = 0.6pt];
\draw [fill=black] (0.9,-0.3) circle [radius = 0.6pt];
\draw[color=black,decorate,decoration={gluon, amplitude=1.2pt,
    segment length=1.8pt, aspect=0.6}] (-0.6,0.6) -- (0.3,0.6);
    \draw (-0.6,0.6) -- (0.3,0.6);
\draw [red,fill=red] (0.0,0.6) circle [radius=0.8pt];
\draw [red,fill=red] (0.35,0.45) circle [radius=0.8pt];
\draw [red,fill=red] (0.35,-0.45) circle [radius=0.8pt];
\draw [red,decorate glaubr] (0.35,0.45) to [bend left=0] (0.35,-0.45);
\draw [red,decorate glaubr] (0.0,0.6) to [bend left=25] (0.5,-0.625);
\draw [red,fill=red] (0.5,-0.625) circle [radius=0.8pt];
\end{tikzpicture}
+ 
\begin{tikzpicture}[baseline={([yshift=-.5ex]current bounding box.center)},scale=1.2]
\draw (-1,-1) -- (0,0);
\draw (-1,1) -- (0,0);
\draw (0,0) -- (0.8,1);
\draw (0,0) -- (0.8,-1);
\draw [fill=\blobcolor] (0,0) circle [radius = 0.3];
\draw [fill=black] (0.9,0.3) circle [radius = 0.6pt];
\draw [fill=black] (0.95,0.0) circle [radius = 0.6pt];
\draw [fill=black] (0.9,-0.3) circle [radius = 0.6pt];
\draw[color=black,decorate,decoration={gluon, amplitude=1.2pt,
    segment length=1.8pt, aspect=0.6}] (-0.6,0.6) -- (0.3,0.6);
    \draw (-0.6,0.6) -- (0.3,0.6);
\draw [red,fill=red] (0.0,0.6) circle [radius=0.8pt];
\draw [red,fill=red] (0.43,-0.5) circle [radius=0.8pt];
\draw [red,fill=red] (0.43,0.5) circle [radius=0.8pt];
\draw [red,decorate glaubr] (0.43,-0.5) to (0.43,0.5);
\draw [red,decorate glaubr] (0.0,0.6) to (0.275,0.35);
\draw [red,fill=red] (0.275,0.35) circle [radius=0.8pt];
\end{tikzpicture}
+ 
\begin{tikzpicture}[baseline={([yshift=-.5ex]current bounding box.center)},scale=1.2]
\draw (-1,-1) -- (0,0);
\draw (-1,1) -- (0,0);
\draw (0,0) -- (0.8,1);
\draw (0,0) -- (0.8,-1);
\draw [fill=\blobcolor] (0,0) circle [radius = 0.3];
\draw [fill=black] (0.9,0.3) circle [radius = 0.6pt];
\draw [fill=black] (0.95,0.0) circle [radius = 0.6pt];
\draw [fill=black] (0.9,-0.3) circle [radius = 0.6pt];
\draw[color=black,decorate,decoration={gluon, amplitude=1.2pt,
    segment length=1.8pt, aspect=0.6}] (-0.6,0.6) -- (0.3,0.6);
    \draw (-0.6,0.6) -- (0.3,0.6);
\draw [red,fill=red] (0.0,0.6) circle [radius=0.8pt];
\draw [red,fill=red]  (0.35,-0.425) circle [radius=0.8pt];
\draw [red,fill=red] (0.35,0.45) circle [radius=0.8pt];
\draw [red,decorate glaubr] (0.0,0.6) to  (0.35,0.45);
\draw [red,decorate glaubr] (0.25,0.325) to [bend left=10] (0.35,-0.425);
\draw [red,fill=red] (0.25,0.325) circle [radius=0.8pt];
\end{tikzpicture} 
  \\
=  \frac{1}{2!} \Big(  (\TT_\ccT \cdot \TT_\ccj)  (\TT_\ccj \cdot \TT_\cck)  +  (\TT_\ccj \cdot \TT_\cck)  (\TT_\ccT \cdot \TT_\ccj)  + (\ccj \leftrightarrow \cck) \Big)    
\Sp^{\zero} \overline{ \cM}^{\zero}   \\
 \times \Big( \frac{ \alpha_s}{2 \pi}  \Big)^2 
\left( i \pi \right)^2   \left(\frac{  4 \pi \mu^2 }{ \vec{p}_{\ccT,\perp}^2} \right)^{2 \ep}    
  \frac{ \Gamma( - \ep) \Gamma(  1 + \ep)  }{  \Gamma(1- 2\ep) } \, 
 \left(  \frac{1}{\ep_\text{UV}} - \frac{1}{\ep} \right)   \\
 \stackrel{\text{double pole}}{=}   \Big(  (\TT_\ccT \cdot \TT_\ccj)  + (\TT_\ccT \cdot \TT_\cck)   \Big)    
\Sp^{\zero}  \,  (\TT_\ccj \cdot \TT_\cck )  \,   \overline{ \cM}^{\zero} \, 
\Big( \frac{ \alpha_s}{2 \pi}  \Big)^2 
\left( i \pi \right)^2  \, \frac{1}{\ep^2}
\end{multline} 
 The sum of these last  four diagrams will cancel the corresponding  terms in the contribution
  $ \Sp^{1,\text{non-fac.}}  \cMb^\one_\text{Glauber loops} $
to \Eq{Sp2GlauberDiagrams}.  

Let us summarize and put together the results for double Glauber diagrams in \Fig{GBox2} and \Fig{GBox3}. 
As expected, these diagrams have no explicit dependence on the momenta of non-collinear partons. 
They are only sensitive to the physical scale associated with the splitting. 
All the non-vanishing double Glauber graphs have a ladder-type topology, where  two vertical rungs represent Glauber interactions ordered in time.  
\begin{align}
& \text{\Fig{GBox2}} +  \text{\Fig{GBox3}} - \Sp^{(1), \text{non-fac.}}  \cMb^\one_\text{Glauber loops} \nn \\
& =  \left\{ \frac12 \Big( -\TT_\ccT \cdot \TT_\ccTH +  \sum_{j=4}^m  \TT_\ccT \cdot \TT_\ccj  \Big)^2  
+   \sum_{j =4}^m i  f_{abc} \,  \TT_\ccT^a \,  \TT_\ccTH^b  \,  \TT_\ccj^c   \right\}  \Sp^{\zero}  \,  \overline{ \cM}^{\zero} \nn   \\
&\quad  \times \Big( \frac{ \alpha_s}{2 \pi}  \Big)^2 
\left( i \pi \right)^2   \left(\frac{  4 \pi \mu^2 }{ \vec{p}_{\ccT,\perp}^2} \right)^{2\ep} \, 
 \left(  \frac{1}{\ep^2} - \frac{\pi^2}{6} + \cO(\ep)   \right)    \\ 
 & \stackrel{\text{double pole}}{=}   \left\{ \frac12 \left[ \frac{\alpha_s}{ 2\pi} \, \frac{ i \pi }{\ep } \Big( -\TT_\ccT \cdot \TT_\ccTH + \sum_{j=4}^m  \TT_\ccT \cdot \TT_\ccj  \Big) \right]^2 
-    \frac{ \alpha_s^2 }{ 4 \ep^2 } \, \sum_{j =4}^m i  f_{abc} \,  \TT_\ccT^a \,  \TT_\ccTH^b  \,  \TT_\ccj^c   \right\}   \Sp^{\zero}  \,  \overline{ \cM}^{\zero}      \label{GBoxPhase}
\end{align} 

The first term in \Eq{GBoxPhase}  comes directly from exponentiating the one-loop Glauber phase given by \Eq{GOnePhase}. 
The second term  with a purely non-abelian color structure $   i  f_{abc} \,  \TT_\ccT^a \,  \TT_\ccTH^b  \,  \TT_\ccj^c  $ corresponds to the anti-hermitian part of  $\Delta_{C}^2(\ep)$ shown in 
 \Eq{delta2real}.
 Thus we find that the real part of the  $1/\ep^2$ IR poles in $\Spnf$ at 2-loops from~\cite{Catani:2011st} 
are exactly reproduced by SCET. The subleading terms and the imaginary part of $\Spnf$ at 2-loops involve graphs other than the double Glauber ones. We leave the complete computation of $\Spnf$ at 2-loops to future work.

\section{Analytic properties of Glauber gluons in SCET \label{sec:nona}}
We have shown that SCET with the addition of Glauber operators as proposed in~\cite{Rothstein:2016bsq} reproduces known results about factorization violating contributions to splitting amplitudes from QCD. For the theory to be consistent
it is critical that the  Glauber operators do not
destroy fractorization in situations where it supposed to hold.

SCET graphs involving Glauber gluons have unusual properties compared to graphs in QCD. For example, the form-factor graph 
vanishes if $p_\ccO \cdot p_\ccT < 0$ but  is non-zero if $p_\ccO \cdot p_\ccT > 0$:
\be
\begin{gathered}
\resizebox{20mm}{!}{
     \fmfframe(0,0)(0,0){
\begin{fmfgraph*}(20,35)
\fmfstraight
	\fmfleft{L1}
	\fmfright{R1,R2}
	\fmf{fermion}{L1,v1}
	\fmf{fermion,label=$p_\ccT$,l.s=right}{v1,R1}
	\fmf{fermion,label=$p_\ccO$,l.s=right}{R2,v2}
	\fmf{fermion}{v2,L1}
	\fmf{dbl_dots, fore=red,tension=0}{v1,v2}
        \fmfv{d.sh=circle, d.f=30,d.si=0.2w}{L1}
\end{fmfgraph*}
}
}
\end{gathered}
=\frac{\alpha_s}{2 \pi} \frac{i \pi}{\ep_{\text{IR}}},
\hspace{2cm}
\begin{gathered}
\resizebox{30mm}{!}{
     \fmfframe(0,00)(0,0){
\begin{fmfgraph*}(30,20)
	\fmfbottom{L1}
	\fmftop{R1,R2}
	\fmf{fermion}{L1,v1}
	\fmf{fermion,label=$p_\ccT $}{v1,R1}
	\fmf{fermion,label=$p_\ccO$,l.s=left}{R2,v2}
	\fmf{fermion}{v2,L1}
	\fmf{dbl_dots, fore=red,tension=0}{v1,v2}
        \fmfv{d.sh=circle, d.f=30,d.si=0.15w}{L1}
\end{fmfgraph*}
}}
\end{gathered}
=0
\label{FFdiag}
\ee
This implies in particular that the Glauber graph is not an analytic function of the external momenta. The non-analyticity comes about through the non-analytic rapidity regulator. This regulator is an essential part of the definition of SCET with Glaubers.

Glauber gluons have an intimate connection to soft gluons. For example, the amplitudes in Eq.~\eqref{FFdiag} are exactly the imaginary part of the corresponding soft graphs. More generally, the Glauber region corresponds to a particular approach to the soft singularity:
 $k^\mu \to 0$ with $k^\pm \lesssim \frac{k_\perp^2}{Q}$ for a hard scale $Q$.  Thus, Glauber gluons can be understood by studying the region
 around the soft pinch surface, as we did in Section~\ref{sec:contain}.
 In that section, we showed that
 when factorization holds, Glauber gluons can be safely ignored. 
 More precisely, we showed that when there is no pinch in the Glauber region the eikonal approximation can be justified
 to reproduce the complete soft singularity.
 While these observations about pinched contours are useful for studying factorization in QCD, they do not immediately translate
 to observations about graphs involving Glauber operator insertions in SCET. 
 
In SCET, when the Glauber operator contribution is entirely contained in the soft contribution, the soft-Glauber correspondence is said to hold~\cite{Rothstein:2016bsq}. When the soft-Glauber correspondance holds, the Glauber contributions can be completely ignored due to the
zero-bin subtraction. More precisely, in~\cite{Rothstein:2016bsq}, Rothstein and Stewart called a soft graph in SCET without Glaubers subtraction a``naive" soft graph denoted by $\widetilde{S}$.  Then the "pure" soft contribution $S$ is
the naive soft contribution with its Glauber limit subtracted off:
$S = \widetilde{S} - S^{(G)}$.
The pure soft graphs have nice properties, such as that they are independent of the direction of the soft Wilson lines; all of the unusual properties
of the Glauber-gluon graphs, such as the non-analytic behavior of Eq.~\eqref{FFdiag} and the necessity of a rapidity regulator are eliminated by this subtraction. When the soft-Glauber correspondence hold, $S^{(G)} = G$. 

What we would like to be true is that, in momenta configurations for which there is no Glauber pinch in QCD,
then the soft-Glauber correspondence holds. This is not easy to show, since there is not a 1-to-1 correspondence between the Glauber limit of graphs in QCD (via the method of regions) and graphs in SCET with Glauber operator insertions. Thus, the check that Glauber operators do not destroy factorization in SCET is non-trivial. Here, we provide a general argument why it should be true in general.

To study factorization violating effects of Glauber gluons, it is not particularly useful to subract off the Glauber limit of each soft graph and add it back in through a pure Glauber contribution. 
The graphs in Eq.~\eqref{FFdiag}   are irrelevant to factorization violation since there is no pinch in the Glauber region for either kinematic configuration. 
Instead, we 
want to start with factorization in SCET without Glaubers, and look at what new effects adding Glaubers will have.
That is, we would like to consider
$G_{\text{non-fact}} = G - G_{\text{fact}}$ where $G$ refers to any graph with Glaubers and $G_{\text{fact}}$ are the Glauber graphs
with double count contributions from the factorized expression in SCET without Glaubers.

A critical property that $G_{\text{non-fact}}$ must have is that it does not spoil factorization in situations where factorization is supposed to hold.
For example, consider the canonical Glauber pinch graph as in Eq.~\eqref{Gpinch}. 
We suppose there are two collinear momenta $p_\ccO \parallel p_\ccT$ and want to look at how the contribution 
changes when $p_\ccT^\mu$ goes from incoming to outgoing:
\be
\begin{gathered}
\begin{tikzpicture}
 \node at (0,0) {
\parbox{60mm}{
\hspace{-2cm}
\resizebox{55mm}{!}{
     \fmfframe(30,0)(-30,0){
     \begin{fmfgraph*}(80,50)	
    \fmfleft{L1,L2,L3,L4,L5}
    \fmfright{R1,R2,R3}
    \fmf{plain,label={\rotatebox{25}{$\nwarrow$}}$~~p_\ccO$,l.s=left,l.d=0.1mm}{L5,v2}
    \fmf{plain}{v2,v}
    \fmf{phantom,tension=2}{L1,v1}
    \fmf{phantom,tension=0.1}{L2,f1,v1}
    \fmf{plain}{v1,v}
    \fmf{phantom,tension=0.5}{v,R2}
    \fmf{phantom}{v2,v3}
    \fmf{phantom}{v3,R3}
    \fmf{phantom}{v1,v4}
    \fmf{phantom}{v4,R1}
    \fmf{gluon}{L4,w1,v2}
    \fmf{phantom,label=$p_\ccT\leftarrow~~~~$,tension=0,l.d=0.4cm}{L4,w1}
    \fmf{dbl_dots,fore=red,tension=0}{f1,w1}
    \fmf{phantom,tension=0.01}{v,x1,x2,x4,R1}
    \fmf{plain,tension=0}{v,x1}
    \fmf{phantom,tension=0.01}{v,y1,y2,y4,R2}
    \fmf{plain,tension=0}{v,y1}
    \fmf{phantom,tension=0.01}{v,z1,z2,z4,R3}
    \fmf{plain,tension=0}{v,z1}
    \fmfblob{0.1w}{v}
\end{fmfgraph*}
}
}
}
};
\draw[black,fill=gray!20] (-1.4,-1.5) ellipse (1.6 and 0.5);
\end{tikzpicture}
\end{gathered}
\hspace{-20mm}
= 0,
\hspace{1cm}
\begin{gathered}
\vspace{-5mm}
\begin{tikzpicture}
 \node at (0,0) {
\parbox{60mm}
{
\resizebox{55mm}{!}{
     \fmfframe(0,0)(0,0){
\begin{fmfgraph*}(80,50)	
	\fmfleft{L1,L2}
	\fmfright{R1,R2,R3}
    \fmf{plain,label=$p_\ccO${\rotatebox{25}{$\nwarrow$}},tension=2,l.d=0.5mm}{L2,v2}
    \fmf{plain}{v1,v}
    \fmf{phantom,tension=2}{L1,v1}
    \fmf{plain}{v2,v}
    \fmf{phantom,tension=0.5}{v,R2}
    \fmf{gluon}{v2,v3}
    \fmf{gluon,label=$p_\ccT${\rotatebox{5}{$\rightarrow$}},l.s=left}{v3,R3}
    \fmf{phantom}{v1,v4}
    \fmf{phantom}{v4,R1}
    \fmf{phantom,tension=0.1}{v3,v8,v9,v4}
\fmffreeze
    \fmf{phantom}{L1,u1,u2,u3,u4,u5,u6,u7,R1}
    \fmf{dbl_dots,fore=red,tension=0.1}{v3,u5}
    \fmf{phantom,tension=0.01}{v,x1,x2,x4,R1}
    \fmf{plain,tension=0}{v,x1}
    \fmf{phantom,tension=0.01}{v,y1,y2,y4,R2}
    \fmf{plain,tension=0}{v,y1}
    \fmf{phantom,tension=0.01}{v,z1,z2,z4,R3}
    \fmf{plain,tension=0}{v,z1}
    \fmfv{d.sh=circle, d.f=30,d.si=0.1w}{v}
\end{fmfgraph*}
}
}
}
};
\draw[black,fill=gray!20] (-0.4,-1.5) ellipse (1.6 and 0.5);
\end{tikzpicture}
\end{gathered}
\hspace{-10mm}
\ne 0
\label{eq:genpin}
\ee
In the configuration on the left, $p_\ccO \cdt p_\ccT >0$, there is no Glauber pinch, and factorization should hold. For this graph,
there is no corresponding soft graph for the Glauber to be contained in.\footnote{
Soft graphs are only sensitive to the net momenta in the collinear sector. The soft graph with a gluon exchanged from the $p_\ccO$ sector has
a Glauber limit given by the graph where the Glauber connects between the two legs closest to the hard vertex, as in Eq.~\eqref{ISG2}.} 
Thus the graph on the left must vanish or else factorization {\it would} be violated.
The right graph, with $p_\ccO \cdt p_\ccT <0$, must reproduce known factorization-violating results from QCD (as we have shown it does).
Note that the corresponding graphs in QCD are analytic functions of momenta and  generically
do not vanish for
either sign of $p_\ccO \cdt p_\ccT$. 

The remarkable non-analytic property of the diagrams in Eq.~\eqref{eq:genpin} as a function of $p_\ccT^\mu$ is achieved through a conspiracy of the power expansion in SCET and the rapidity regulator. The power expansion sequesters all of the $k^+$ and $k^-$ dependence:
\be
\begin{gathered}
\vspace{-5mm}
\begin{tikzpicture}
 \node at (0,0) {
\parbox{60mm}
{
\resizebox{55mm}{!}{
     \fmfframe(0,0)(0,0){
\begin{fmfgraph*}(80,50)	
	\fmfleft{L1,L2}
	\fmfright{R1,R2,R3}
    \fmf{plain,label=$p_\ccO${\rotatebox{25}{$\nwarrow$}},tension=2,l.d=0.5mm}{L2,v2}
    \fmf{plain,label=$k^- \sim 0${\rotatebox{7}{$\nearrow$}},l.s=left,l.d=0.1mm}{v1,v}
    \fmf{phantom,tension=2}{L1,v1}
    \fmf{plain,label=$k^+ \sim 0${\rotatebox{-5}{$\nwarrow$}},l.s=right,l.d=0.1mm}{v2,v}
    \fmf{phantom,tension=0.5}{v,R2}
    \fmf{gluon,label=$k^+ \sim 0${\rotatebox{5}{$\rightarrow$}},l.s=left}{v2,v3}
    \fmf{gluon,label=$p_\ccT${\rotatebox{5}{$\rightarrow$}},l.s=left}{v3,R3}
    \fmf{phantom}{v1,v4}
    \fmf{phantom}{v4,R1}
    \fmf{phantom,tension=0.1}{v3,v8,v9,v4}
\fmffreeze
    \fmf{phantom}{L1,u1,u2,u3,u4,u5,u6,u7,R1}
    \fmf{dbl_dots,fore=red,tension=0,l.s=left,label=$\downarrow k^\pm \sim 0$}{v3,u5}
    \fmf{phantom,tension=0.01}{v,x1,x2,x4,R1}
    \fmf{plain,tension=0}{v,x1}
    \fmf{phantom,tension=0.01}{v,y1,y2,y4,R2}
    \fmf{plain,tension=0}{v,y1}
    \fmf{phantom,tension=0.01}{v,z1,z2,z4,R3}
    \fmf{plain,tension=0}{v,z1}
    \fmfv{d.sh=circle, d.f=30,d.si=0.1w}{v}
\end{fmfgraph*}
}
}
}
};
\draw[black,fill=gray!20] (-0.4,-1.5) ellipse (1.6 and 0.5);
\end{tikzpicture}
\end{gathered}
\hspace{-10mm}
 \sim \int \frac{d^4 k}{(2\pi)^4} \frac{f(k_\perp,p_\cci)}{k^- - \delta_\ccO - i\fme} \frac{1}{k^- - \delta_\ccT + i\fme}\frac{1}{k^+ - \delta_\ccj + i\fme}  
 \label{lpform}
\ee
~\\
\noindent As indicated in the figure, the power expansion lets us drop certain components of the loop momentum $k^\mu$ on each leg that it
flows through.
 The propagator for the red, dotted Glauber leg, with momentum $k^\mu$, only depends on the largest component, which is $\vec{k}_\perp$ according to Glauber scaling. This propagator has been absorbed into the $f(k_\perp,p_\cci)$ function in the numerator. 
 Similarly, the $k^+$ component is dropped when $k^\mu$ is added to $p_\ccO^\mu$ or $p_\ccT^\mu$
and the $k^-$ component is dropped when $k^\mu$ is added to whatever momentum $q^\mu$ flows into the hard vertex from the rest of the diagram. Note that this momentum $q^\mu$ must be lightlike or there is no pinch and the entire diagram is not infrared sensitive and can be
dropped at leading power. The leading power expansion also forces the locations of the poles $\delta_\cci$ and the numerator to depend only on the largest components of $k^\mu$, namely $k_\perp$. That is, $\delta_\ccO, \delta_\ccT,\delta_\ccj$ and $f$ depend only on 
$k_\perp$ and components of the external momenta $p_\cci$, but not on $k^+$ or $k^-$. An explicit example is given in Eq.~\eqref{IGej}.

In the case where $p_\ccT^\mu$ is incoming the diagram has the same form but  the $\delta_\ccT$ pole crosses the real axis. 
That is, we make the replacement $k^- - \delta_\ccT + i\fme \to k^- - \delta_\ccT' - i\fme$ in Eq.~\eqref{lpform}.  This flip removes the pinch from the Glauber region. In fact, it naively seems that since both the poles in the $k^+$ plane are on the same side of the axis, then Eq.~\eqref{lpform} vanishes, as we expect for $p_\ccT^\mu$ incoming. Unfortunately, things are not that simple: if the integral vanishes for $p_\ccT^\mu$ incoming and is an analytic function of momenta, then it must also vanish for $p_\ccT^\mu$ outgoing. In fact, we cannot conclude
that it vanishes simply because the $k^-$ integral seems to give zero. The problem is that the power expansion has made the $k^-$ integral infinite. Although $k^-$ has nothing to do with the pinch in $k^+$ in the Glauber region, we need to regulate the whole integral to make the calculation well-defined.

After adding a factor of $|k_z|^{-\eta}$ from the rapidity regulator it is natural to change variables from $(k^-,k^+)$ to $(k^-,k_z)$.
Doing the $k^-$ integral in Eq.~\eqref{lpform} then gives
\be
\int\frac{ dk_z d^2 k_\perp}{(2\pi)^3} 
\frac{1}{|k_z|^\eta}
\frac{f(k_\perp,p_\cci)}{2 k_z + \delta_\ccO(k_\perp) - \delta_\ccj(k_\perp) - i\fme} \frac{1}{\delta_\ccO(k_\perp) - \delta_\ccT(k_\perp)}
\ne 0
\ee
For the spacelike splitting case, the $k^-$ poles are on the same side of the real axis and the integral gives zero.

This general 1-loop example is all that is required to show that Glauber operator contributions do not destroy factorization for 
timelike splittings in general.
In the more general $n$-loop case, the infrared sensitive region has all the loop momenta near the pinch surface. Thus we can focus on a single loop, over a momentum $k^\mu$, with the other momenta placed on the pinch surface ($k_i^\mu = 0$ or $k_i^\mu$ proportional to some external momentum). 
 For timelike splittings, the energies and large light-cone components of
all the momenta in each collinear sector have the same sign. This places all the poles in poles in $k^-$ on the same side of the real axis. Therefore, once the integral is regulated with the rapidity regulator, the integral over $k^-$ will give zero just as in the 1-loop case.

In this way, the Glauber contribution with the rapidity regulator remarkably produces diagrams with the right properties: they vanish when factorization holds but contribution when factorization is violated. If a QCD diagram does not have a pinch in the Glauber region, then a corresponding diagram with Glauber gluon exchange in SCET with a rapidity regulator will vanish. This is a non-trivial consistency check on the SCET-Glauber formulation, requiring both the power expansion and the rapdity regulator. 
It is only possible because Glauber contributions in SCET are non-analytic functions of external momenta. 

The above arguments suggest that there may be a way to identify the contribution from operators in SCET with properties of amplitudes computed in QCD. Since the Glauber contributions are non-analytic and vanish when $p_\ccT^\mu \to - p_\ccT^\mu$, we might identify the Glauber contribution as $G= \cM{(p_\ccT,p_\ccj)} - \cM{(-p_\ccT,p_\ccj)}$. At 1-loop order, this is equivalent to half the discontinuity across the cut on the real $p_\ccT\cdt p_\ccj$ axis. Beyond 1-loop, taking the discontinuity across the cut can only reproduce the imaginary part of the amplitude, not terms like $( i \pi)^2$ coming from double-Glauber exchange; flipping the sign of $p_\ccT$ could get all of the multi-Glauber effects correct. It would certainly be interesting to investigate the connection between Glauber contributions in SCET and analyticity of QCD amplitudes in greater detail.

\section{Summary and conclusions\label{sec:conclusion}}

In this paper we have studied factorization-violation in collinear splittings from the effective field theory point of view.  The first few sections of the paper discussed situations where factorization holds. In particular, the importance of Glauber scaling was reviewed. We discussed how factorization requires application of the eikonal approximation to preserve all of the singularities in a small ball around the soft pinch surface. This requirement fails when there is a pinch in the Glauber region. Understanding the interplay between factorization, the eikonal approximation, and Glauber pinches allowed us to extend the precise amplitude-level formulation of factorization developed in~\cite{FS1,FS2,Feige:2015rea} to situations where there are colored particles in the initial state. In particular, as long as no incoming direction and outgoing direction are collinear, strict factorization holds. This result, although implicit in much of the early literature on factorization, has never been stated explicitly or proven to our knowledge, so we include it here for completeness. Understanding where and why factorization {\it does} hold is a firm starting point for analysis of factorization violation.

Regarding factorization violation, it had been shown from full QCD that in spacelike splittings (as in initial state radiation) the splitting amplitude is different from 
timelike splittings (as in final state radiation)~\cite{Catani:2011st,Forshaw:2012bi}. In particular, for spacelike splittings, strict factorization is violated, in that the collinear splitting amplitude depends on the colors and kinematics of non-collinear partons. These results were derived in QCD by looking at the IR divergences of amplitudes with $n+m$ well-separated partons and taking the limit where $m$ of the partons become collinear. 
We showed that these results can be reproduced using SCET with the inclusion of Glauber operators as proposed in~\cite{Rothstein:2016bsq}. In
particular, we confirmed the divergent and finite factorization-violating terms at 1-loop and the leading real divergent part at 2-loops, for $m=1$. 
These calculations are non-trivial in SCET and require careful use of the rapidity regulator and the power expansion.

In the SCET approach, the splitting is computed from emissions off an amplitude with $n$ collinear sectors, rather than by taking limits of $m+n$ parton amplitudes. This conceptual difference might be advantageous in studying physical implications of factorization-violation, for example, by sequestering factorization-violating effects to certain operator matrix elements. However, this is not yet possible as it is not clear how the Glauber contributions in SCET can be disentangled from the factorization-preserving soft and collinear contributions. 

In~\cite{Rothstein:2016bsq} it was shown that much of the time the contribution from Glauber operators is identical to the Glauber limit of the soft contribution. This equivalence, $G=S^{(G)}$ was called the ``soft-Glauber correspondance" in~\cite{Rothstein:2016bsq}.
It is important to understand when the soft-Glauber correspondance holds, as the soft-Glauber overlap (as well as the collinear-Glauber overlap) must be zero-bin subtracted to avoid overcounting.
Unfortunately, it seems very hard to establish the soft-Glauber correspondence to all orders. Some examples were given in~\cite{Rothstein:2016bsq} and suggestive general arguments. In this paper, we connected the soft-Glauber correpondance (a feature of SCET) to situations in which integration contours can be deformed out of the Glauber region into the eikonal region in full QCD. When this deformation is possible, 
 as in situations where no incoming parton is collinear to an outgoing parton,
 the soft-Glauber correspondance {\it must} hold.

One intriguing feature of the SCET-Glauber contributions is that they produce necessarily non-analytic functions of external momentum. 
Non-analyticity is critical for the Glauber contributions both to vanish when a momentum is outgoing ($E >0$) and to not vanish when a momentum
is incoming ($E<0$). In QCD,  amplitudes are analytic functions of momenta (up to poles and branch cuts)
but in SCET they are not. An example of how this works is the 1-loop Sudakov form factor, where QCD gives a $\frac{1}{\ep} \ln( - p_\ccO \cdt p_\ccT - i\fme)$ term, which is analytic, while the Glauber contribution gives 
just the discontinuity of this result, $- \frac{i\pi}{\ep} \theta(p_\ccO \cdt p_\ccT)$, which is non-analytic. For this form factor, the Glauber contribution is not factorization-violating, as the soft-Glauber correspondence holds, but the same non-analyticity is critical
in factorization-violating cases. Indeed, the 1-loop 1-emission Glauber graphs are also non-analytic functions of a momentum $p_\ccT$, as they must vanish when $p_\ccT$ is outgoing (so as not to spoil factorization when it holds) and reproduce factorization-violating results from QCD when $p_\ccT$ is incoming. The SCET formalism achieves this through a combination of the power expansion, which sequesters all the dependence on certain momentum components into certain parts of the Feynman diagrams so that Glauber graphs can exactly vanish when factorization holds, and the rapidity regulator, which is non-analytic. 

These observations, summarized in Section~\ref{sec:nona} are suggestive that the factorization-violating Glauber contributions may be identified with a soft of generalized discontinuity of the QCD amplitude: they reproduce the difference between
 $\ket{\cM( p_\ccT , p_\ccj)}$ and $\ket{\cM(-p_\ccT,p_\ccj)}$. Another corollary of these observations is that the rapidity divergences in the Glauber graphs {\it must} be regulated with a non-analytic regulator. The non-analyticity is helpful, in that it allows for the Glauber graphs to isolate the factorization-violating effects, but it also makes computing Glauber contributions beyond 1-loop order more challenging than for graphs where dimensional regularization can be used.

\section{Acknowledgements}
We would like to thank I. Stewart, I Feige and J. Collins for helpful conversations. This work is supported in part by the U.S. Department of Energy, under grant DE-SC0013607, and by the Office of Nuclear Physics of the U.S. Department of Energy
under Contract No. DE-SC0011090.
\label{sec:ackn}

\appendix

\section{Double Glauber integrals \label{app}}

In the Appendix we give more details of some representative 2-loop double-Glauber-exchange diagrams.

First consider \Fig{GBox2}(c) with two parallel Glauber rungs between $p_\ccT$ and $p_\ccTH$:
 \begin{align}
 \text{\Fig{GBox2}(c)} &=
 \begin{tikzpicture}[baseline={([yshift=-.5ex]current bounding box.center)},scale=1.3]
\draw (-1,-1) -- (0,0);
\draw (-1,1) -- (0,0);
\draw (0,0) -- (0.8,1);
\draw (0,0) -- (0.8,-1);
\draw [fill=\blobcolor] (0,0) circle [radius = 0.3];
\draw [fill=black] (0.7,0.3) circle [radius = 0.6pt];
\draw [fill=black] (0.75,0.0) circle [radius = 0.6pt];
\draw [fill=black] (0.7,-0.3) circle [radius = 0.6pt];
\draw[color=black,decorate,decoration={gluon, amplitude=1.2pt,
    segment length=1.8pt, aspect=0.6}] (-0.6,0.6) -- (0.3,0.6);
    \draw (-0.6,0.6) -- (0.3,0.6);
\draw [red,fill=red] (-0.5,0.6) circle [radius=0.8pt];
\draw [red,fill=red] (-0.5,-0.5) circle [radius=0.8pt];
\draw [red,decorate glaubr] (-0.5,0.6) -- (-0.5,-0.5);
\draw [red,fill=red] (-0.325,0.6) circle [radius=0.8pt];
\draw [red,fill=red] (-0.325,-0.35) circle [radius=0.8pt];
\draw [red,decorate glaubr] (-0.325,0.6) -- (-0.325,-0.35);
 \draw [-{Latex[length=3pt]}] (-0.6,0.2)--(-0.6,0.0)node[below, xshift=-2, scale=0.4]{$\tiny{ l-k}$};
 \draw [-{Latex[length=3pt]}] (-0.4,0.2)--(-0.4,0.0)node[below, scale=0.4]{$\tiny{ k}$};
 \draw [-{Latex[length=3pt]}] (-0.8,1) node[above,  scale=0.5]{$\ccO$}--(-0.6,0.8);
 \draw [-{Latex[length=3pt]}] (-0.8,-1) node[below,  scale=0.5]{$\ccTH$}--(-0.6,-0.8);
 \draw [-{Latex[length=3pt]}] (0.1,0.7)--(0.4,0.7) node[above, scale=0.5]{$\ccT$};
\end{tikzpicture} \\
& =  4  g_s^5  
\, (\TT_\ccT^b   \TT_\ccT^c ) ( \TT_\ccTH^c \TT_\ccTH^b  ) \, \TT_\ccO \, 
 \overline{ \cM}^{\zero}  
\nn \\
& \times \int  \frac{d^{d} \ell}{(2 \pi)^d} \, \frac{d^{d} k}{(2 \pi)^d}\,  |2 (\ell_z -   k_z) |^{-\eta}  |2 k_z|^{-\eta} \,
  N^\mu  ( p_\ccO, p_\ccT, \ell_\perp ) \pol_{ \mu} (p_\ccT) \,   \frac{1}{  (\vec{\ell}_\perp- \vec{k}_\perp)^2 }  \,  
 \frac{1}{ \vec{k}_\perp^2  }  \nn \\  
& \times  
 \frac{ 1 }{  k^- - \delta'_\ccT + i\fme} 
  \frac{ -1 }{ (\ell^+ -  k^+ ) - \delta'_\ccTH  + i\fme}  
  \frac{ 1}{ \ell^- - \delta_\ccT + i\fme} 
  \frac{ -1 }{ \ell^+  - \delta_\ccTH  + i\fme}  
   \,\frac{ 1  }{  -\ell^- - \delta_{\ccO} + i\fme }   \\
 &= 
 4  g_s^5  
\, (\TT_\ccT^b   \TT_\ccT^c ) ( \TT_\ccTH^c \TT_\ccTH^b  ) \, \TT_\ccO \, 
 \overline{ \cM}^{\zero}  
 \nn \\
 & \times \int  \frac{d^{d-2} \ell_\perp }{(2 \pi)^{d-2}}  \frac{d^{d-2} k_\perp}{ (2 \pi)^{d-2}} 
  N^\mu ( p_\ccO, p_\ccT, \ell_\perp ) \pol_{ \mu} (p_\ccT)\,   \frac{1}{ ( \vec{\ell}_\perp- \vec{k}_\perp)^2 }  \,  
 \frac{1}{ \vec{k}_\perp^2  }\,  \frac{ 1  }{  \delta_\ccT  - \delta_\ccO  }     \nn \\  
 & \times  
  \int \frac{d \ell^z}{ 2 \pi} \, \frac{d k^z}{2 \pi}    |2 (\ell_z -   k_z) |^{-\eta}  |2 k_z|^{-\eta} \, 
 \frac{ 1 }{ 2 (\ell^z-k^z)   - \delta_\ccO  - \delta'_\ccT- \delta'_\ccTH +  i\fme} \, 
 \frac{ 1 }{  2 \ell^z   - \delta_\ccO - \delta_\ccTH + i\fme}  
 \end{align}
where 
\begin{align} \label{deltaOTTH}
&  \delta_{\ccO} = - \frac{  (\vec{p}_{\ccT,\perp} +\vec{\ell}_\perp)^2 }{ Q- p_\ccT^{+}} -p_\ccT^-, \quad     \delta_{\ccT} =  \frac{  (\vec{p}_{\ccT,\perp} +\vec{\ell}_\perp)^2 }{ p_\ccT^{+}} -p_\ccT^-  ,   \quad  \delta'_\ccT= - \frac{  ( \vec{p}_{\ccT \perp} + \vec{k}_\perp )^2 }{ Q- p_\ccT^{+}}
  -p_\ccT^- , \nn \\ &
 \delta_\ccTH= \frac{ \vec{\ell}_\perp^2}{ Q_\ccTH } ,  \quad  \delta'_\ccTH= \frac{ (\vec{\ell}_\perp - \vec{k}_\perp)^2}{ Q_\ccTH } ,
\quad  N^\mu(p_\ccO, p_\ccT, l) =    \frac{n\slash}{2}    \Big[  - \frac{2 (p_{\ccT,\perp} + \ell_\perp)^\mu}{ p_\ccT^+ }  + \frac{( p_\ccT\slash_{ \perp} + l \slash_\perp) \gamma_{\perp}^{\mu}}{ -Q + p_\ccT^+ } \Big] \frac { \bar{n}\slash }{ 2 } v_n(p_\ccO)
\end{align} 
The $\ell^z$ and $k^z$ integrals can be conveniently carried out in position space. 
After Fourier transforming the light-cone propagators, the integral becomes integrals of light-cone coordinates $x$ and $y$,
\begin{align}
& \text{\Fig{GBox2}(c) }
 = 4  g_s^5  
\, (\TT_\ccT^b   \TT_\ccT^c ) ( \TT_\ccTH^c \TT_\ccTH^b  ) \, \TT_\ccO \, 
 \overline{ \cM}^{\zero} 
  \left[  - \frac{2 \pol_{\perp, \mu}}{  p_\ccT^+  }  -  \frac{ \gamma_{\perp, \mu}  \pol\slash_{\perp} }{ Q - p_\ccT^+ }  \right]   v_n (p_\ccO)  \label{PBoxPosition}   \\
&  \times   \frac{p_\ccT^{+} (Q - p_\ccT^+ )}{ Q }
  \int  \frac{d^{d-2} \ell_\perp}{(2 \pi)^{d-2}} \frac{d^{d-2} k_\perp}{ (2 \pi)^{d-2}}
   \frac{   p_{\ccT,\perp}^\mu + \ell_\perp^\mu  }{  ( \vec{p}_{\ccT,\perp} + \vec{\ell}_\perp )^2  } 
     \frac{1}{  ( \vec{\ell}_\perp- \vec{k}_\perp)^2 }  
    \frac{1}{\vec{k}_\perp^2}  \nn \\
&  \times \frac{1}{4} \left(  \kappa_{\eta} \frac\eta2  \right)^2 \int dx dy \, \theta(x-y)  \theta(y) \, \frac{ 1}{|x|^{1+\eta}} \frac{1}{ |y|^{1+\eta} }  e^{ -i (x -y )( \delta'_\ccTH+ \delta'_\ccT + \delta_{\ccO})/2 -i y ( \delta_\ccTH+ \delta_\ccO)/2   } \nn \\
&=  (i)^2 g_s^5  
\, (\TT_\ccT^b   \TT_\ccT^c ) ( \TT_\ccTH^c \TT_\ccTH^b  ) \, \TT_\ccO \, 
 \overline{ \cM}^{\zero}  
  \left[  - \frac{2 \pol_{2\perp, \mu}}{  p_\ccT^+  }  -  \frac{ \gamma_{\perp, \mu}  \pol\slash_{2\perp} }{ Q - p_\ccT^+ }  \right]   v_n (p_\ccO)     \\
&  \times   \frac{ p_\ccT^{+} ( Q - p_\ccT^+ )}{ Q }
 \frac{d^{d-2} \ell_\perp}{(2 \pi)^{d-2}} \frac{d^{d-2} k_\perp}{ (2 \pi)^{d-2}}
   \frac{   p_{\ccT,\perp}^\mu + \ell_\perp^\mu  }{  ( \vec{p}_{\ccT,\perp} + \vec{\ell}_\perp )^2  } 
     \frac{1}{  ( \vec{\ell}_\perp- \vec{k}_\perp)^2 }  
    \frac{1}{\vec{k}_\perp^2}  \,  \times \frac{1}{4\times 2!}   \left(   1+ \cO(\eta) \right)    \nn
\end{align}
where $\kappa_\eta = 2^{-\eta} \Gamma(1 - \eta) \sin(\pi \eta/2)/(\pi \eta/2) = 1 + \cO ( \eta )$.
The effective diagram with the $\eta-$regulator 
preserves the physical property of Glauber interactions: they are instantaneous Coulomb interactions that are ordered in time.  
The $\theta$-functions in \Eq{PBoxPosition} guarantee that the 
Glauber exchanges take place at light-cone time $-x< -y<0$, both earlier than the hard interaction.  
Time ordering between the two Glaubers produces a $1/2!$ symmetry factor.

The $\ell_\perp, k_\perp$ integral  contains an $1/\pol^2$ divergence,
\begin{align}
& \mu^{2(4-d)}\int \frac{d^{d-2} \ell_\perp}{(2 \pi)^{d-2}} \frac{d^{d-2} k_\perp}{ (2 \pi)^{d-2}}
   \frac{   p_{\ccT,\perp}^\mu + \ell_\perp^\mu  }{  (\vec{p}_{\ccT,\perp} + \vec{\ell}_\perp )^2  }
     \frac{1}{  (\vec{\ell}_\perp- \vec{k}_\perp)^2 }  
    \frac{1}{\vec{k}_\perp^2}     \nn  \\
 & = \left(  \frac{1}{4 \pi}  \right)^2  
 \frac{  p_{\ccT,\perp}^\mu  }{ \vec{p}_{\ccT,\perp}^2} \left(\frac{  4 \pi \mu^2 }{ \vec{p}_{\ccT,\perp}^2} \right)^{2 \ep}    
 [\Gamma(- \ep)]^2 \frac{  \Gamma(1-\ep) \Gamma(1 + 2 \ep)}{ \Gamma(1- 3\ep)}  \nn  \\
 & = \left( \frac{1}{4 \pi} \right)^2    \frac{  p_{\ccT,\perp}^\mu  }{ \vec{p}_{\ccT,\perp}^2}  \, 
 \left(\frac{  4 \pi e^{-\gamma_{E}} \mu^2 }{ \vec{p}_{\ccT,\perp}^2} \right)^{2\ep}   \left( \frac{1}{\ep^2} - \frac{\pi^2}{6} + \cO(\ep)  \right)
\end{align}

As expected, the two-loop Glauber diagram has the same spin structure as the tree-level diagram, and thus proportional to the $\Sp^\zero$,
\begin{align}
\text{\Fig{GBox2}(c)}
 &= \frac{1}{2!} \Big( \frac{ i g_s^2}{2}  \Big)^2 g_s  
\,   (\TT_\ccT^b   \TT_\ccT^c ) ( \TT_\ccTH^c \TT_\ccTH^b  ) \, \TT_\ccO \, 
 \overline{ \cM}^{\zero} 
  \left[  - \frac{2 \pol_{\perp, \mu}}{  p_\ccT^+  }  -  \frac{ \gamma_{\perp, \mu}  \pol\slash_{\perp} }{ Q - p_\ccT^+ }  \right]   v_n (p_\ccO)    \nn  \\
&  \times   \frac{ p_\ccT^{+} (Q - p_\ccT^+ )}{ Q } \frac{  p_{\ccT,\perp}^\mu  }{ \vec{p}_{\ccT,\perp}^2}  \,
\left( \frac{1}{4 \pi} \right)^2   \left(\frac{  4 \pi \mu^2 }{ \vec{p}_{\ccT,\perp}^2} \right)^{2 \ep}    
 [\Gamma(- \ep)]^2 \frac{  \Gamma(1-\ep) \Gamma(1 + 2 \ep)}{ \Gamma(1- 3\ep)} \nn\\
 & =  (\TT_\ccT^b   \TT_\ccT^c ) ( \TT_\ccTH^c \TT_\ccTH^b  ) \, 
\Sp^{\zero} \, \overline{ \cM}^{\zero}  \\
& \times \frac{1}{2!} \Big( \frac{ \alpha_s}{2 \pi}  \Big)^2 
\left( i \pi \right)^2   \left(\frac{  4 \pi \mu^2 }{ \vec{p}_{\ccT,\perp}^2} \right)^{2 \ep}    
 [\Gamma(- \ep)]^2 \frac{  \Gamma(1-\ep) \Gamma(1 + 2 \ep)}{ \Gamma(1- 3\ep)} \nn
\end{align}

\vspace{12pt}

Next we consider  diagrams with two Glaubers gluons exchanged between  $p_\ccT^\mu$ and two  non-collinear partons  $p_\cci^\mu$  and  $p_\ccj^\mu$, $ ( \cci, \ccj  = 3 ,\cdots ,m)  $.  
In these diagrams we insert Glauber operators $\cO_{n_\ccO} \frac{1}{\cP_\perp^2} \cO_{n_\cci} $ and $\cO_{n_\ccO} \frac{1}{\cP_\perp^2} \cO_{n_\ccj} $, where $ n_\cci^\mu$ and $n_\ccj^\mu$
 are not generically related to $n_\ccO^\mu$. 
Stictly speaking, the transverse label momentum $\cP_\perp$ should be replaced by 
$\cP^\perp_{\ccO \ccj }$, where  $p_{\perp,\ccO \ccj}^\mu$ to extract the components of  a 4-vector $p^\mu$ in the plane transverse to $n_\ccO^\mu$ and $n_\ccj^\mu$ in Minkowski space. 
Take $  1\r    [ \ccj  $ and $  1]  \l \ccj  $   to be an orthogonal basis in the transverse plane with respect to $n_\ccO^\mu$ and $n_\ccj^\mu$. 
Basis with different choice of  light-cone directions  can be related through the following equations
\begin{align}
\frac{ 1\rangle [ j}{[j1]} &=   1 \rangle [1   \, \frac{ [ji]}{ [1i]  [j1]} + \frac{ 1\rangle [i  }{[i1] } \nn \\
\frac{ j \rangle [ 1}{ \l 1 j \r } &=  1 \rangle [1  \, \frac{ \langle ij \rangle}{    \langle i1 \rangle  \langle 1j \rangle} + \frac{ i \rangle [1  }{ \langle 1i \rangle }
\end{align}
Projecting a $4-$vector $p^\mu$ onto transverse direction $1\rangle [ j$ and $ j \rangle [ 1$, we have
\begin{align} 
p^{\perp,\mu}_{\ccO \ccj } = \frac{n_1\cdot p }{ n_\ccO \cdot n_\ccj } (n_\ccj)^{\perp,\mu}_{\ccO \cci}  + p^{\perp,\mu}_{\ccO \cci }
\end{align} 
If $p^\mu$ has collinear or Glauber scaling, $(n_\ccO \cdot p) \sim \lambda^2$ while $p^\perp_{\ccO \cci } \sim \lambda$, 
 then at leading power we can drop $(n_\ccO \cdot p)  $ with respect to $p_\perp$. Thus,
 Thus,
 \begin{align} \label{LPPerp}
p^{\perp,\mu}_{\ccO \cci } \LPeq  p^{\perp,\mu}_{\ccO \ccj },  \quad (p^\mu \text{  Glauber  or collinear to }n_\ccO^\mu) 
\end{align}
The operator $\cP_\perp^2$ sandwiched between $\cO_{n_\cci} $ and $\cO_{n_\ccj}$ 
will pull out the virtuality of the exchanged Glauber gluon,  which takes the same form no matter  what  light-cone coordinates we choose.

We will show the explicit calculation of \Fig{GBox3}(a) and \Fig{GBox3}(c) in the following. 
In \Fig{GBox3}(a), one Glauber connecting parton $p_\ccT$ and an incoming parton $p_\ccTH$ has virtual momentum $\ell-k$, the other Glauber with virtual  momentum $k$ connects $p_\ccT$ and outgoing parton $p_\ccj$. 
As we argued above,  at leading power  both Glaubers  
can be treated in the same way as those exchanged between back-to-back jets. 
 The integrand is the following, where the propagators still depend linearly on 
 the light-cone components of loop momenta, 
\begin{align} 
 & \text{\Fig{GBox3}}(a)  = 
\begin{tikzpicture}[baseline={([yshift=-.5ex]current bounding box.center)},scale=1.2]
\draw (-1,-1) -- (0,0);
\draw (-1,1) -- (0,0);
\draw (0,0) -- (0.8,1);
\draw (0,0) -- (0.8,-1);
\draw [fill=\blobcolor] (0,0) circle [radius = 0.3];
\draw [fill=black] (0.9,0.3) circle [radius = 0.6pt];
\draw [fill=black] (0.95,0.0) circle [radius = 0.6pt];
\draw [fill=black] (0.9,-0.3) circle [radius = 0.6pt];
\draw[color=black,decorate,decoration={gluon, amplitude=1.2pt,
    segment length=1.8pt, aspect=0.6}] (-0.6,0.6) -- (0.3,0.6);
    \draw (-0.6,0.6) -- (0.3,0.6);
\draw [red,fill=red] (0.0,0.6) circle [radius=0.8pt];
\draw [red,fill=red] (0.5,-0.625) circle [radius=0.8pt];
\draw [red, decorate glaubr] (0.0,0.6) to [bend left=20] (0.5,-0.625);
\draw [red,fill=red] (-0.3,0.6) circle [radius=0.8pt];
\draw [red,fill=red] (-0.5,-0.5) circle [radius=0.8pt];
\draw [red, decorate glaubr] (-0.3,0.6) to [bend right=0] (-0.5,-0.5);
\end{tikzpicture}\\
&=  4  g_s^5  
\, (\TT_\ccT^b   \TT_\ccT^c ) \TT_\ccTH^c   \TT_\ccj^b \, \TT_\ccO \, 
 \overline{ \cM}^{\zero}   
\nn \\
& \int  \frac{d^{d} \ell}{(2 \pi)^d} \, \frac{d^{d} k}{(2 \pi)^d}\,  |n_1 \cdot (\ell-k) - n_2 \cdot (\ell-k) |^{-\eta}  |n_1 \cdot k - n_j \cdot k |^{-\eta} \,
  N^\mu ( p_\ccO, p_\ccT , \ell_\perp ) \pol_\mu (p_\ccT) \,   \frac{1}{  (\vec{\ell}_\perp- \vec{k}_\perp )^2 }  \, 
 \frac{1}{ \vec{k}_\perp^2  }\,   \nn \\  
& \times  \kappa_{\ccO \ccTH}  \kappa_{\ccO\ccj} \, 
 \frac{ 1 }{  n_\ccO \cdot k  - \delta'_\ccT + i\fme} 
  \frac{ -1 }{ n_\ccTH \cdot (\ell-k)  - \delta'_\ccTH  + i\fme}  
    \frac{ 1 }{  - (n_\ccj \cdot k ) - \delta_\ccj  + i\fme}  
  \frac{ 1}{ n_\ccO \cdot \ell  - \delta_\ccT + i\fme} 
   \,\frac{ 1  }{  - n_\ccO \cdot \ell  - \delta_\ccO + i\fme }   \label{fig4a1} 
 \end{align} 
Here, $\delta_\ccO, \delta_\ccT, \delta_\ccT' ,  \delta'_\ccTH$ are  the same as those defined in \Eq{deltaOTTH}. 
We also define $\delta_{\ccj}= \vec{k}_\perp^2/Q_\ccj$ and $\delta'_{\ccj}= ( \vec{\ell}_\perp- \vec{k}_\perp)^2/Q_\ccj$, with 
$p_\ccj^\mu \equiv  \frac12 Q_\ccj n_\ccj^\mu$,  for any outgoing non-collinear parton $\ccj$.   
 In \Eq{fig4a1}, $\kappa_{\cci \ccj}  \equiv  (n_\cci \cdot n_\ccj)/2 $, which is equal to 1 only for back-to-back directions.  
 These factors are inserted at the operator level to guarantee the RPI-III invariance of the $\text{SCET}_\text{G}$ Lagrangian\cite{Rothstein:2016bsq}.
Let  us choose integration variables to be
\be 
 \ell^- = n_\ccO \cdot \ell,\quad  k^- = n_\ccO \cdot k,\quad  k_1^z = \frac{n_\ccTH \cdot (\ell-k) -    n_\ccO \cdot (\ell-k) }{2}, \quad 
 k_2^z =   \frac{n_\ccj \cdot k -    n_\ccO \cdot k }{2}  
 \ee
  so that 
 \begin{align} 
  \int   \dbar^{d} \ell \, \dbar^{d} k\,   \rightarrow    \frac{1}{ \kappa_{\ccO \ccTH} \kappa_{\ccO\ccj}} \int   
  \dbar \ell^- \,  \dbar k^- \,  \dbar k_1^z  \, \dbar k_2^z\,  \dbar^{d-2} \ell_\perp  \dbar^{d-2} k_\perp 
 \end{align} 
 After change of integration variables, the $\kappa_{\cci \ccj}$ terms drops out and the integrand looks independent of the direction of $p_\ccj$. 
After integrating over $\ell^-$ and $k^-$, 
each collinear propagator depends linearly on $k_1^z, k_2^z$. Hence we 
can easily transform into position space 
\begin{align} 
\text{\Fig{GBox3}}(a) &= 
      4  g_s^5 
\,  (\TT_\ccT^b   \TT_\ccT^c )  (- \TT_\ccTH^c )  \,  \TT_\ccj^b  \,  \TT_\ccO  \,  
 \overline{ \cM}^{\zero}  
 \nn \\
 & \times \int  \frac{d^{d-2} \ell_\perp}{(2 \pi)^{d-2}}  \frac{d^{d-2} k_\perp}{(2 \pi)^{d-2}}
  N^\mu ( p_\ccO, p_\ccT, \ell_\perp ) \pol_\mu (p_\ccT)\,  \frac{1}{  (\vec{\ell}_\perp- \vec{k}_\perp )^2 }  \, 
 \frac{1}{ \vec{k}_\perp^2  }\,    \frac{ 1  }{  \delta_\ccT  - \delta_\ccO  }     \nn \\  
 & \times  
  \int \frac{d k_1^z}{2 \pi} \, \frac{d k_2^z}{2 \pi}    |2   k_1^z |^{-\eta}  |2 k_2^z|^{-\eta} \, 
 \frac{ 1 }{ 2 k_1^z  - \delta_\ccO - \delta'_\ccT - \delta'_\ccTH+  i\fme} \, 
 \frac{ 1 }{  - 2 k_2^z  -  \delta_\ccT'  - \delta_\ccj + i\fme}   \\
 & =     4  g_s^5 
\,  (\TT_\ccT^b   \TT_\ccT^c )  (- \TT_\ccTH^c )  \,  \TT_\ccj^b  \,  \TT_\ccO  \,  
 \overline{ \cM}^{\zero} 
  \nn \\
 & \times   \frac{ p_\ccT^{+} (p_\ccO^+ - p_\ccT^+ )}{ p_\ccO^{+} }  \int  \frac{d^{d-2} \ell_\perp}{(2 \pi)^{d-2}}  \frac{d^{d-2} k_\perp}{(2 \pi)^{d-2}}
  N^\mu ( p_\ccO, p_\ccT, \ell_\perp ) \pol_\mu (p_\ccT)\,  \frac{1}{  (\vec{\ell}_\perp- \vec{k}_\perp )^2 }  \, 
 \frac{1}{ \vec{k}_\perp^2  }\,    \frac{ 1  }{  (\vec{\ell}_\perp  + \vec{p}_{\ccT,\perp})^2  }   \nn \\ 
 &\times \frac{1}{4} \left(  \kappa_{\eta} \frac\eta2  \right)^2 \int dx dy \, \theta(x) \theta(y) \, \frac{ 1}{|x|^{1+\eta}} \frac{1}{ |y|^{1+\eta} }  e^{ -i x ( \delta'_\ccTH+ \delta'_\ccT+ \delta_{\ccO})/2 -i y ( \delta_\ccj+ \delta_\ccT')/2   }  \\ 
 & =(i)^2 g_s^5 
\,  (\TT_\ccT^b   \TT_\ccT^c )  (- \TT_\ccTH^c )  \,  \TT_\ccj^b  \,  \TT_\ccO  \,  
 \overline{ \cM}^{\zero}   
 \nn \\
&  \times \frac{ p_\ccT^{+} (Q - p_\ccT^+ )}{ Q }
 \int  \frac{d^{d-2} \ell_\perp}{(2 \pi)^{d-2}}  \frac{d^{d-2} k_\perp}{(2 \pi)^{d-2}}
   \frac{    N^\mu( p_\ccO, p_\ccT, \ell_\perp )   }{  (\vec{p}_{\ccT,\perp} + \vec{\ell}_\perp )^2  } 
     \frac{1}{  ( \vec{\ell}_\perp- \vec{k}_\perp)^2 }  
    \frac{1}{\vec{k}_\perp^2}    \left( \frac{1}{2} \right)^2 \left(   1+ \cO(\eta) \right)   
\end{align} 
The position-space picture describes Glauber exchange  before and after the hard interaction, taking place at  light-cone time $-x$ and $y$ with $-x <0 < y$. 
The integration region is the  quarter $(x,y)-$plane.
At $\cO(\eta^0)$, the $d x$ and $dy$ integrals are symmetric, the result being twice of the parallel box diagrams in \Fig{GBox2}, 
 \begin{align} 
\text{\Fig{GBox3}}(a) &= - (\TT_\ccT \cdot \TT_\ccTH) (\TT_\ccT \cdot \TT_\ccj) \, \Sp^{\zero} \, \overline{ \cM}^{\zero} \nn  \\
& \times \Big( \frac{ \alpha_s}{2 \pi}  \Big)^2 
\left( i \pi \right)^2   \left(\frac{  4 \pi \mu^2 }{ \vec{p}_{\ccT,\perp}^2} \right)^{2 \ep}    
 [\Gamma(- \ep)]^2 \frac{  \Gamma(1-\ep) \Gamma(1 + 2 \ep)}{ \Gamma(1- 3\ep)} 
 \end{align}

\Fig{GBox3}(c)  has two Glaubers connecting $p_\ccT$ with  two outgoing partons $p_\ccj$ and $p_\cck$, with virtual momenta $k$ and $\ell- k$, respectively.  
Choose integration variables to be
 \be
 \ell^- = n_\ccO \cdot \ell,  \quad k^- = n_\ccO \cdot k,  \quad k_1^z = \frac{n_\ccj \cdot (\ell-k) -    n_\ccO \cdot (\ell-k) }{2},  \quad  k_2^z =   \frac{n_\cck \cdot k -    n_\ccO \cdot k }{2}  
 \ee 
 then 
\begin{align} 
& \text{\Fig{GBox3}}(c) = 
\begin{tikzpicture}[baseline={([yshift=-.5ex]current bounding box.center)},scale=1.2]
\draw (-1,-1) -- (0,0);
\draw (-1,1) -- (0,0);
\draw (0,0) -- (0.8,1);
\draw (0,0) -- (0.8,-1);
\draw [fill=\blobcolor] (0,0) circle [radius = 0.3];
\draw [fill=black] (0.9,0.3) circle [radius = 0.6pt];
\draw [fill=black] (0.95,0.0) circle [radius = 0.6pt];
\draw [fill=black] (0.9,-0.3) circle [radius = 0.6pt];
\draw[color=black,decorate,decoration={gluon, amplitude=1.2pt,
    segment length=1.8pt, aspect=0.6}] (-0.6,0.6) -- (0.3,0.6);
    \draw (-0.6,0.6) -- (0.3,0.6);
\draw [red,fill=red] (0.0,0.6) circle [radius=0.8pt];
\draw [red,fill=red] (0.5,-0.625) circle [radius=0.8pt];
\draw [red,decorate glaubr] (0.0,0.6) to [bend left=10] (0.5,-0.625);
\draw [red,fill=red] (-0.3,0.6) circle [radius=0.8pt];
\draw [red,fill=red] (0.325,0.4) circle [radius=0.8pt];
\draw [red,decorate glaubr] (-0.3,0.6) to  (0.325,0.4);
\end{tikzpicture}\\
 & = 4  g_s^5 
\,  (\TT_\ccT^b   \TT_\ccT^c ) \, \TT_\ccj^b \,   \TT_\cck^c  \,  \TT_\ccO \, 
 \overline{ \cM}^{\zero}   
 \nn \\
 & \times \int  \frac{d^{d-2} \ell_\perp}{(2 \pi)^{d-2}}  \frac{d^{d-2} k_\perp}{(2 \pi)^{d-2}} \, 
  \frac{ d \ell^-}{2 \pi} \,  \frac{ d k^-}{2 \pi} \,  \frac{d k_1^z}{2 \pi}  \, \frac{d  k_2^z}{2 \pi}\,    |2 k_1^z |^{-\eta}  |2 k_2^z|^{-\eta} \,  
 N^\mu ( p_\ccO, p_\ccT, \ell_\perp ) \pol_\mu (p_\ccT) \,   \frac{1}{  ( \vec{\ell}_\perp- \vec{k}_\perp)^2 }  \, 
 \frac{1}{ \vec{k}_\perp^2  }\,   \nn  \\  
 & \times  
  \frac{ 1 }{ - \ell^- + k^- - 2 k_1^z  - \delta'_\ccj  + i\fme}  
    \frac{ 1 }{  - k^- - 2 k_2^z - \delta_\cck  + i\fme}  
    \frac{ 1 }{  k^-  - \delta'_\ccT + i\fme} 
  \frac{ 1}{  \ell^-  - \delta_\ccT + i\fme} 
   \frac{ 1  }{  - \ell^-  - \delta_{\ccO} + i\fme }    \\
 & =    4  g_s^5 
\,  (\TT_\ccT^b   \TT_\ccT^c ) \, \TT_\ccj^b \,   \TT_\cck^c  \,  \TT_\ccO \, 
 \overline{ \cM}^{\zero}   
 \nn \\
 & \times   \frac{ p_\ccT^{+} (Q - p_\ccT^+ )}{ Q } \int  \frac{d^{d-2} \ell_\perp}{(2 \pi)^{d-2}}  \frac{d^{d-2} k_\perp}{(2 \pi)^{d-2}} \, 
  N^\mu ( p_\ccO, p_\ccT, \ell_\perp ) \pol_\mu (p_\ccT) \,   \frac{1}{  (\vec{\ell}_\perp- \vec{k}_\perp)^2 }  \,  
 \frac{1}{ \vec{k}_\perp^2  }\,  \frac{ 1  }{  (\vec{\ell}_\perp  + \vec{p}_{\ccT,\perp})^2  }   \nn \\ 
 &\times \frac{1}{4} \left(  \kappa_{\eta} \frac\eta2  \right)^2 \int dx dy \, \theta(x) \theta(y-x) \, \frac{ 1}{|x|^{1+\eta}} \frac{1}{ |y|^{1+\eta} }  e^{ -i x ( \delta'_\ccj+ \delta_\ccT- \delta_{\ccT}' )/2 -i (y-x) ( \delta_\cck+ \delta_\ccT')/2   } \nn \\ 
 & =(i)^2 g_s^5 
\,  (\TT_\ccT^b   \TT_\ccT^c ) \, \TT_\ccj^c \,   \TT_\cck^b  \,  \TT_\ccO \, 
 \overline{ \cM}^{\zero}   
 \nn \\
&  \times \frac{ p_\ccT^{+} (Q - p_\ccT^+ )}{ Q }
 \int  \frac{d^{d-2} \ell_\perp}{(2 \pi)^{d-2}}  \frac{d^{d-2} k_\perp}{(2 \pi)^{d-2}} \, 
   \frac{    N^\mu ( p_\ccO, p_\ccT, \ell_\perp )   }{  (\vec{p}_{\ccT,\perp} + \vec{\ell}_\perp )^2  } 
     \frac{1}{  ( \vec{\ell}_\perp- \vec{k}_\perp)^2 }  
    \frac{1}{\vec{k}_\perp^2}  \times \frac{1}{2!}  \left( \frac{1}{2} \right)^2 \left(   1+ \cO(\eta) \right)   
\end{align} 
Here the $\theta$-functions
ensure that both Glaubers are produced after the hard interaction with a particular ordering.  
The time-ordering between the two Glaubers gives a $\frac{1}{2!}$ symmetry factor.
 \begin{align} 
\text{\Fig{GBox3}}(c) &=  \frac{1}{2!} (\TT_\ccT \cdot \TT_\ccj) (\TT_\ccT \cdot \TT_\cck) \, \Sp^{\zero} \, \overline{ \cM}^{\zero} \nn  \\
& \times \Big( \frac{ \alpha_s}{2 \pi}  \Big)^2 
\left( i \pi \right)^2   \left(\frac{  4 \pi \mu^2 }{ \vec{p}_{\ccT,\perp}^2} \right)^{2 \ep}    
 [\Gamma(- \ep)]^2 \frac{  \Gamma(1-\ep) \Gamma(1 + 2 \ep)}{ \Gamma(1- 3\ep)} 
 \end{align}

\bibliography{split}

\bibliographystyle{utphys}

\end{fmffile}

\end{document}